\documentclass[12pt]{article}
\usepackage[utf8]{inputenc}

\usepackage{comment}

\usepackage{mathtools, amsthm,amsmath,amssymb,amsfonts,bm,fancyhdr}  
\usepackage[margin=1in]{geometry}
\usepackage{tikz}
\usepackage{color}
\usepackage[OT1]{fontenc}
\usepackage{threeparttable}
\usepackage{hyperref, cleveref}
\usepackage{multirow}
\usepackage{hyperref}
\usepackage[section]{placeins}
\usepackage{soul}
\usepackage{float}
\usepackage{lscape}
\usepackage{bbm}
\usepackage{rotating}
\usepackage{makecell}
\usepackage{tabularx}
\usepackage{caption}
\usepackage[round,authoryear, ]{natbib}
\usepackage{graphicx,booktabs,subcaption, adjustbox}
\graphicspath{{./figures/}}
\usepackage{arydshln}
\usepackage{algorithm} 
\usepackage{algpseudocode} 
\usepackage{setspace}

\newcommand{\by}{ \bm{y}}
\newcommand{\bY}{ \bm{Y}}
\newcommand{\bx}{ \bm{x}}
\newcommand{\bX}{ \bm{X}}

\newcommand{\bzeros}{ \bm{0}}
\newcommand{\bOmega}{ \bm{\Omega}}
\newcommand{\bSigma}{ \bm{\Sigma}}

\newcommand{\bb}{\bm{b}}
\newcommand{\bB}{\bm{B}}

\title{Bayesian Covariate-Dependent Graph Learning with a Dual Group Spike-and-Slab Prior}

\author{Zijian Zeng, Meng Li and Marina Vannucci}

\date{%
    Department of Statistics, Rice University, Houston, Texas, USA\\[2ex]%
    \today
}

\begin{document}

\maketitle

\begin{abstract}
Covariate-dependent graph learning has gained increasing interest in the graphical modeling literature for the analysis of heterogeneous data. This task, however, poses challenges to modeling, computational efficiency, and interpretability. The
parameter of interest can be naturally represented as a three-dimensional array with elements that can be grouped according to two directions, corresponding to node level and covariate level, respectively. In this article, we propose a novel dual group spike-and-slab prior that enables multi-level selection at covariate-level and node-level, as well as individual (local) level sparsity. We introduce a nested strategy with specific choices to address distinct challenges posed by the various grouping directions. For posterior inference, we develop a full Gibbs sampler for all parameters, which mitigates the difficulties of parameter tuning often encountered in high-dimensional graphical models and facilitates routine implementation. Through simulation studies, we demonstrate that the proposed model outperforms existing methods in its accuracy of graph recovery. We show the practical utility of our model via an application to microbiome data where we seek to better understand the interactions among microbes as well as how these are affected by relevant covariates.
\end{abstract}


\section{Introduction}
\label{sec:intro}
Gaussian graphical models have been applied in a wide variety of fields to recover the dependence structure among data \citep{Lauritzen1996,Maathuis2018}. The idea dates back to \cite{Dempster1972}, who proposed the covariance selection method that estimates conditional independencies based on the inverse covariance matrix (a.k.a., precision matrix or concentration matrix) by linking the absence of an edge in an undirected graph to a zero entry in the precision matrix. Expanding upon this idea, \cite{Meinshausen2006} showed that neighborhood selection for each node in the graph is equivalent to perform variable selection in a Gaussian linear model, turning the edge detection problem into variable selection for independent regressions. This approach has inspired numerous studies with a focus on using different selection methods to recover edges within a graph \citep{Peng2009,Leday2017,Liu2017}. 

Recent work has demonstrated the value of incorporating covariates in the modeling of subject-specific graphs via Gaussian graphical regression models, in particular for characterizing and discovering interactions in complex biological systems and diseases such as cancer \citep{ni2019bayesian, Zhang2022, Wang2022, niu2023covariate}. Most of the existing literature has focused on covariate-adjusted mean structures in Gaussian graphical models, with either constant graphs across subjects or group-specific graphs depending on \textit{discrete} covariates; for a comprehensive review, see~\cite{Zhang2022} and  Section 1.3 of~\cite{Wang2022}, with \cite{Osborne2022} providing a recent example. In this article, instead, we focus on modeling the dependence of the precision matrix on covariates, a framework referred to as \textit{precision-on-scalar} regression. This covariate-dependent graph learning task is comparatively much less studied and poses challenges to modeling, computational efficiency, and interpretability.  Partition-based Bayesian approaches to model covariate-dependent graphs are explored by~\cite{niu2023covariate}, 
while \cite{Wang2022} consider an edge regression model for undirected graphs, which estimates conditional dependencies as a function of subject-level covariates, and employs shrinkage priors.  
\cite{Zhang2022} introduce bi-level sparsity, where element- and group-wise sparsity are encouraged by lasso and group lasso, respectively. Also, \cite{ni2019bayesian, ni2022jmlr} allow the graph structure to vary with covariates and induce local sparsity by thresholding; in particular, \cite{ni2019bayesian} assume that the hierarchical ordering of the nodes is known, a prior knowledge that might not always be available in real-world applications, and also focus on the final subject-specific graph, providing limited inference about how each covariate impacts the graph.

Precision-on-scalar regression models are characterised by an ultra high-dimensional parameter space, which can be viewed according to more than one grouping direction, e.g. node or covariate. It is desirable to have both node-level and covariate-level group sparsity, in addition to individual (local) level sparsity. This simultaneous sparsity at the local level and the two group levels is crucial for interpretable graphical models, particularly in the presence of many nodes and covariates.
The majority of the existing works on heterogeneous graphs fail to model such structured sparsity, as they typically group parameters in one direction only or focus on local sparsity without grouping structure, and there is a lack of efficient estimation strategies %with easy parameter tuning 
to address the daunting computational challenges. Also, in the work of \cite{Zhang2022}, the authors use lasso and group lasso to induce covariate-level sparsity by imposing node-level sparsity. However, relying on one group level to induce the sparsity of another restricts the ability to flexibly capture interactions between the two group levels. Overall, we are not aware of any work in the Bayesian literature addressing multiple covariates with the aforementioned structured sparsity. 

Here, we introduce a novel dual-group spike-and-slab prior as a general framework to encode group sparsity at both the covariate and the node level. At the covariate level, we allow for group (global) and individual (local) sparsity. Even though this general prior is complementary to a wide range of existing priors and empowers them into dual-group variants, modeling the two grouping directions in the context of graphical models has distinct challenges. To this end, we propose to use particular choices tailored to each grouping direction, leading to a dual-group spike-and-slab prior well suited for graphical models. We complete our proposed modeling construction with full posterior sampling, that aids model interpretability.  Through simulations, we demonstrate that the proposed model outperforms the method of \cite{Zhang2022} in its accuracy of graph recovery, when the sparsity of the parameter space is introduced by the grouping at the covariate level. We also compare performances to the Bayesian sparse group selection method of \cite{Xu2015}.

As an illustration of the utility of our method, we consider an application to multivariate data arising from microbiome studies.
The human microbiome has been implicated in many diseases including colorectal cancer, inflammatory bowel disease, and immunologically mediated skin diseases. Here, we apply the proposed method to real data from the Multi-Omic Microbiome Study-Pregnancy Initiative (MOMS-PI) study, to estimate the interaction between microbes in the vagina, as well as the interplay between vaginal cytokines and microbial abundances, providing insight into mechanisms of host-microbial interaction during pregnancy. These factors influence the microbiome by introducing new organisms, changing the abundance of metabolites, or altering the pH of their environment. Identifying factors that lead to the prevalence of different microbes can improve the understanding of the importance and the function of the microbiome.  Our method identifies a large number of microbiome interactions (edges) that are simultaneously influenced by multiple cytokines. It also highlights a subnetwork of multiple microbes that belong to the same family (phylum) and that appear to be consistently detected as having covariate-dependent interactions for various cytokines, which aligns with previous findings.

The rest of the paper is organized as follows. In Section~\ref{sec:method}, we introduce the proposed prior construction and the sampling procedure. In Section~\ref{sec:sim}, we conduct simulations and compare the proposed approach with existing methods. In Section~\ref{sec:app}, we apply the proposed model to a human microbiome study. In Section \ref{sec:final} we provide some concluding remarks.

\section{Methods}
\label{sec:method}
\subsection{Gaussian Graphical Regression Models with Covariates}
\label{sec:reg}

Let $\bY = (Y^1, \ldots, Y^p)$ be a $p$-dimensional outcome vector
and $\bX = (X^1, \ldots, X^q)$ a $q$-dimensional covariate vector. We denote $N$ independent and identically distributed observations by $\by_n = \left(y^1_n, \ldots, y^p_n \right)$ and $\bx_n = \left( x^1_n, \ldots, x^q_n \right)$, for $n = 1, \ldots, N$.  For simplicity, we assume that the outcomes have been centered with zero mean. The covariate-dependent Gaussian graphical model can be written as
\begin{equation}
\label{eq:GGM}
\by_n | \bx_n \sim N_p\left( \bzeros, \left[\bOmega\left(\bx_n\right) \right] ^{-1}\right),
\end{equation}
where $\bOmega\left(\bx_n\right) = \left( \omega^{ij}\left( \bx_n\right) \right)_{i,j=1}^p$. % and covariance matrix $\bSigma\left( \bx\right) = \left[\bOmega\left(\bx\right) \right] ^{-1}$. % being symmetric and positive definite.  
Similarly to the typical covariate-free setting studied in the Gaussian graphical model literature~\citep{Lauritzen1996}, the covariate-dependent precision matrix $\bOmega\left(\bX \right)$ encodes independence for node $i$ and node $j$ given the other nodes $\bm{Y}^{-(i,j)}$ but in a covariate dependent manner as: 
$$
    \omega^{ij}\left(\bm{X}\right) = 0 \Longleftrightarrow Y^i \perp Y^j | \bm{Y}^{-(i,j)}, \bX. 
$$
This adds flexibility to modeling the dependence structure of $\bY$. %where $E$ denotes the set of edges in graph $\mathcal{G}(\bX)$.}}

Under the Gaussian assumption \eqref{eq:GGM} the elements of the precision matrix, $\omega^{ij}\left( \bx_n\right)$, are related to the coefficients in the linear regression of $y^i_n$ on the other $y^j_n$, $1 \le i \ne j \le p$ as
$$
	y^i_n = \sum_{j  \ne i}^{p} \theta^{ij}\left( \bx_n\right) y^j_n + \epsilon^i_n \quad \epsilon^i_n \sim N\left( 0, \sigma^2_i(\bx_n)\right),
$$
where $\theta^{ij}(\bx_n) = - \frac{\omega^{ij}\left( \bx_n\right)}{\omega^{ii}(\bx_n)}$, $\sigma^2_i(\bx_n) = \frac{1}{\omega^{ii}(\bx_n)}$. This model generalizes the standard treatment of Gaussian graphical models \citep{Meinshausen2006,Peng2009} to a covariate-dependent regime. 

With particular interest in learning how covariates influence graph structures, and in alignment with previous works \citep{Wang2022,Zhang2022}, we assume linear structures $\theta^{ij}(\bx_n) =  -\frac{1}{\omega^{ii}} \sum^{q}_{k=1} \omega^{ij}(x^k_n) =\sum^q_{k=1} \beta^{ij}_k x^k_n$ for covariate dependent sub-networks, with covariate-independent variance $\sigma^2_i = \frac{1}{\omega^{ii}}$. 
These structures linearly decompose the precision matrix into precision coefficients for each covariate:
%\begin{equation*}
$\bOmega(\bx_n) = \sum^{q}_{k=1} \bOmega(x^k_n) + \bOmega_0,$
%\end{equation*}
where $\left(\bOmega(x^k_n)\right)^{ij} = \omega^{ij}(x^k_n) = -\omega^{ii} \beta^{ij}_k x^k_n$ models the effect of covariate $x^k_n$ on the sub-network $\bOmega(x^k_n)$, with $\bOmega_0 = \text{diag}\left(\omega^{11}, \ldots, \omega^{pp}\right)$ the diagonal element. Correspondingly, we have the linear system
\begin{equation}
\label{eq:likelihood}
	y^i_n = \sum_{j \ne i}^p \sum^q_{k=1} \beta^{ij}_k x^k_n y^j_n + \epsilon^i_n, \quad \epsilon^i_n \sim N\left( 0, \sigma^2_i\right), \quad 1 \le i \le p, \quad n= 1, ..., N,
\end{equation}
which models the effect of the covariate $x^k_n$ on edge $(i,j)$ via the coefficient $\beta^{ij}_k$. Sparsity in the coefficients, expressed as $\beta^{ij}_k = 0$, implies $\omega^{ij}(x^k_n) = 0$, indicating that covariate $x^k_n$ has no impact on the element $(i,j)$ of the sub-network. Therefore, sparsity of the regression coefficients $\beta^{ij}_k$ induces both sparsity in the covariate-specific subnetwork $\bOmega(x^k_n)$ for a given covariate $k$ and sparsity in the node-specific connections $\omega^{ij}(\bx_n)$ for the pair of nodes $(i, j)$.

To enforce this sparsity structure, we employ a discrete spike-and-slab prior on the coefficient $\beta^{ij}_k$ and extend the prior to multi-level selection. The discrete spike-and-slab prior \citep{Mitchell1988,George1993, Brown1998} combines a ``spike", a point mass distribution at zero to induce sparsity, with a ``slab", a normal distribution for non-zero values, enabling variable selection in regression problems. This prior has been recently extended to
%utilized in both linear regression with scalar responses \citep{Brown1998, Xu2015} and 
vector-on-scalar regression by \cite{Zeng2024}. For a comprehensive review of this class of priors with their applications, we refer readers to \cite{Vannucci2021}.

In this work, we exploit the fact that the linear system in Eq.~\eqref{eq:likelihood} can be interpreted as both types of regressions from different grouping perspectives. This dual perspective enables the usage of prior structures at both levels, adapting effectively to the high-dimensional $p$-by-$(p-1)$-by-$q$ set of coefficients of the entire linear system.

\subsection{Dual group spike-and-slab prior}
\label{sec:prior}

Let us collect the coefficients $\beta^{ij}_k$ in Eq.~\eqref{eq:likelihood} into a $p$-by-$p$-by-$q$ array $\mathcal{B} = \left( \beta^{ij}_k\right),$ for $i, j = 1, \dots, p$ and $k = 1, ..., q$, with diagonal elements $\beta^{ii}_k = 0$. The elements of $\mathcal{B}$ can be grouped in different ways, i.e., as node-level and covariate-level groupings. 
Simultaneous sparsity at the two group levels, as well as locally at the individual level, can improve the interpretability and estimability of covariate-dependent graphical models, particularly in the case of many nodes and covariates~\citep{Zhang2022}. 

Here, we introduce a novel dual-group spike-and-slab prior as a general framework to encode group sparsity at both the covariate and the node level. At the covariate level, we allow for group (global) and individual (local) sparsity. We complete our proposed modeling construction with Gibbs posterior sampling, that aids model interpretability. Even though this general prior is complementary to a wide range of existing priors and empowers them into dual-group variants, the two grouping directions in the context of graphical models have distinct challenges. 
In our construction, we allow covariate-dependent directed effects between two nodes to be asymmetric; symmetrizing $\beta^{ij}_k$, if desired, can be achieved by enforcing the constraint $\beta^{ij}_k = \beta^{ji}_k$ as in \cite{Wang2022}, or via posterior summary, as commonly done in the literature~\citep{Meinshausen2006, Zhang2022}. 
We start with a conventional spike-and-slab prior of the type:
\begin{equation}
\beta^{ij}_k | \delta^{ij}_k, \sigma^{ij}_k \sim \delta^{ij}_k  N\left[ 0, \left(\sigma^{ij}_k \right)^2\right] + ( 1 - \delta^{ij}_k) \delta_0, 
\end{equation}
where $\delta^{ij}_k \in \{ 0, 1 \}$ is the overall selection indicator for a combination of nodes $(i,j)$ and covariate $k$, $\sigma^{ij}_k$ represents the prior variance of the slab distribution, and $\delta_0$ is the Dirac mass at $0$.
To encode sparsity, we decompose the selection indicator $\delta^{ij}_k$ into three parts: $\delta^{ij}_k = \delta^{ij} \times \delta_k \times \gamma^{ij}_k$, where $\delta_k$ is the covariate-level indicator, $\delta^{ij}$ is the node-level indicator, and $\gamma^{ij}_k$ represents the local-level indicator. For each $(i, j, k)$, the marginal prior on $\beta^{ij}_k$ is %This yields a class of dual-group spike-and-slab priors: 
\begin{equation}
\label{eq:GSS}
\beta^{ij}_k | \delta^{ij}, \delta_k, \gamma^{ij}_k, \sigma^{ij}_k \sim \delta^{ij} \delta_k \gamma^{ij}_k N\left[ 0, \left(\sigma^{ij}_k \right)^2\right] + \left( 1 - \delta^{ij} \delta_k \gamma^{ij}_k  \right) \delta_0.
\end{equation}
One particular challenge in this \textit{dual-group} approach is to build interpretable group structures into the prior on $\beta^{ij}_k$ \textit{jointly} across $(i, j, k)$ beyond the marginal specification in~\eqref{eq:GSS}, which not only should encode two group structures but also account for the distinct challenges posed by high-dimensional graphical models. 

The two sets of group indicators $(\delta^{ij})$ and $(\delta_k)$ are symmetric in~\eqref{eq:GSS} in that no particular order between the two groups is enforced when combined with the local-level indicator $\gamma^{ij}_k$. The spike-and-slab specification allows us to consider any sequential order of them, leading to a notion of nested decomposition of the two groups. Without loss of generality, below we focus on the sequence of first $\delta^{ij}$ then $\delta_k$, and discuss the different corresponding model structures along with our specification of each group sparsity. To this end, 
we let $\tau^{ij}_k = \delta_k \gamma^{ij}_k\sigma^{ij}_k$ and reparameterize Eq.~\eqref{eq:GSS} as
\begin{align}
    \beta^{ij}_k | \delta^{ij}, \tau^{ij}_k & \sim \delta^{ij} N\left[ 0, \left(\tau^{ij}_k \right)^2\right] + \left( 1 - \delta^{ij} \right) \delta_0, \label{eq:SSP1}
\end{align}
 which mirrors a typical spike-and-slab specification using $\delta^{ij}$ as the sole indicator and implies that $\beta^{ij}_k$ follows $N\left[ 0, \left(\sigma^{ij}_k \right)^2\right]$ if $(\delta^{ij}, \delta_k, \gamma^{ij}_k) = (1, 1, 1)$, and $\delta_0$ otherwise, as defined in the prior specified in Eq.~\eqref{eq:GSS}. We 
describe our dual-group structured prior specification below. 

\subsubsection{Node-level group sparsity: outer-layer structured prior for scalar response}

For a given pair of nodes $(i,j)$, let the coefficient vector $\bm{B}^{ij} = (\beta^{ij}_k)_{1\le k \le q}$ indicate the coefficients grouped based on the paired indices $(i,j)$. The model in Eq.~\eqref{eq:likelihood} leades to
$$
    y^i_n = \sum_{j\ne i }^p  (\bx_n^T \bB^{ij} ) y^j_n + \varepsilon^i_n, 
$$
where $\bB^{ij} = \bzeros$ implies that node $j$ has no effect on node $i$. 
The group sparsity at the node-level is described by the sparsity of vector $\bB^{i} = (\bB^{ij})^{1 \le j \le p}_{j \neq i}$ with group label $j$. One challenge in defining a prior on $\bB^{ij}$ is the need to achieve sparsity at both the group level and individual level, with the added difficulty when addressing dual group sparsity. 

Note that the model above is a \textit{high-dimensional linear regression model} with standard scalar response, for which a rich menu of group priors has been proposed, such as \cite{Stingo2011, Xu2015,Bai2022}. We propose to use the multivariate spike-and-slab prior for group sparsity in the outer layer of our model:  
\begin{equation}
    \bB^{ij}  = \text{diag}\left(\tau^{ij}_1, ..., \tau^{ij}_q \right) \bb^{ij},  \quad  1 \le i \ne j \le p,    \label{eq:CSS}
\end{equation}
where $\bb^{ij} = \left(b^{ij}_1, ..., b^{ij}_q \right)^T$ follows 
\begin{align}
     \begin{cases}
    			  \bm{b}^{ij} | \delta^{ij}   \sim  \delta^{ij} MVN\left( \bm{0}_q, \bm{I}_q \right) + \left( 1 - \delta^{ij}\right) \delta_{\bm{0}_q} , \\
    			  \delta^{ij} | \pi^{i}   \sim  \text{Bernoulli}\left( \pi^{i}\right), \quad  \pi^{i}  \sim  \text{Beta}\left( a^{i}, b^{i}\right).
    \end{cases} \label{eq:node_level} 
\end{align}
Conditional on $\tau^{ij}_k$ for $k = 1, \ldots, q$, Eqs.~\eqref{eq:CSS} and~\eqref{eq:node_level} provide node-level selection for the paired indices $(i,j)$ for all $k$. The node-level indicator $\delta^{ij}$ models the group effect of node $j$ on node $i$ through all covariates. For every $j$, if $\delta^{ij} = 0$ then effects $\bm{b}^{ij}$ will be excluded from the model, implying that node $j$ does not affect node $i$ through any of the covariates, i.e., $\bB^{ij} = \bzeros$. The parameter $\pi^i$ can be interpreted as the prior probability that $y^i$ is affected by the other nodes.

\subsubsection{Covariate-level group and local sparsity: inner-layer structured prior for multivariate response}

For a given covariate $x^k$, let the coefficient matrix $\bm{B}_k = (\beta^{ij}_k)^{1\le i,j \le p}$ indicate the coefficients grouped based on the index $k$. One challenge in modeling the sparsity of matrix $\bB_k$ is 
to achieve simultaneous group sparsity and individual sparsity in an interpretable manner. This calls for a strategy different from the treatment of node-level sparsity. To see this, the independent regression system from Eq.~\eqref{eq:likelihood} has the representation
 $$
        \by_n = \sum^q_{k=1} \left(\bB_k \by_n \right) x^k_n +\bm{\varepsilon}_n,
$$
which, unlike the preceding node-level representation, is a high-dimensional \textit{vector-on-scalar regression} model with multivariate response, where $\bm{B}_k = {\bf 0}$ implies that covariate $x^k$ has no influence on the precision matrix.
We propose to jointly model $(\delta_k, \gamma_k^{ij})$ by  
\begin{equation} \label{eq:cov.joint}
            \delta_k = I(\pi_k \ge d_k), \quad
             \gamma^{ij}_k | \pi_k  \sim   \text{Bernoulli}( \pi_k ), \quad
            \pi_k \sim  \text{Beta}( a_k, b_k). 
  % \vspace{-0.59em}
\end{equation} 
This {\it global-local} structure has been recently advocated by \cite{Zeng2024} in a different setting when studying image-on-scalar regression, which is the inner layer of our prior. 
The global level indicator $\delta_k$ represents the covariate-level selection, as $\delta_k = 0$ zeros out $\tau^{ij}_k$ for any $1 \le i \ne j \le p$, eliminating the covariate from the model. That is, $\delta_k = 0$ implies that covariate $x^k$ has no influence on any of the edges, hence the whole graph. At the local level $\gamma^{ij}_k$ refers to the influence of the covariate on the pair of nodes $(i,j)$. The importance of a covariate is characterized by the total number of pairs influenced by that covariate. The parameter $\pi_k$, which can be interpreted as the probability that $x^k$ has an influence on the graph, is called \textit{participation rate} in \cite{Zeng2024} and is estimated under the assumption that the covariate affects all pairs independently. The participation rate $\pi_k$ also informs the selection by excluding those covariates expected to affect less than $d_k \times 100\%$ pairs, saying ${\bm{\tau}}_k = \bm{0}$, if $\pi_k < d_k$, leading to $\bB_k = \bm{0}$. This hard-threshold provides a probability-based selection rule and uses the local-level selection to inform the global-level selection.  Without prior domain knowledge, \cite{Zeng2024} recommend $d_k = 0.05$ (i.e., 5\%) as a conventional probability threshold for sparse models.

Eqs.~\eqref{eq:CSS}, ~\eqref{eq:node_level}, and~\eqref{eq:cov.joint} lead to a dual-group spike-and-slab prior. The encoded sparsity in this prior can be obtained by calculating the expectation of 
the selection indicator $\delta^{ij}_k$: %= [\delta_k \times \gamma^{ij}_k] \times  \delta^{ij} = \left[I(\pi_k \ge d_k)\times \gamma^{ij}_k\right] \times \left[ \delta^{ij} \right] $: 
\begin{align}
        \nonumber E[\delta^{ij}_k] & = E\left[I(\pi_k \ge d_k)\gamma^{ij}_k \delta^{ij} \right] =  E_{\pi_k} \left[ E\left[ I(\pi_k \ge d_k)\gamma^{ij}_k \right] | \pi_k \right] E_{\pi^{i}} \left[ E\left[  \delta^{ij} \right] | \pi^{i}\right] \\
        \nonumber & = E_{\pi_k} \left[  I(\pi_k \ge d_k)\pi_k \right]  E_{\pi^{i}}\left[ \pi^{i} \right]  = \int^1_{d_k} \pi_k \frac{1}{B(a_k, b_k)} \pi_k^{a_k -1 } (1-\pi_k)^{b_k -1 } d \pi_k  \frac{a^i}{a^i + b^i}\\
        & =   E\left[\pi_k \right] \left[ 1 - F_{\text{Beta}_k}(d_k)\right]E\left[ \pi^{i} \right] =   \frac{a^i}{a^i + b^i} \frac{a_k}{a_k+b_k} \left[ 1 - F_{\text{Beta}_k}(d_k)\right]   , \label{eq:prior.sparsity}
\end{align}
where $F_{\text{Beta}_k}(\cdot)$ is the cumulative distribution function of $\text{Beta}(a_{k}+1, b_{k})$.  Eq.~\eqref{eq:prior.sparsity} describes how sparsity at multiple prior levels influences the sparsity of $\delta_{k}^{ij}$, and offers a flexible way to incorporate prior beliefs on the graphical structure. For example, the model of \cite{Zhang2022} corresponds to a prior belief that the graph consists of a dense population level and a sparse covariate level, which corresponds to an always-included intercept term with imbalanced penalization for the intercept and the covariates. Even though our model does not specifically incorporate an intercept, it can mimic this belief by incorporating $x^1_n = 1$ for all $n$ and adjusting the prior parameters $d_k, a_k$ and $b_k$, i.e., $d_1 = 0$ and $ \frac{a_1}{a_1+b_1} >  \frac{a_k}{a_k+b_k}$ for $k \neq 1$. Similarly, prior parameters $a^i$ and $b^i$ can be adjusted to adapt to beliefs on node sparsity. To prevent the influence of prior knowledge through these prior parameters, we use the non-informative prior Beta$(1,1)$ for all $k$ in simulations and applications.
% Without any prior information, a non-informative prior can be used, allowing the model to learn from the data.

\subsubsection{Complete conjugate prior}

We place a prior $\mathcal{F}^{ij}_k$ with positive support on $\sigma^{ij}_k$. In particular, we specify $\mathcal{F}^{ij}_k = N^+(0, s^2_k)$, ie. a truncated normal on the positive line. Combining this with Eq.~\eqref{eq:cov.joint}, our prior on $\tau^{ij}_k$ is 
\begin{align}
   & \begin{cases}
			  \tau^{ij}_k =  \tilde{\tau}^{ij}_k I\left( \pi_k \ge d_k \right) \\
			\tilde{\tau}^{ij}_k | \gamma^{ij}_k  \sim  \gamma^{ij}_k N^{+}\left( 0, s^2_k\right) + \left( 1 - \gamma^{ij}_k\right) \delta_0 \\
			\gamma^{ij}_k | \pi_k  \sim   \text{Bernoulli}( \pi_k ),  \quad \pi_k \sim  \text{Beta}( a_k, b_k).
\end{cases}  \label{eq:cov_level} 
\end{align}
Finally, we assign conjugate priors $s^2_k \sim \text{InvGamma}(1,t)$, with $t  \sim \text{Gamma} \left(a_t, b_t\right)$, for $k=1\ldots, q,$ and inverse gamma priors on the error variances, $\sigma^2_i \sim \text{InvGamma}( a^i_\sigma, b^i_\sigma)$, for $i=1, \ldots, p$. We set $a_t = b_t = 0$, which leads to a commonly used flat and improper prior on $t$, although other values can be used. These specifications, along with Eqs.~\eqref{eq:CSS}, ~\eqref{eq:node_level} and~\eqref{eq:cov_level}, complete our prior model and facilitate a full Gibbs sampler for posterior inference. 

We conclude this section by noting that in the presentation of our proposed dual-group spike-and-slab prior we have used the sequential order of nesting a vector-on-scalar regression into a scalar-on-scalar regression. In practice, one can also vary this order and substitute particular choices for each module with alternative structures, following the same nesting strategy to address multi-level sparsity.  For the same joint prior structure of $\mathcal{B}$, altering the grouping order only determines whether covariate-level or node-level parameters are updated first during each iteration of posterior computation, with minimal impact on computational complexity or results. 

\subsection{Posterior Inference}
\label{sec:sampler}
We derive a full Gibbs sampler for inference in the proposed model, which combines blocked Gibbs strategies based on the samplers used in \cite{Xu2015} and \cite{Zeng2024}. We describe the updates of the parameters below and provide  pseudo code along with detailed derivations in the Supplementary Materials.

\begin{itemize}
	\item \textbf{Update the covariate-level selection parameters $\left\{ \tau^{ij}_k, \tilde{\tau}^{ij}_k, \gamma^{ij}_k, \pi_k\right\}$} \\
	We rewrite the distribution of response node $i$ in Eq.~\eqref{eq:likelihood} by separating the parameters to be sampled as follows: 
	\begin{equation}
	\label{eq:cov_cond_likelihood}
	 y^i_n | - \sim N \left(  \underbrace{ \sum_{s \ne k }  \sum_{l\ne i} \beta^{il}_s   y^l_n x^s_n}_{  \substack{  \text{ conditional on $s \neq k$} \\ \text{denoted as } c^{1, ijk}_{n} }}  +  \underbrace{\sum_{j' \notin \{i, j\} } \beta^{ij'}_k y^{j'}_n x^k_n }_{  \substack{   \text{ conditional on } j' \notin \{i,j\} \\ \text{denoted as } c^{2,ijk}_n }} + \beta^{ij}_k y^j_n x^k_n,  \sigma^2_i \right).
	\end{equation}
	By denoting $ y^{ijk}_n = y^i_n - c^{1,ijk}_n - c^{2,ijk}_n$.  Eqs.~\eqref{eq:cov_level} and~\eqref{eq:cov_cond_likelihood} lead to the following conditional probabilities  given the latent coefficients and indicators,
    \begin{eqnarray}
	\label{eq:cov_LLM}
	    y^{ijk}_n | \gamma^{ij}_k = 1, -  \sim  N\left( \tilde{\tau}^{ij}_{k}  b^{ij}_k y^j_n x^k_n ,  \sigma^2_i  \right), \quad 
	    y^{ijk}_n |  \gamma^{ij}_k = 0, -  \sim N\left( 0,  \sigma^2_i  \right) 
    \end{eqnarray}
	% \begin{eqnarray}
	% \label{eq:cov_LLM}
	%     y^{ijk}_n | \tilde{\tau}^{ij}_{k}, \gamma^{ij}_k = 1, - & \sim & N\left( \tilde{\tau}^{ij}_{k}  b^{ij}_k y^j_n x^k_n ,  \sigma^2_i  \right) \nonumber \\
	%     \tilde{\tau}^{ij}_k | \gamma^{ij}_k = 1 & \sim & N^{+}\left( 0, s^2_k\right) \\
	%     y^{ijk}_n |  \gamma^{ij}_k = 0, - & \sim & N\left( 0,  \sigma^2_i  \right) \nonumber
 %    \end{eqnarray}
    with corresponding  posterior probability ratio, integrating out $\tilde{\tau}^{ij}_k$, given as
    \begin{equation*}
    \begin{aligned}
    \theta^{ij}_k   = \frac{  p\left(  y^{ijk}_{\cdot} | \gamma^{ij}_k = 0, \tilde{\tau}^{ij}_k = 0 \right) \times \left( 1 - \pi_k \right) }{   \int p\left( y^{ijk}_\cdot | \gamma^{ij}_k = 1, \tilde{\tau}^{ij}_k \right)p\left( \tilde{\tau}^{ij}_k \right)d \tilde{\tau}^{ij}_k \times \pi_k }  = \frac{  1 - \pi_k }{ 2 \left(  s^2_k \right)^{-\frac{1}{2}}\times \left( \tilde{\nu}^2_{ijk} \right)^{\frac{1}{2}} \exp\left\{  \frac{1}{2}  \frac{ \tilde{m}^2_{ijk} }{ \tilde{\nu}^2_{ijk}} \right\}  \Phi\left( \frac{ \tilde{m}_{ijk}}{\tilde{\nu}_{ijk}}\right) \times \pi_k}, 
    \end{aligned}
    \end{equation*}
    % \begin{equation*}
    % \begin{aligned}
    % & \theta^{ij}_k   = \frac{  p\left(  y^{ijk}_{\cdot} | \gamma^{ij}_k = 0, \tilde{\tau}^{ij}_k = 0 \right) \times \left( 1 - \pi_k \right) }{   \int p\left( y^{ijk}_\cdot | \gamma^{ij}_k = 1, \tilde{\tau}^{ij}_k \right)p\left( \tilde{\tau}^{ij}_k \right)d \tilde{\tau}^{ij}_k \times \pi_k }  \\
    % & = \frac{  1 - \pi_k }{ 2 \left(  s^2_k \right)^{-\frac{1}{2}}\times \left( \tilde{\nu}^2_{ijk} \right)^{\frac{1}{2}} \exp\left\{  \frac{1}{2}  \frac{ \tilde{m}^2_{ijk} }{ \tilde{\nu}^2_{ijk}} \right\}  \Phi\left( \frac{ \tilde{m}_{ijk}}{\tilde{\nu}_{ijk}}\right) \times \pi_k}, 
    % \end{aligned}
    % \end{equation*}
     where $y^{ijk}_{\cdot}$ denotes $\left\{ y^{ijk}_n\right\}^N_{n=1}$, $\Phi(\cdot)$ denotes the cumulative distribution function of the standard Normal distribution, and
      \begin{eqnarray*}
		\tilde{\nu}^2_{ijk}  =  \left( \sum^N_{n=1} \left(y^j_n x^k_n \right)^2 \left(b^{ij}_k\right)^2 / \sigma^2_i   + 1/ s^2_k \right)^{-1} \text{ and } \tilde{m}_{ijk}  =  \tilde{\nu}^{2}_{ijk}  b^{ij}_k \sum^N_{n=1} y^j_n x^k_n y^{ijk}_n / \sigma^2_i.
      \end{eqnarray*}
      This posterior probability ratio allows to sample the local indicators $\gamma^{ij}_k$ from 
	 $$
	    \gamma^{ij}_k | - \sim \text{Bernoulli}\left( \frac{1}{1+ \theta^{ij}_k} \right).
	 $$
    If $\gamma^{ij}_k = 1$, Eq.~\eqref{eq:cov_LLM} leads to the update $\tilde{\tau}^{ij}_{k}  | - \sim N^+\left(  \tilde{m}_{ijk}, \tilde{\nu}^2_{ijk}\right)$; else if $\gamma^{ij}_k = 0$, we set $\tilde{\tau}^{ij}_{k}  = 0$. \\
    After updating all indicators for covariate $x^k$, we update
	$$
	    \pi_k| - \sim \text{Beta}\left(  a_k + \sum_{ 1\le i\ne j\le p } \gamma^{ij}_{k}, b_k + p(p-1) - \sum_{ 1\le i\ne j\le p } \gamma^{ij}_{k}\right).
	$$
     leading to $\tau^{ij}_k = \tilde{\tau}^{ij}_k \delta_k = \tilde{\tau}^{ij}_k I\left( \pi_k \ge d\right)$. This joint update of parameters and selection indicators avoids reversible jump \citep{Savitsky2011}.
 \item \textbf{Update the node-level selection parameters $\left\{ \bm{b}^{ij}, \delta^{ij}, \pi^i\right\}$ together with $\{\beta^{ij}_k\}$} \\
	We rewrite the distribution of response node $i$ in Eq.~\eqref{eq:likelihood} by separating the parameters to be sampled as follows: 
	 \begin{eqnarray}
	 \label{eq:edge_cond_likelihood}
    		 y^i_n | - \sim N\left(  \sum_{ j' \notin \left\{ i,j \right\} } \sum_{k = 1}^q \beta^{ij'}_k y^{j'}_n x^k_n  +  \sum_{k = 1}^q \beta^{ij}_k y^j_n x^k_n,, \sigma^2_i \right). 
    \end{eqnarray}
    Denoting $z^{ij}_n = y^i_n -   \sum_{ j' \notin \left\{ i,j \right\} } \sum_{k=1}^q \beta^{ij'}_k y^{j'}_n x^k_n   $, 
    eqs.~\eqref{eq:node_level} and~\eqref{eq:edge_cond_likelihood} lead to the following conditional probability distributions given the latent coefficients and indicators
	\begin{eqnarray}
	\label{eq:edge_LLM}
	 z^{ij}_n  | \delta^{ij} = 1, -  \sim  N\left(  \left(\bm{X}^{ij}_n\right)^T \bm{V}^{ij} \bm{b}^{ij}, \sigma^2_i \right), \quad 
    z^{ij}_n | \delta^{ij} = 0, -   \sim  N\left( 0, \sigma^2_i \right)  
    \end{eqnarray}
 %    \begin{equation}
	% \begin{eqnarray}
	% \label{eq:edge_LLM}
	%  z^{ij}_n  | \delta^{ij} = 1, - & \sim & N\left(  \left(\bm{X}^{ij}_n\right)^T \bm{V}^{ij} \bm{b}^{ij}, \sigma^2_i \right) \nonumber \\
	%  \bm{b}^{ij} | \delta^{ij} = 1 & \sim &MVN\left( \bm{0}_q, \bm{I}_q \right) \\
 %    z^{ij}_n | \delta^{ij} = 0, -   & \sim & N\left( 0, \sigma^2_i \right)  \nonumber
 %    \end{eqnarray}
    where $\bm{X}^{ij}_n =  \left( y^j_n x^1_n , ..., y^j_n x^q_n\right)^T$ and $\bm{V}^{ij} = \text{diag}\left( \tau^{ij}_1, \ldots, \tau^{ij}_q\right)$. In addition, we denote $\bm{Z}^{ij} = \left( z^{ij}_1, \ldots, z^{ij}_n\right)^T$ and $\bX^{ij} =\left( \bX^{ij}_1,..., \bX^{ij}_n \right)^T$. The  posterior probability ratio of edge $(i,j)$ can be obtained by integrating out $\bb^{ij}$,
    \begin{equation*}
    \begin{aligned}
     \theta^{ij}  = \frac{  p\left(  \bm{Z}^{ij} | \delta^{ij} = 0, \bm{b}^{ij} = \bm{0}  \right) \times \left( 1 - \pi^{i} \right) }{   \int p\left( \bm{Z}^{ij} | \delta^{ij} = 1, \bm{b}^{ij} \right)p\left( \bm{b}^{ij} \right)d \bm{b}^{ij} \times \pi^{i} } = \frac{  1 - \pi^{i} }{\left| \tilde{\bm{\Sigma}}^{ij} \right|^{\frac{1}{2}} \exp \left\{ \frac{1}{2} \left( \tilde{\bm{\mu}}^{ij}\right)^T\left( \tilde{\bm{\Sigma}}^{ij}  \right)^{-1} \tilde{\bm{\mu}}^{ij} \right\} \times \pi^{i}},
    \end{aligned}
    \end{equation*}
%     \begin{equation*}
% \begin{aligned}
%  \theta^{ij}   &  = \frac{  p\left(  \bm{Z}^{ij} | \delta^{ij} = 0, \bm{b}^{ij} = \bm{0}  \right) \times \left( 1 - \pi^{i} \right) }{   \int p\left( \bm{Z}^{ij} | \delta^{ij} = 1, \bm{b}^{ij} \right)p\left( \bm{b}^{ij} \right)d \bm{b}^{ij} \times \pi^{i} } \\
%  & = \frac{  1 - \pi^{i} }{\left| \tilde{\bm{\Sigma}}^{ij} \right|^{\frac{1}{2}} \exp \left\{ \frac{1}{2} \left( \tilde{\bm{\mu}}^{ij}\right)^T\left( \tilde{\bm{\Sigma}}^{ij}  \right)^{-1} \tilde{\bm{\mu}}^{ij} \right\} \times \pi^{i}},
% \end{aligned}
% \end{equation*}
where
$\tilde{\bm{\Sigma}}^{ij}   = \left( \frac{1}{\sigma^2_i}\left( \bm{X}^{ij}\bm{V}^{ij} \right)^T\left(\bm{X}^{ij}\bm{V}^{ij}\right) + \bm{I}_q \right)^{-1}$ and $\tilde{\bm{\mu}}^{ij}  =  \left( \frac{1}{\sigma^2_i} \left(\bm{Z}^{ij}\right)^T \bm{X}^{ij} \bm{V}^{ij} \tilde{\bm{\Sigma}}^{ij} \right)^T$. 
This  posterior probability ratio allows to sample the indicators $\delta^{ij}$ as 
 $$
    \delta^{ij} | - \sim \text{Bernoulli}\left( \frac{1}{1+ \theta^{ij}} \right).
 $$
 Then, if $\delta^{ij} = 1$, we update $\bm{b}^{ij} | - \sim MVN\left(  \tilde{\bm{\mu}}^{ij}, \tilde{\bm{\Sigma}}^{ij} \right)$; otherwise if $\gamma^{ij} = 0$, we set $\bm{b}^{ij} = \bm{0}$. The update of $(\beta^{ij}_k)_{1\le k \le q}$ is followed by $\bm{B}^{ij} = \bm{V}^{ij}\bb^{ij}$. 
 
 After all indicators for node $i$ are updated, the probability $\pi^{i}$ can be updated by 
$$
	\pi^{i}| - \sim \text{Beta}\left( a^{i} + \sum_{ j\ne i  }^{} \delta^{ij}, b^i + (p-1) - \sum_{ j\ne i } \delta^{ij}\right).
$$
\item \textbf{Update the variances $\left\{ \sigma^2_i \right\}$}\\
This is a conjugate update
 $$
		\sigma^2_i | - \sim \text{InvGamma}\left(  \frac{N}{2} + a^i_{\sigma},  \frac{1}{2} \sum^N_{n=1} \left( y^i_n - \sum_{  j\ne i }\sum_{k=1}^q \beta^{ij}_k y^j_n x^k_n\right)^2 + b^i_\sigma\right).
$$
    \item \textbf{Update $\left\{ s^2_k \right\}$ and $t$} \\
    These are also conjugate updates. For each $k \in [q]$, we sample $s^2_k$ from 
    $$ s^2_k | -   \sim  \text{InvGamma}\left( 1 +  \frac{1}{2} \sum_{ 1 \le i \ne j \le p } \gamma^{ij}_k, t + \frac{1}{2}  \sum_{  1 \le i \ne j \le p } \left( \tilde{\tau}^{ij}_k \right)^2 \right).
    $$
    % \begin{eqnarray*}
    % 	 s^2_k | -   \sim  \text{InvGamma}\left( 1 +  \frac{1}{2} \sum_{ 1 \le i \ne j \le p } \gamma^{ij}_k, t + \frac{1}{2}  \sum_{  1 \le i \ne j \le p } \left( \tilde{\tau}^{ij}_k \right)^2 \right).
    % \end{eqnarray*}
    After all $s_k$ are updated, we sample $t$ from  $t | - \sim \text{Gamma}\left( q + 1,  \sum_{k=1}^q \frac{1}{s^2_k}\right)$.
    % $$
    %     t | - \sim \text{Gamma}\left( q + 1,  \sum_{k=1}^q \frac{1}{s^2_k}\right).
    % $$
\end{itemize}

 %We provide a pseudo-code-style description of the sampler in the supplementary material. 
The computational bottleneck of the sampler lies in the iterative update of each local indicator, accounting for the complex dependencies among the indicators. Given the linear system Eq.~\eqref{eq:likelihood}, the computational complexity is approximately $O(p^2q)$ for the covaraite-level group and $O(pq)$ for the node-level group. An investigation of the scalability of the sampler is included in the Supplementary Materials, demonstrating how the computation time increases as $N,  p,$ and $q$ grow. 
%Meanwhile, working with linear representation Eq.~\eqref{eq:likelihood} instead of the Gaussian likelihood Eq.~\eqref{eq:GGM}, this sampler does not require inversion of the precision matrix, which is often considered to have a complexity of $O(p^3)$ for a graph with $p$ nodes, or even $O(Np^3)$ when the graph varies across subjects. Additionally, unlike Metropolis-based samplers, the full Gibbs sampler does not require additional tuning for target acceptance rates, thus avoiding the challenging task of tuning, particularly when the linear system is of high dimensionality.

Given the MCMC samples, we perform posterior inference  in a bottom-up manner, 
first calculating the marginal posterior probabilities of inclusion (MPPIs) for the indicators $\delta^{ij}_k$ and then propagating these inclusion decisions to the two group levels. Following the median probability model \citep{Barbieri2004, Zeng2024}, we define the inclusion indicator $\kappa^{ij}_k = 1$ if the MPPIs of $\delta^{ij}_k$ is greater than $0.5$. To infer the edges in the undirected graph, we use the ``OR" rule \citep{Meinshausen2006, Zhang2022},  which concludes that the edge $(i,j)$ is affected by covariate $x_k$ if either $\kappa^{ij}_k  = 1$ OR $\kappa^{ji}_k =1$. At the two group level, we conclude that the edge $(i,j)$ exists if either $\sum_k \kappa^{ij}_k \ne 0$ OR $\sum_k \kappa^{ji}_k \ne 0$, and that the covariate $x^k$ is influential if $\sum_{ij} \kappa^{ij}_k \ne 0$.  Bayesian false discovery rate control methods can also be utilized in determining the $\kappa^{ij}_k$'s \citep{Newton2004}.

\section{Simulation Study}
\label{sec:sim}
In this section, we conduct simulations and compare the proposed approach with selected covariate-dependent Gaussian graphical regression approaches.  Motivated by the dual grouping perspective and real data applications, we observe that the sparsity of a linear system, as described in Eq.~\eqref{eq:likelihood}, arises not only at the node level, i.e., from a sparse graph $\bOmega(\bx_n)$, but also at the covariate level. Specifically, this sparsity can result from a collection of dense sub-networks $\bOmega(x^k_n)$ coexisting with a set of empty ones $\bOmega(x^{k'}_n)=\bzeros$. We aim to demonstrate that existing methods, which rarely consider the latter cause of system sparsity, may produce suboptimal results in graph recovery. This, in turn, supports the importance of addressing sparsity induced by covariates.

\subsection{Data Generation}
We generate data from Eq~\eqref{eq:GGM} using $\bOmega\left(\bx_n\right) = \left( \omega^{ij}\left( \bx_n\right) \right)_{i,j=1}^p = \left( \sum^q_{k=1} \beta^{ij}_k x^k_n \right)_{i,j=1}^p$. We set the number of nodes to $p = 25$ and the number of covariates to $q = 10$, and introduce sparsity as follows. First, we generate a  preliminary  $25$-by-$25$ random graph with sparsity-level of $0.4$ and randomly partition the edges into four groups for $\bB_{1-4}$, with zeros added elsewhere. We set $\bB_{5-10} = \bm{0}$ for the empty covariates.   This results in a sparsity level of approximately $0.4/10 = 4\%$ for $\mathcal{B}$ of main interest.  The values of the non-zero entries $\beta^{ij}_k$ are sampled from uniform distributions supported on the intervals $[-0.5,-0.35]\cup[0.35,0.5]$. To generate a valid precision matrix, we follow \cite{Zhang2022} by first rescaling each row $i$ by dividing by $\frac{1}{2} \sum_{j} \sum_{k} |\beta^{ij}_k | $ and then averaging $\beta^{ij}_k$ and $\beta^{ji}_k$ to fill each entry of $(i,j,k)$. We set $X_1 = 1$ as the intercept and sample $X_{2-10}$ from uniform$[0,1]$. We use two different sample sizes $N = 200, 500$ and repeatedly generate data $50$ times for each size to evaluate the model performances.   Results on scalability for higher-dimension settings, with $N$ ranging from 200 to 2000, $p$ from 25 to 200, and $q$ from 10 to 200, are provided in the Supplementary Materials (Section S2).

\subsection{Prior specification} 
In all simulations, without prior domain knowledge, we use non-informative priors, allowing the model to learn from the data. Specifically, we set $ a^i_\sigma = b^i_\sigma = 0.1$ for the error variance, $a^i = b^i = 1$ for the node-level sparsity,  and $a_k = b_k = 1$ for the local-level sparsity, for $i = 1, \ldots, p$ and $k = 1, \ldots, q$. For the covariate-level threshold, we adhere to the practice outlined by \cite{Zeng2024}, using $d_k = 0.05$ (i.e., $5\%$) as a conventional probability threshold for sparse models.  A sensitivity analysis on the influential sparsity prior is included in the Supplementary Materials (Section S3), showing that although the results may slightly vary as the sparse prior changes, the main conclusion remains unchanged.

\subsection{Comparison Study}
\label{sec:Reg_with_Cov_res}
We compare performances of the proposed method, for which we use the acronym DGSS, to Lasso regression \citep{Tibshirani1996}, the GMMReg method of \cite{Zhang2022} and the Bayesian sparse group selection method with spike and slab prior of \cite{Xu2015}. We implement Lasso by the R-package \texttt{glmnet} and GMMReg by the %\href{https://drive.google.com/file/d/13XEN_sqjKPhPMjKRBrS4bBgfHyRNL20b/view}
Matlab code from the authors' page. For both Lasso and GMMReg, we tune the regularization parameters by cross-validation. We denote the Bayesian sparse group selection with spike and slab prior as \texttt{BSGSSS}, implemented by the R-package \texttt{MBSGS}. Lasso only considers local sparsity, whereas the other methods also take group sparsity into account. Specifically, GMMReg uses node-level sparsity to induce covariate-level sparsity, BSGSSS accounts for node-level sparsity only, and the proposed DGSS decouples node-level and covariate-level sparsity, modeling them with the dual group spike-and-slab prior. For DGSS, we run $20,000$ MCMC iterations  with a burn-in of $10,000$. % We specify non-informative priors as $\pi_k \sim \text{Beta}(1,1)$, $\pi^i \sim \text{Beta}(1,1)$ and use the conventional sparsity threshold $d_k = 0.05$ for all $k = 1, \ldots, q$. 
For BSGSSS, we increase the MCMC iterations from default $10,000$ with a burn-in of $5,000$ to $20,000$ with a burn-in of $10,000$ and keep all other parameters set to their default values, which also includes an additional MCEM step with $10,000$ iterations.

We consider the covariate-dependent edge detection as a classification task with the presence of an edge within each precision coefficient being treated as a positive signal. The total number of parameters is $p(p-1)q = 6,000$, and on average, $p(p-1)\times 0.4 = 240$ of them are signals, which may vary due to random graph generation. The covariate-dependent edge further provides inference to the node-level and covariate-level selections, as described in Section~\ref{sec:sampler}. We report the following four metrics for comparison: True Positive Rate (TPR), False Positive Rate (FPR), F1 score (F1), and Matthews correlation coefficient (MCC). These metrics are defined by:
\begin{align*}
\text{TPR}  &= \frac{\text{TP}}{\text{TP} + \text{FN}}, \quad \text{FPR} = \frac{\text{FP}}{\text{FP} +\text{TN}}, \quad \text{F1} = \frac{\text{TP}}{\text{TP}+1/2\times (\text{FP} + \text{FN})}, \\
\text{MCC} & = \frac{\text{TP}\times\text{TN}- \text{FP}\times \text{FN}}{\sqrt{ \text{TP}+\text{FP}} \times \sqrt{ \text{TP}+\text{FN}}\times \sqrt{ \text{TN}+\text{FP}}\times \sqrt{ \text{TN}+\text{FN}}  },
\end{align*}
% \begin{align*}
% \text{TPR}  & = \frac{\text{TP}}{\text{TP} + \text{FN}},\\ \text{FPR} &= \frac{\text{FP}}{\text{FP} +\text{TN}},\\ \text{F1} &= \frac{\text{TP}}{\text{TP}+1/2\times (\text{FP} + \text{FN})}, \\
% \text{MCC} & = \frac{\text{TP}\times\text{TN}- \text{FP}\times \text{FN}}{\sqrt{ \text{TP}+\text{FP}} \times \sqrt{ \text{TP}+\text{FN}}\times \sqrt{ \text{TN}+\text{FP}}\times \sqrt{ \text{TN}+\text{FN}}  },
% \end{align*}
where TP, FP, TN and FN represent numbers of True Positive, False Positive, True Negative and False Negative, respectively.

\begin{figure}[!thb]
    \centering
    \begin{subfigure}[c]{1\linewidth}        
    \begin{subfigure}[b]{1\textwidth}
        \centering
        \includegraphics[width=0.09\textwidth]{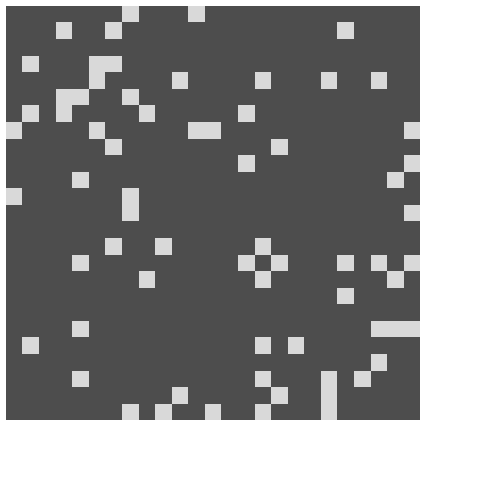}
        \includegraphics[width=0.09\textwidth]{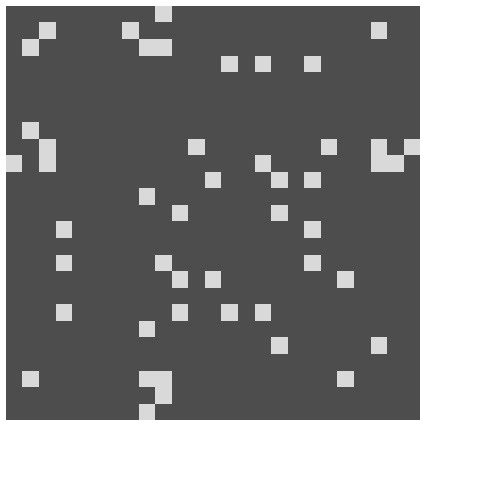}
        \includegraphics[width=0.09\textwidth]{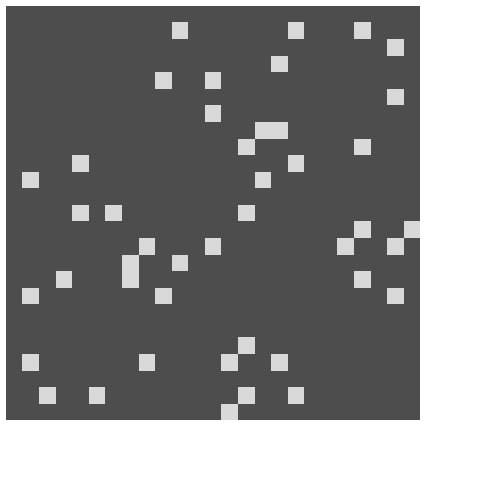}
        \includegraphics[width=0.09\textwidth]{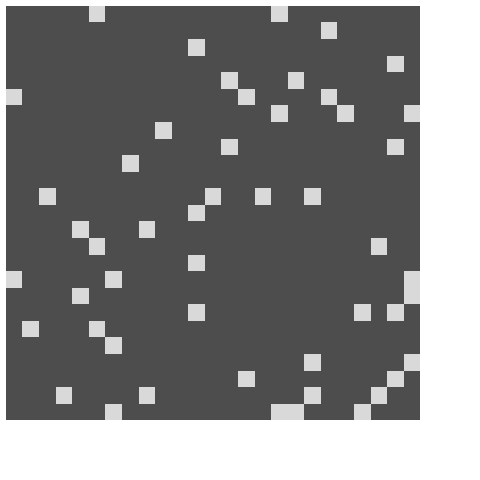}
        \includegraphics[width=0.09\textwidth]{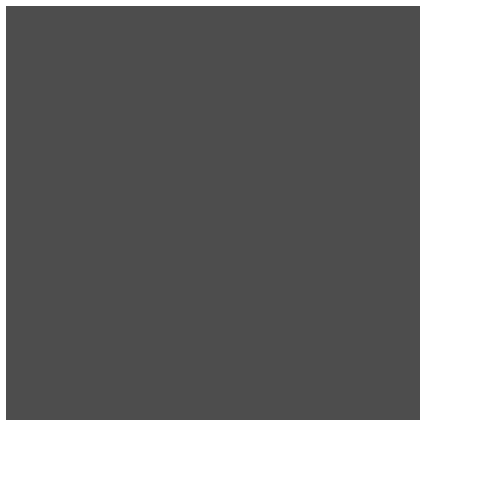}
        \includegraphics[width=0.09\textwidth]{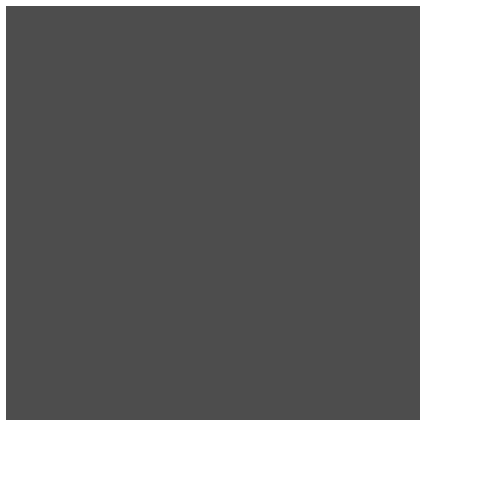}
        \includegraphics[width=0.09\textwidth]{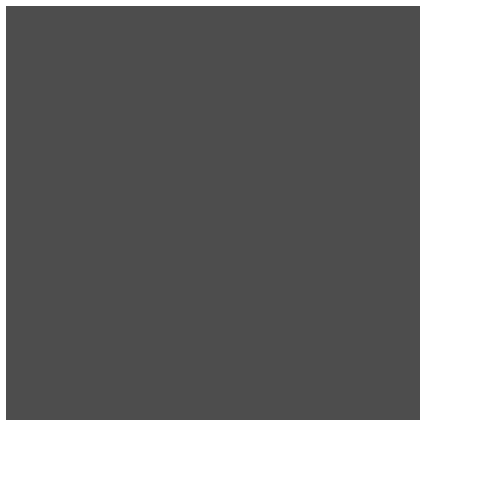}
        \includegraphics[width=0.09\textwidth]{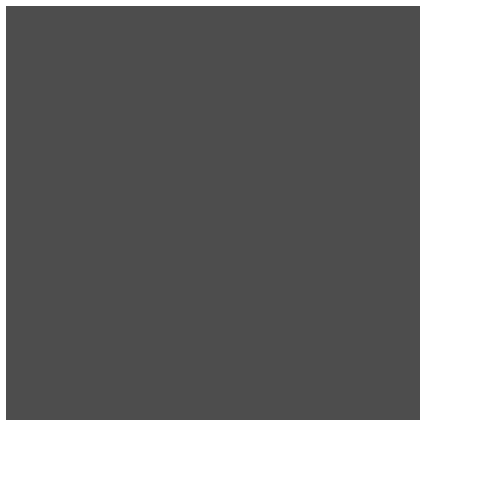}
        \includegraphics[width=0.09\textwidth]{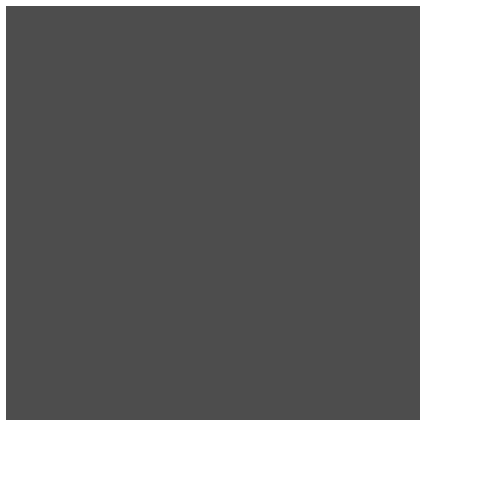}
        \includegraphics[width=0.09\textwidth]{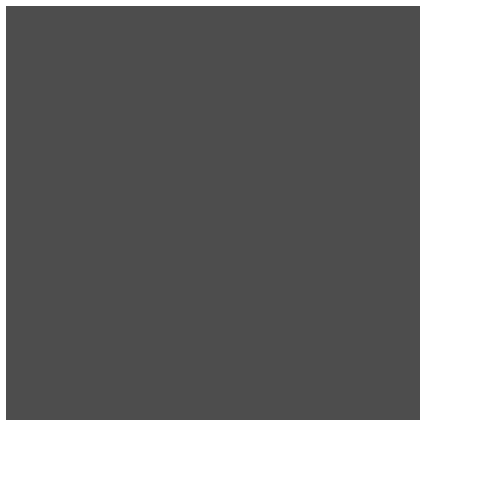}
    \end{subfigure}
    \begin{subfigure}[b]{1\textwidth}
        \centering
        \includegraphics[width=0.09\textwidth]{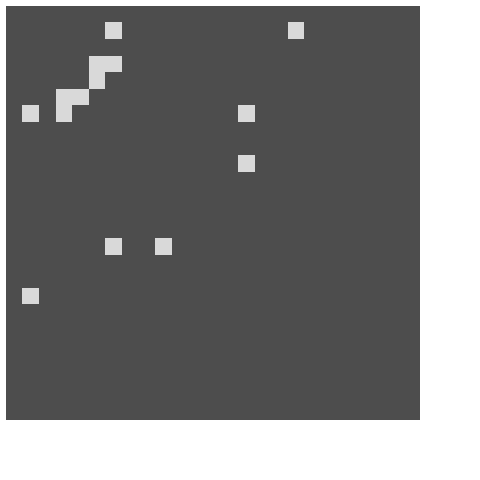}
        \includegraphics[width=0.09\textwidth]{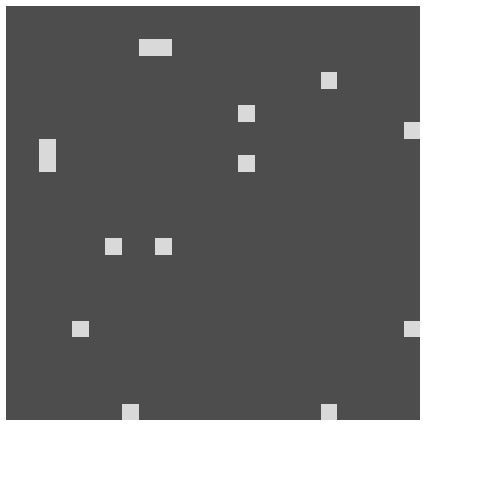}
        \includegraphics[width=0.09\textwidth]{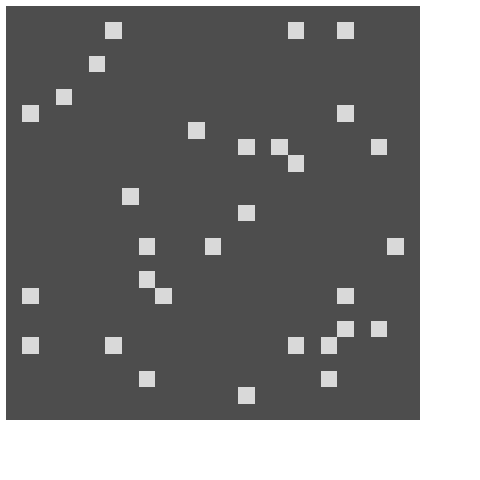}
        \includegraphics[width=0.09\textwidth]{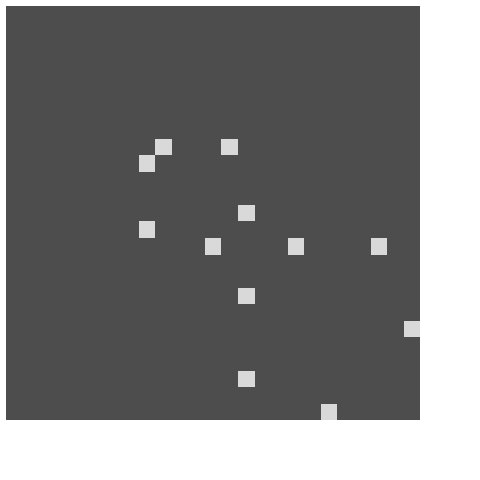}
        \includegraphics[width=0.09\textwidth]{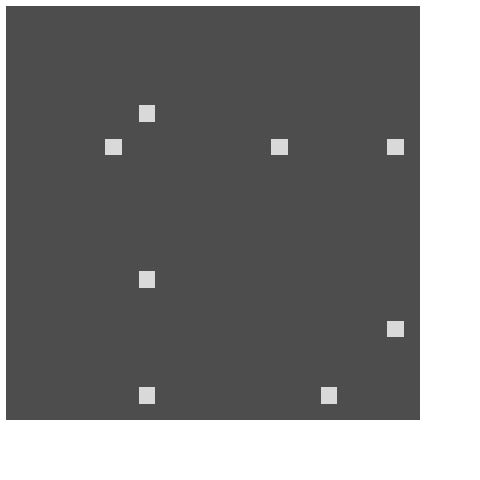}
        \includegraphics[width=0.09\textwidth]{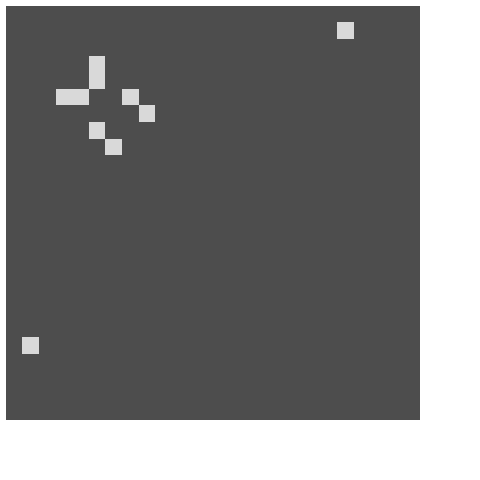}
        \includegraphics[width=0.09\textwidth]{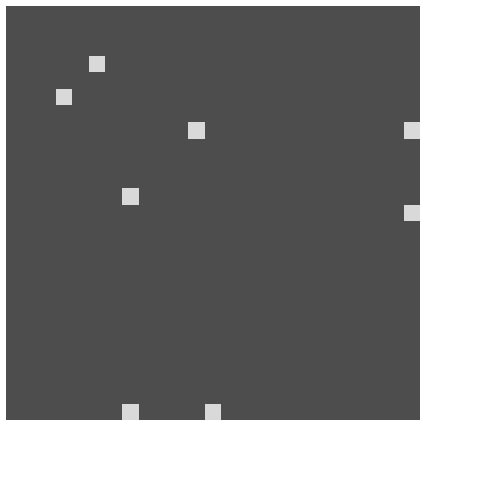}
        \includegraphics[width=0.09\textwidth]{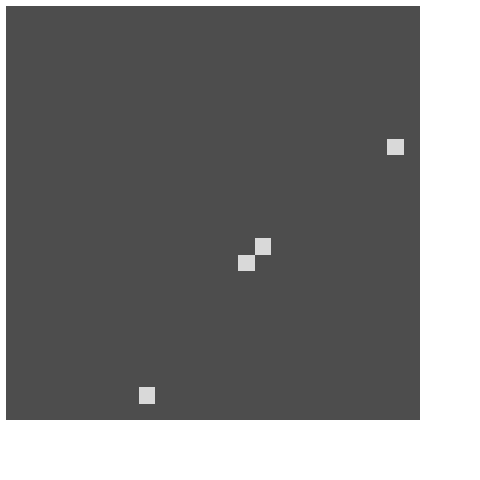}
        \includegraphics[width=0.09\textwidth]{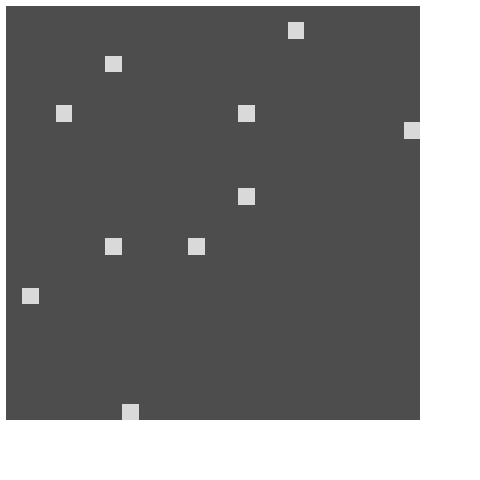}
        \includegraphics[width=0.09\textwidth]{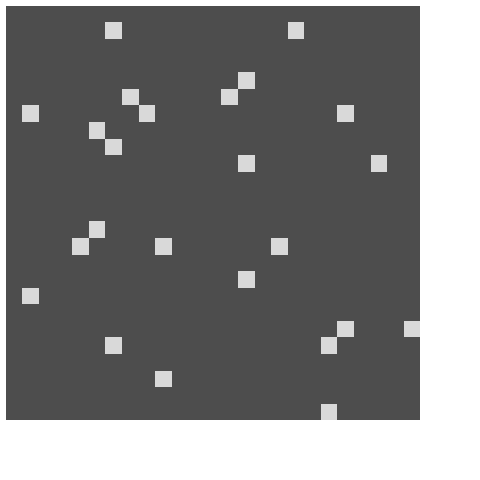}
    \end{subfigure}
    \begin{subfigure}[b]{1\textwidth}
        \centering
        \includegraphics[width=0.09\textwidth]{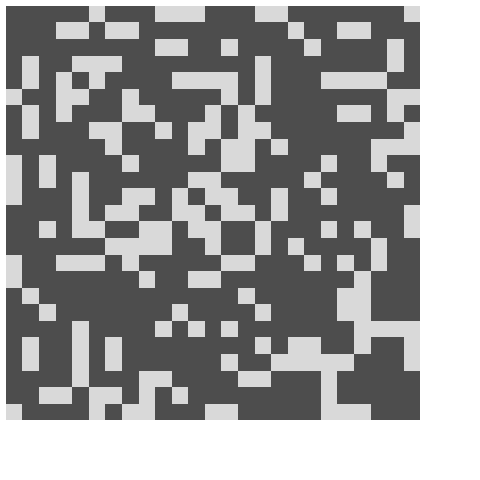}
        \includegraphics[width=0.09\textwidth]{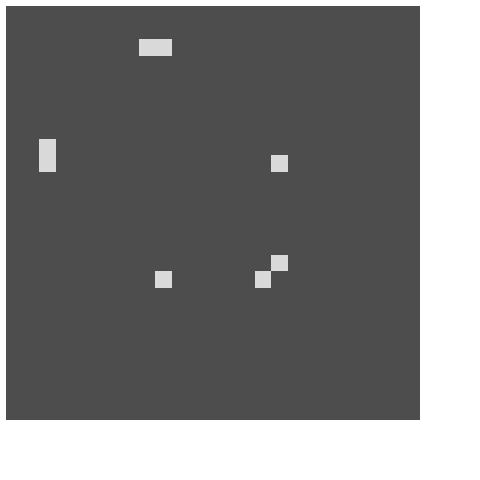}
        \includegraphics[width=0.09\textwidth]{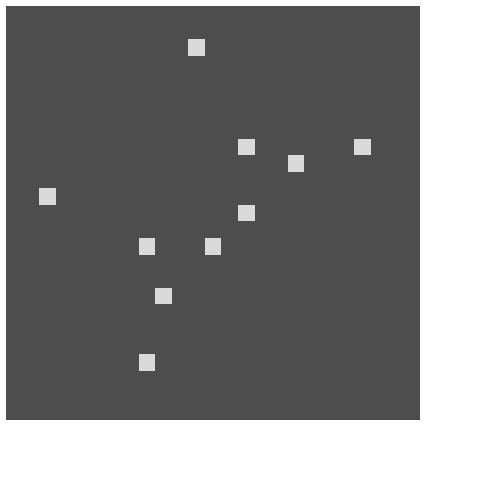}
        \includegraphics[width=0.09\textwidth]{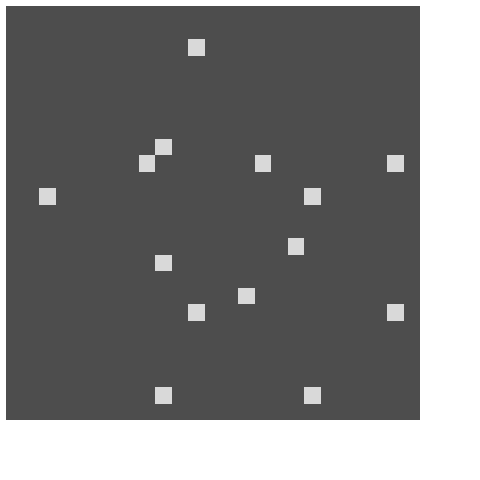}
        \includegraphics[width=0.09\textwidth]{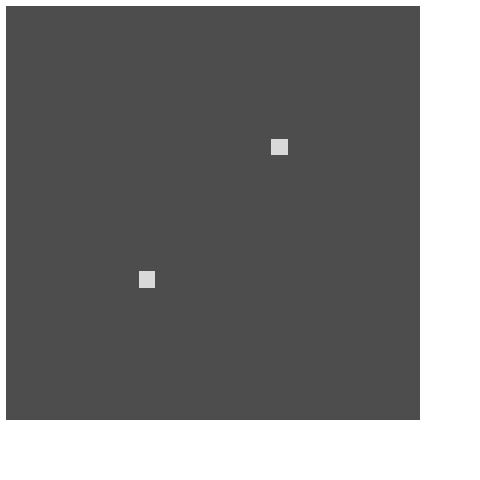}
        \includegraphics[width=0.09\textwidth]{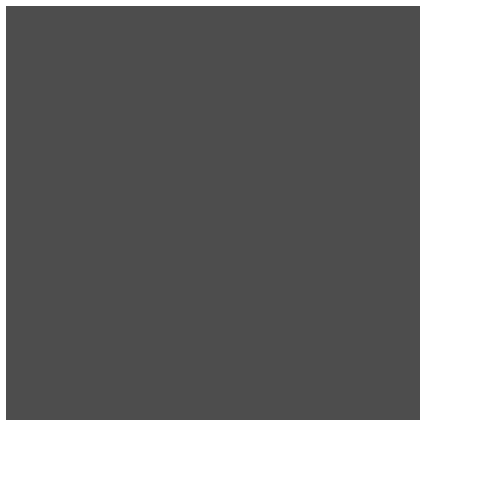}
        \includegraphics[width=0.09\textwidth]{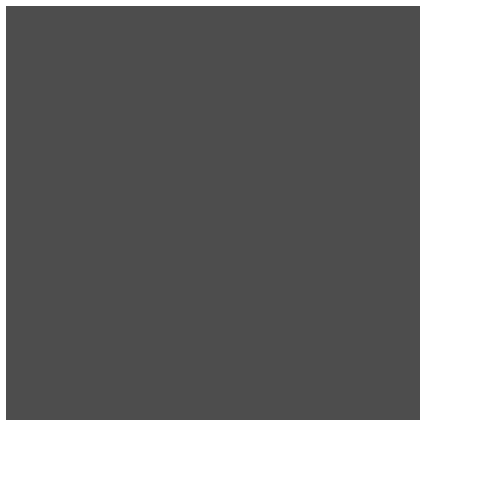}
        \includegraphics[width=0.09\textwidth]{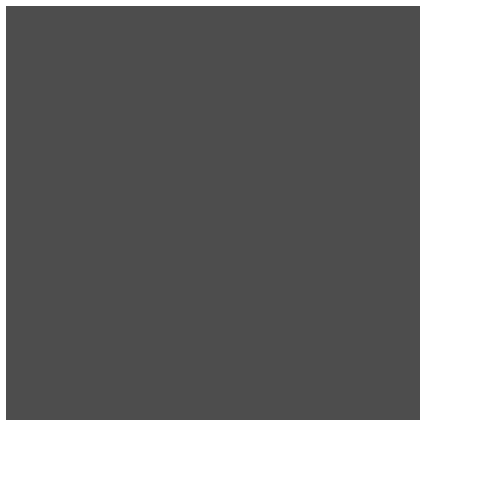}
        \includegraphics[width=0.09\textwidth]{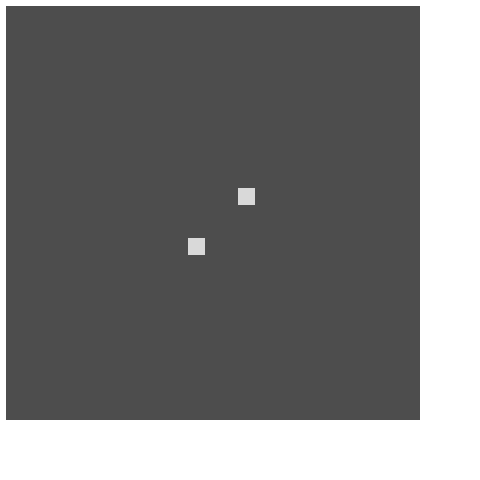}
        \includegraphics[width=0.09\textwidth]{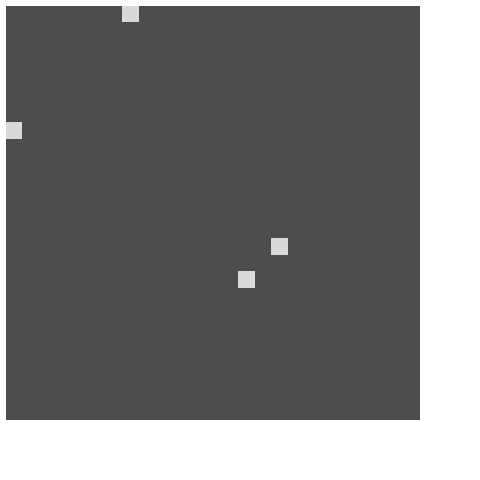}
    \end{subfigure}
    \begin{subfigure}[b]{1\textwidth}
        \centering
        \includegraphics[width=0.09\textwidth]{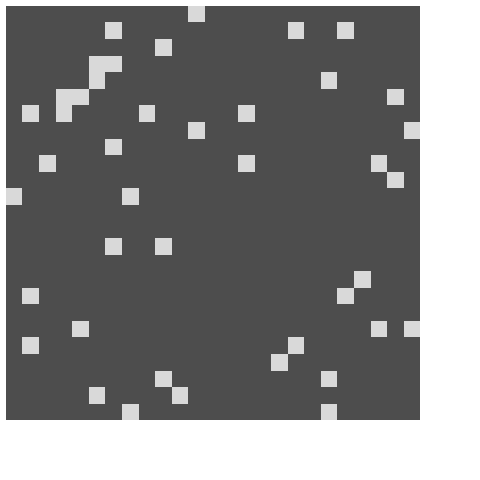}
        \includegraphics[width=0.09\textwidth]{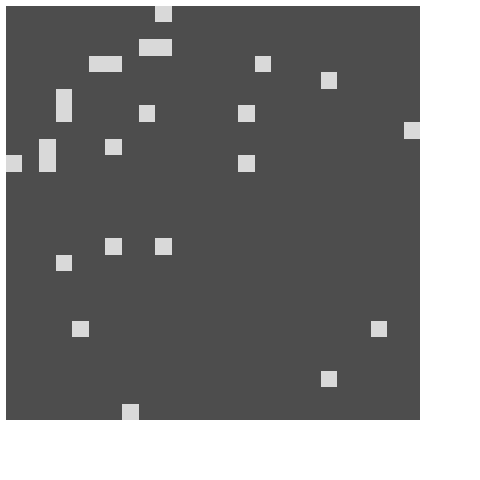}
        \includegraphics[width=0.09\textwidth]{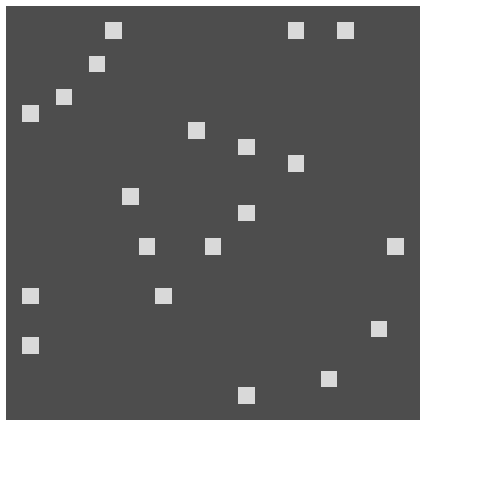}
        \includegraphics[width=0.09\textwidth]{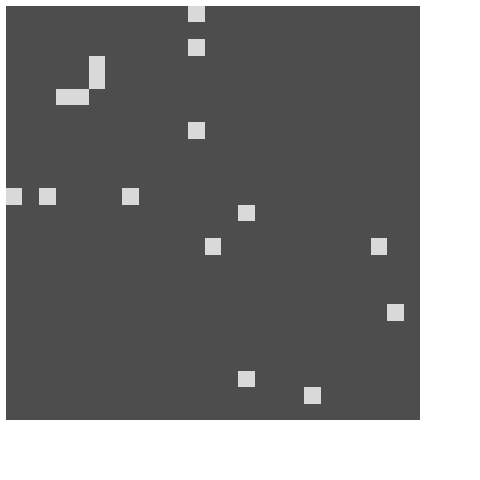}
        \includegraphics[width=0.09\textwidth]{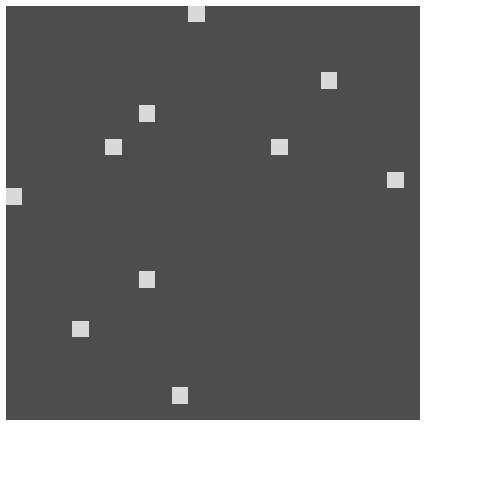}
        \includegraphics[width=0.09\textwidth]{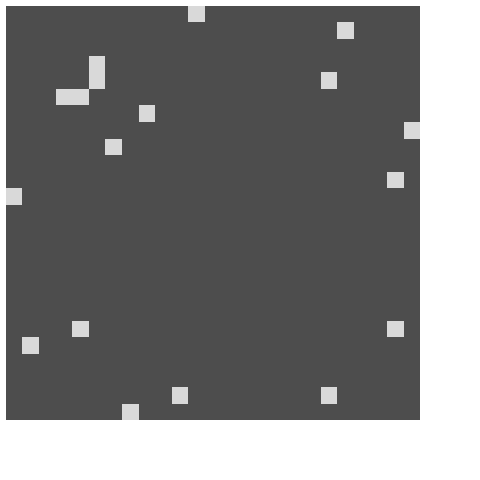}
        \includegraphics[width=0.09\textwidth]{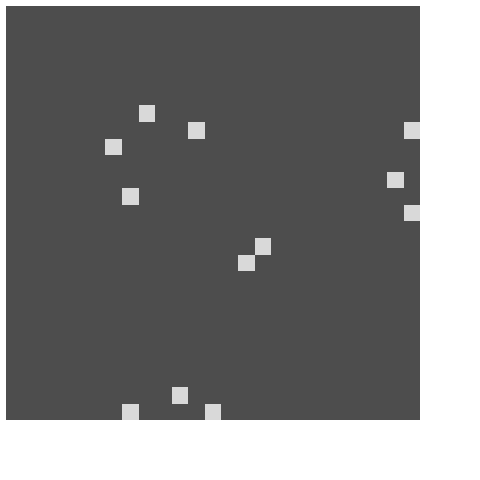}
        \includegraphics[width=0.09\textwidth]{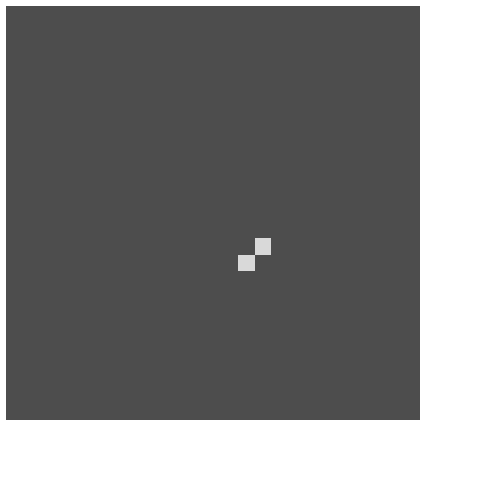}
        \includegraphics[width=0.09\textwidth]{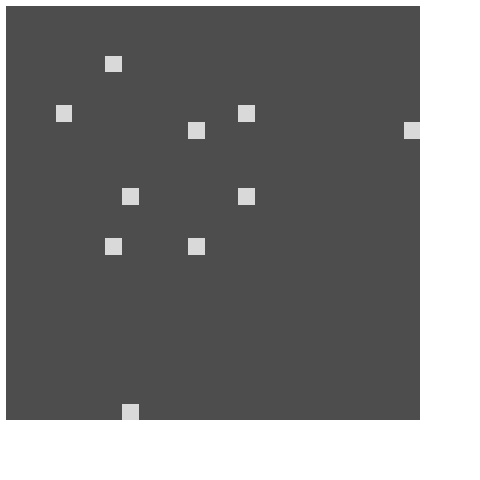}
        \includegraphics[width=0.09\textwidth]{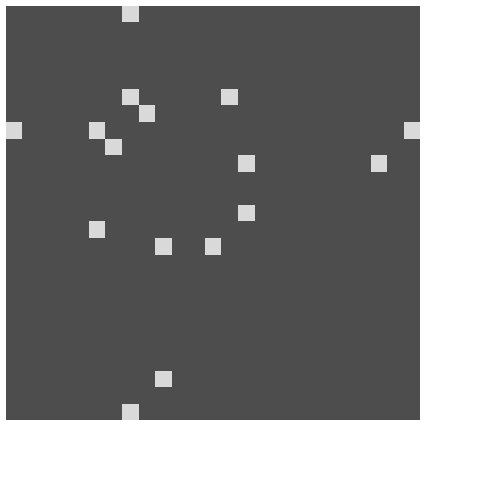}
    \end{subfigure}
    \begin{subfigure}[b]{1\textwidth}
        \centering
        \includegraphics[width=0.09\textwidth]{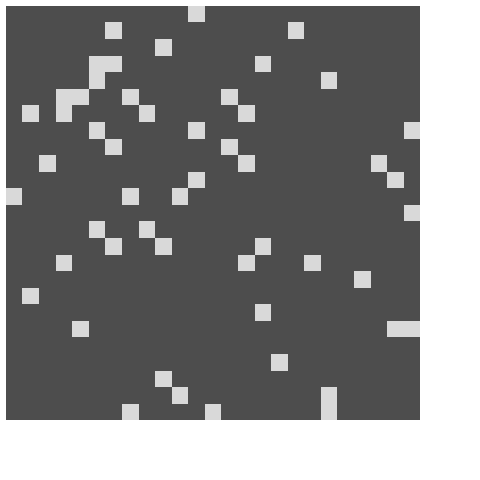}
        \includegraphics[width=0.09\textwidth]{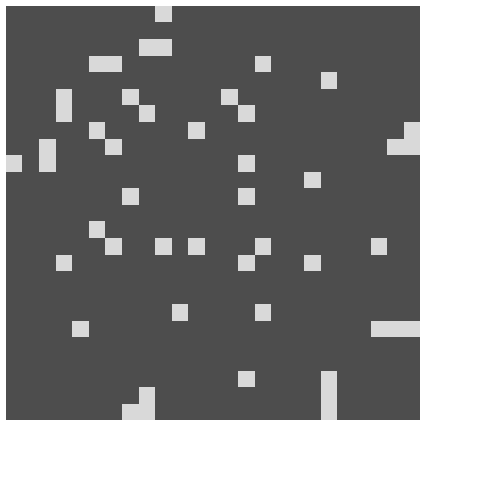}
        \includegraphics[width=0.09\textwidth]{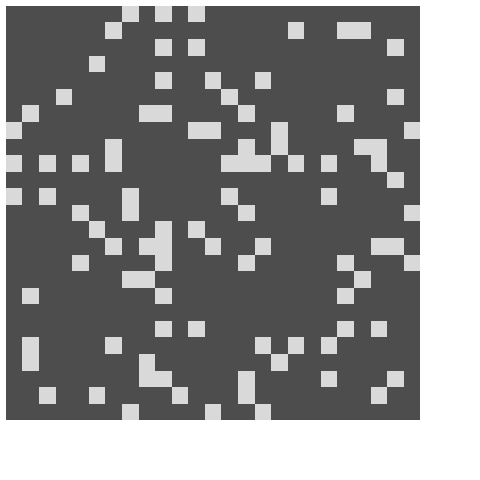}
        \includegraphics[width=0.09\textwidth]{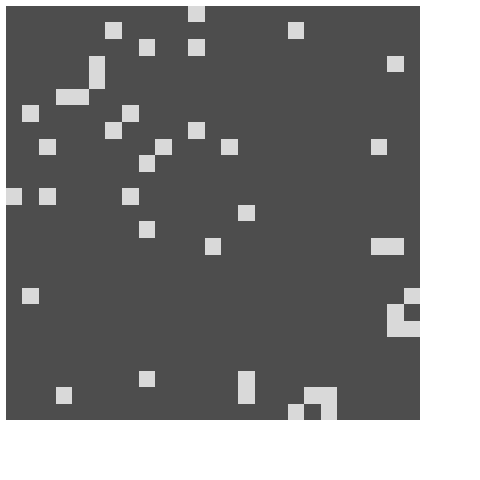}
        \includegraphics[width=0.09\textwidth]{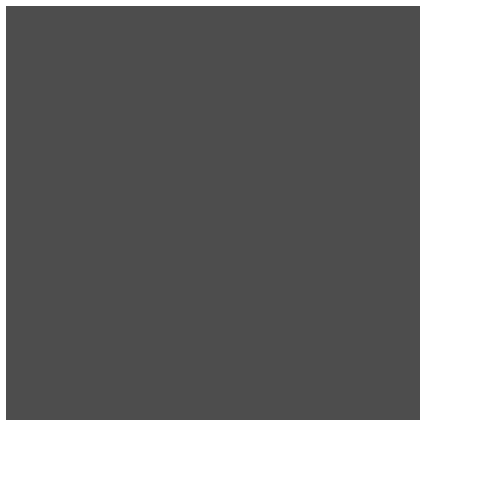}
        \includegraphics[width=0.09\textwidth]{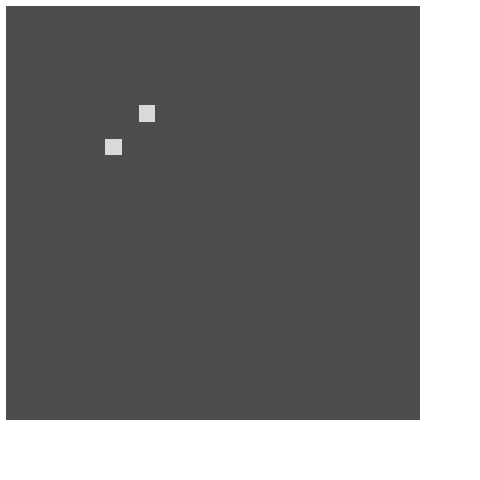}
        \includegraphics[width=0.09\textwidth]{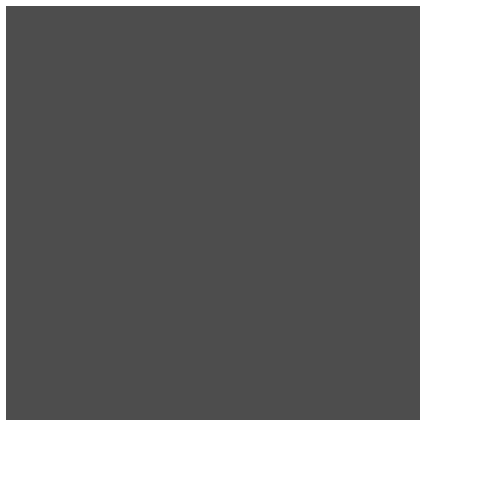}
        \includegraphics[width=0.09\textwidth]{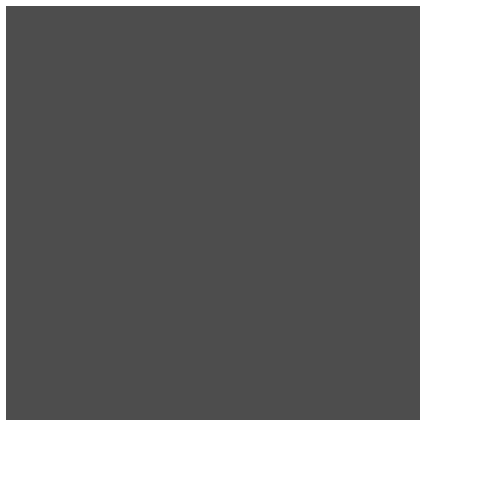}
        \includegraphics[width=0.09\textwidth]{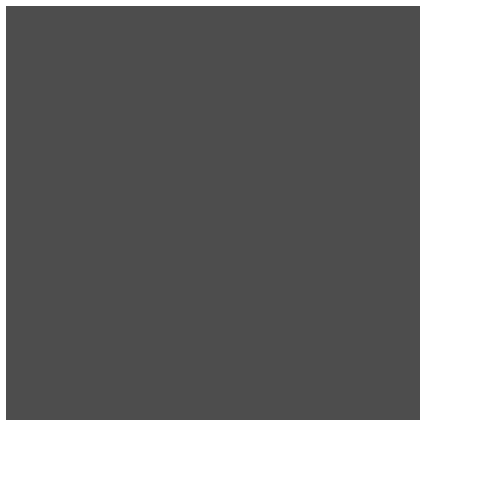}
        \includegraphics[width=0.09\textwidth]{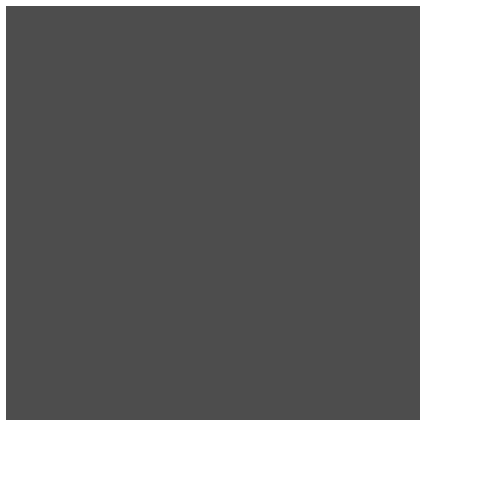}
    \end{subfigure}
     \caption{$N = 200$
     \label{fig:sim.pcoef_200}}
    \end{subfigure}
    \begin{subfigure}[c]{1\linewidth}
    \begin{subfigure}[b]{1\textwidth}
        \centering
        \includegraphics[width=0.09\textwidth]{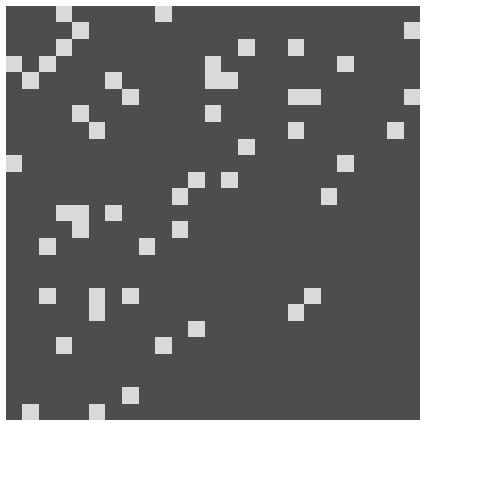}
        \includegraphics[width=0.09\textwidth]{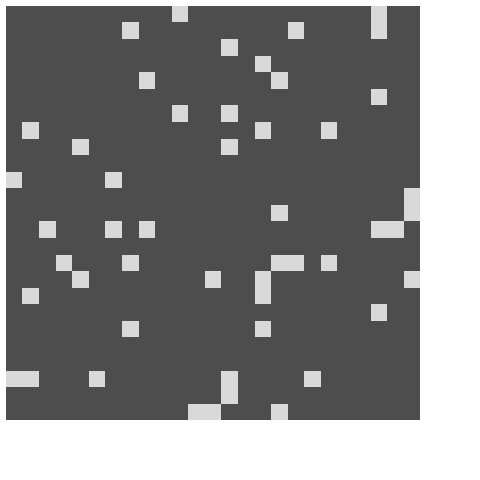}
        \includegraphics[width=0.09\textwidth]{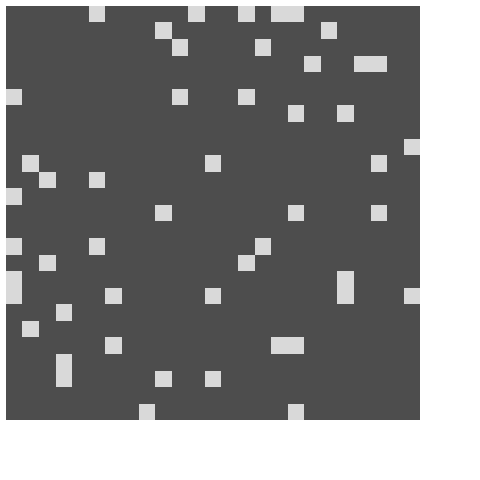}
        \includegraphics[width=0.09\textwidth]{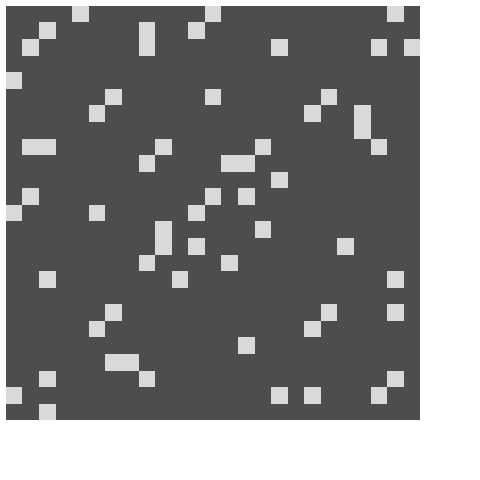}
        \includegraphics[width=0.09\textwidth]{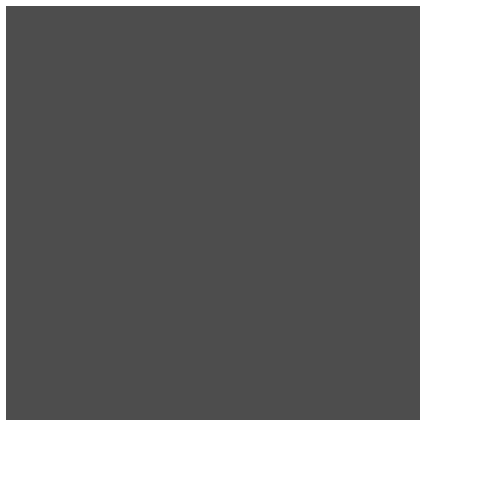}
        \includegraphics[width=0.09\textwidth]{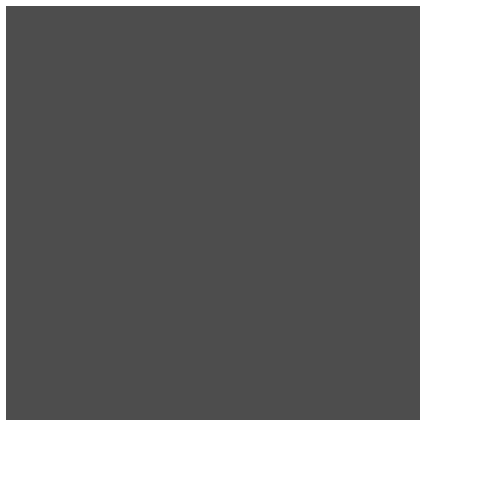}
        \includegraphics[width=0.09\textwidth]{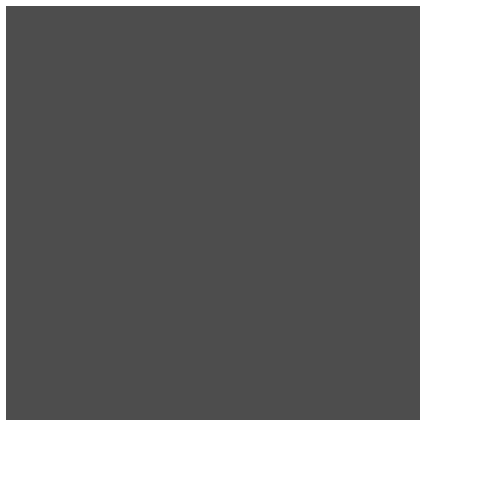}
        \includegraphics[width=0.09\textwidth]{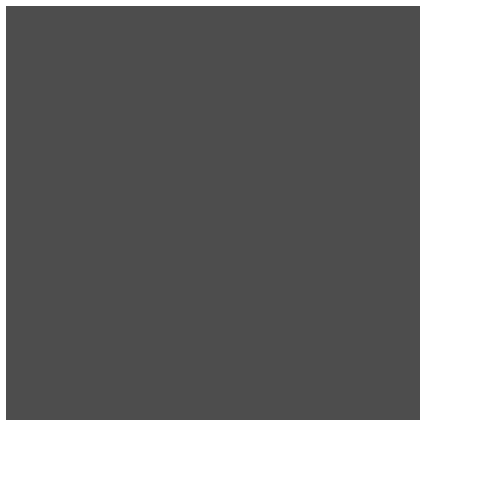}
        \includegraphics[width=0.09\textwidth]{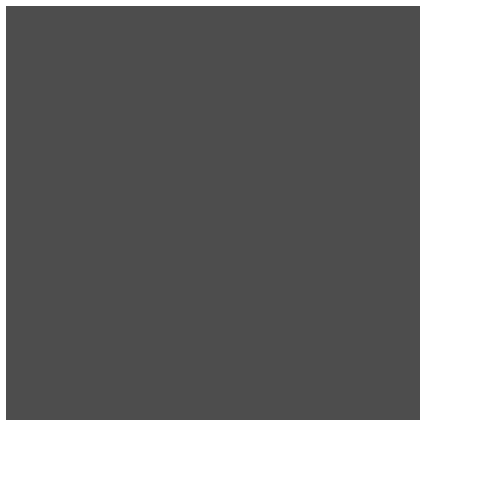}
        \includegraphics[width=0.09\textwidth]{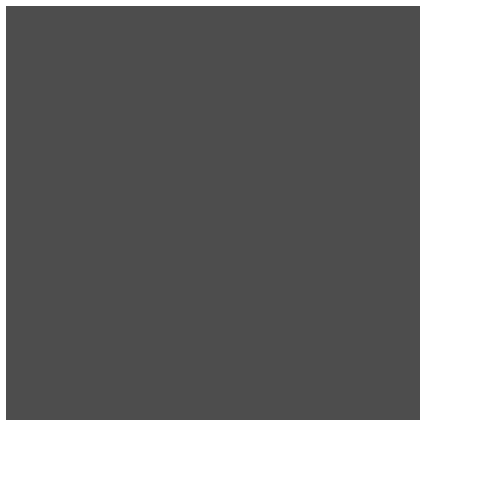}
    \end{subfigure}
    \begin{subfigure}[b]{1\textwidth}
        \centering
        \includegraphics[width=0.09\textwidth]{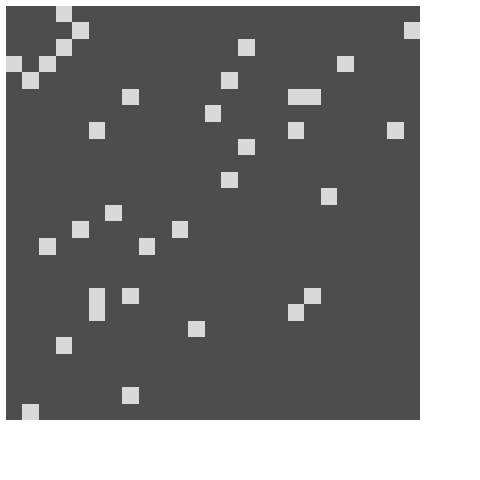}
        \includegraphics[width=0.09\textwidth]{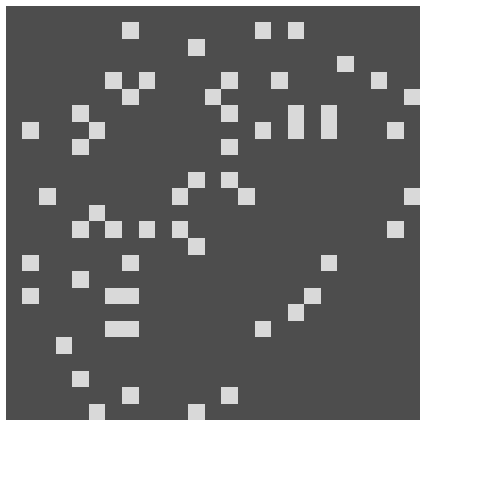}
        \includegraphics[width=0.09\textwidth]{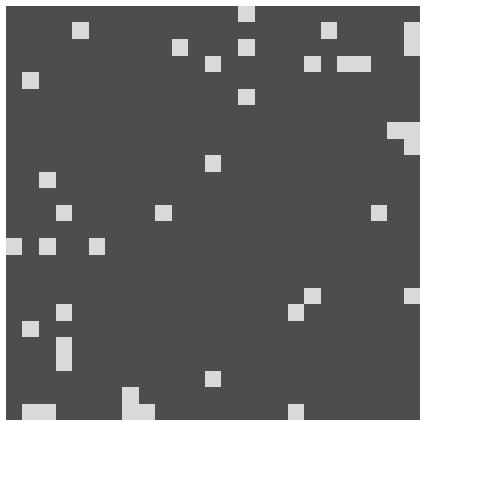}
        \includegraphics[width=0.09\textwidth]{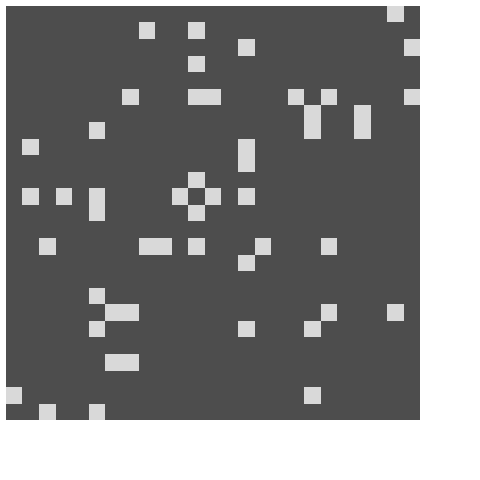}
        \includegraphics[width=0.09\textwidth]{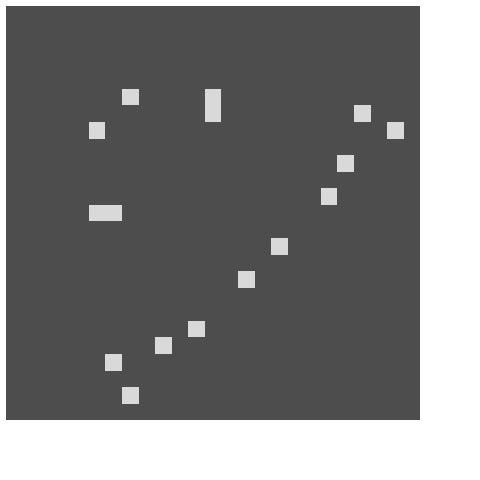}
        \includegraphics[width=0.09\textwidth]{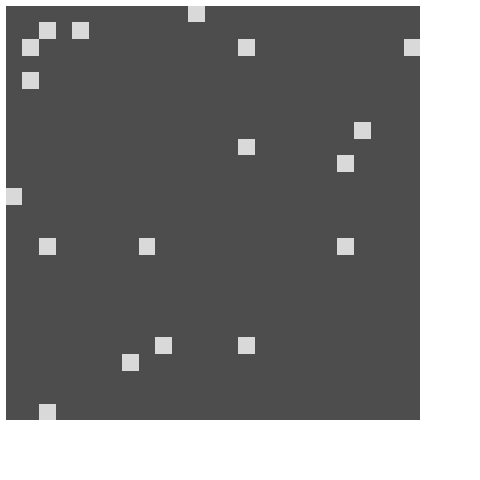}
        \includegraphics[width=0.09\textwidth]{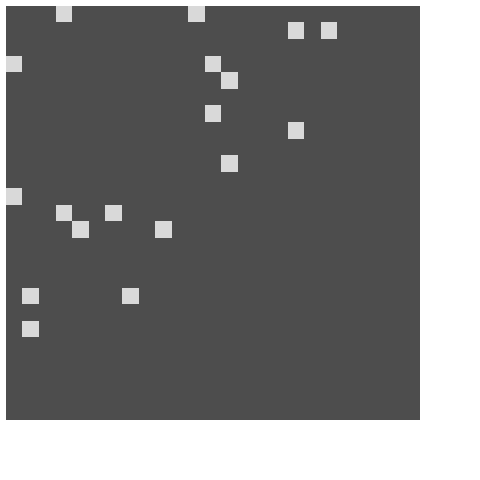}
        \includegraphics[width=0.09\textwidth]{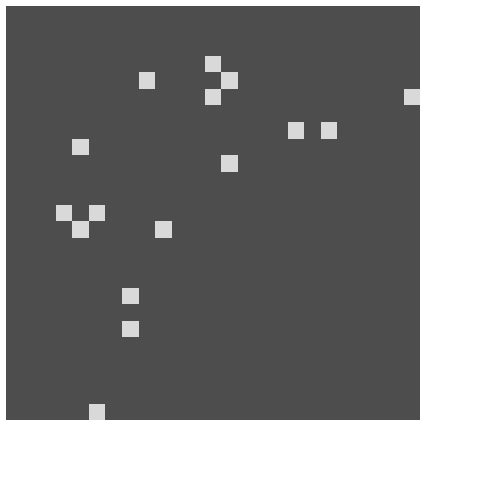}
        \includegraphics[width=0.09\textwidth]{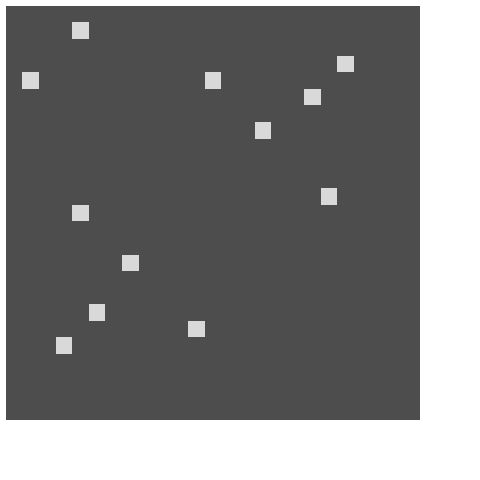}
        \includegraphics[width=0.09\textwidth]{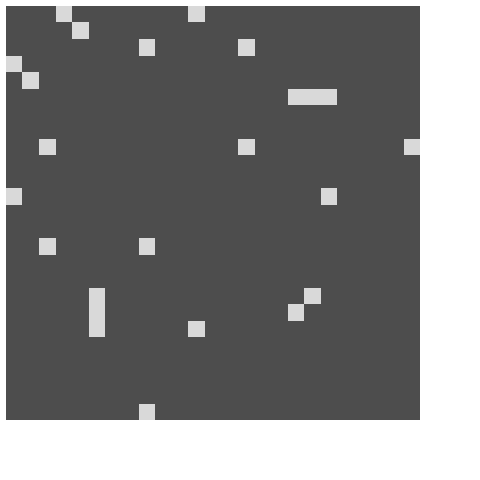}
    \end{subfigure}
    \begin{subfigure}[b]{1\textwidth}
        \centering
        \includegraphics[width=0.09\textwidth]{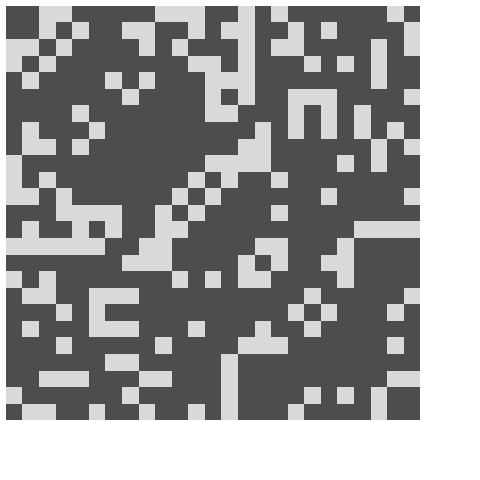}
        \includegraphics[width=0.09\textwidth]{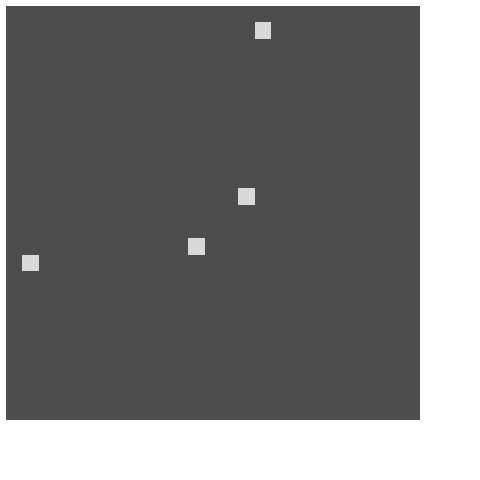}
        \includegraphics[width=0.09\textwidth]{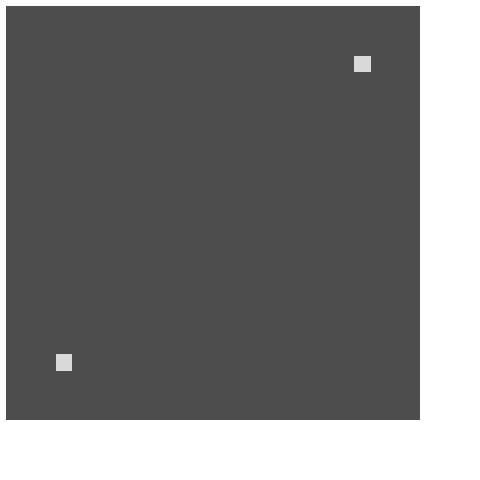}
        \includegraphics[width=0.09\textwidth]{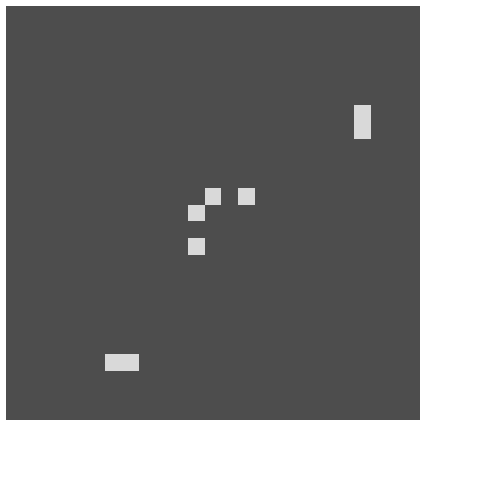}
        \includegraphics[width=0.09\textwidth]{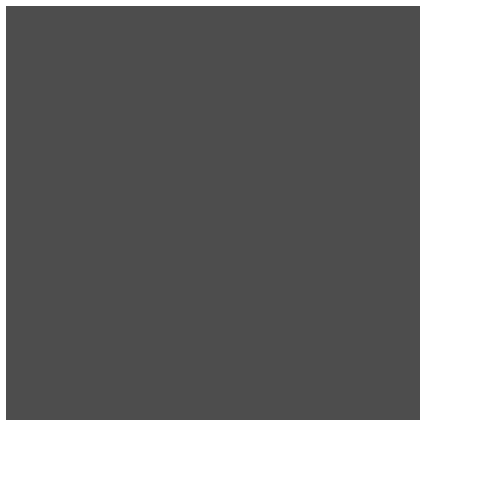}
        \includegraphics[width=0.09\textwidth]{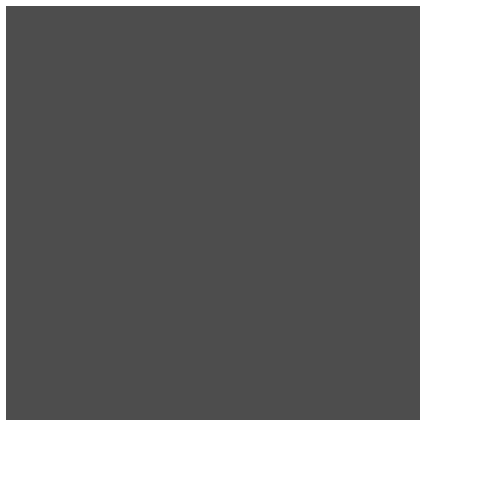}
        \includegraphics[width=0.09\textwidth]{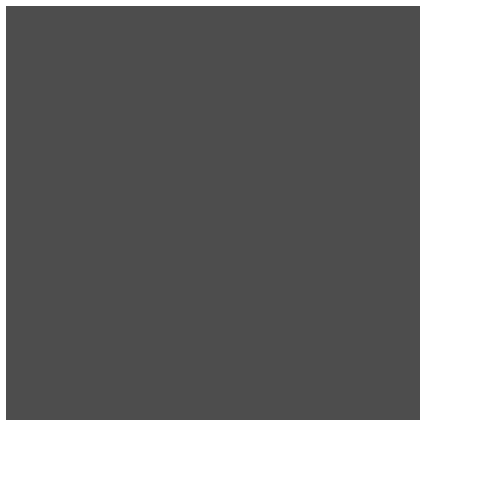}
        \includegraphics[width=0.09\textwidth]{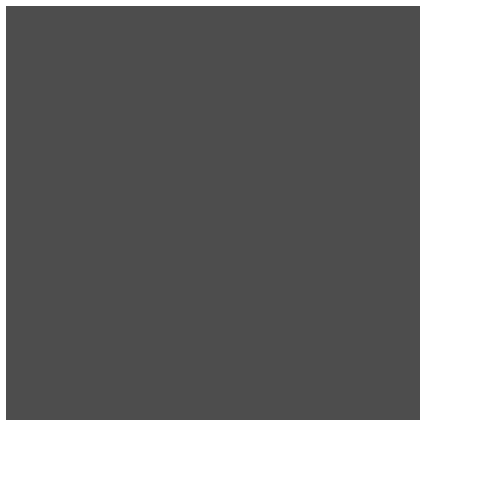}
        \includegraphics[width=0.09\textwidth]{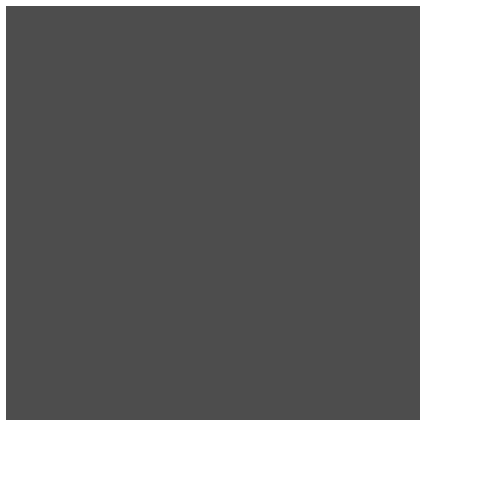}
        \includegraphics[width=0.09\textwidth]{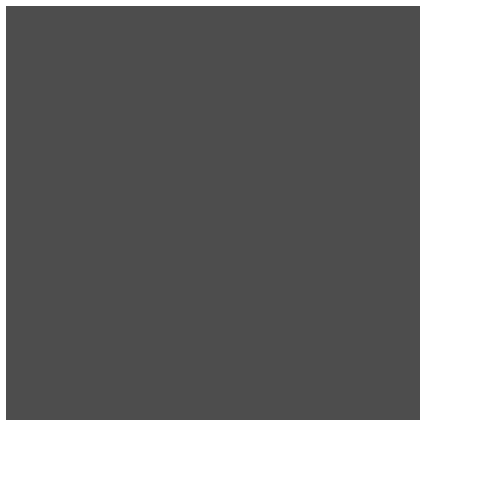}
    \end{subfigure}
    \begin{subfigure}[b]{1\textwidth}
        \centering
        \includegraphics[width=0.09\textwidth]{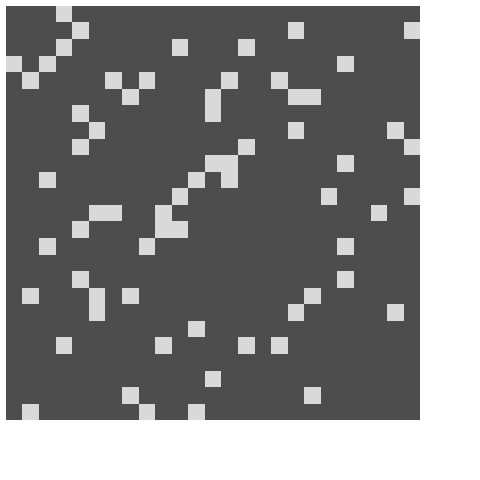}
        \includegraphics[width=0.09\textwidth]{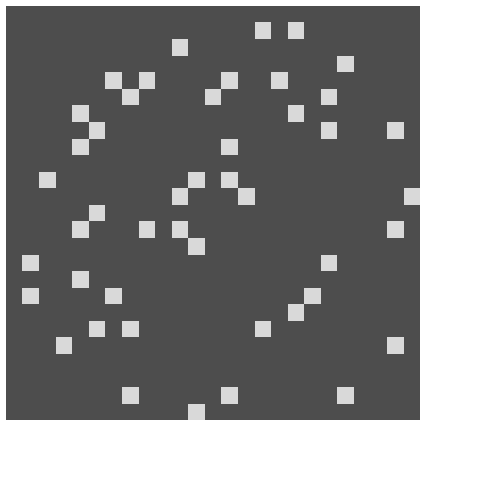}
        \includegraphics[width=0.09\textwidth]{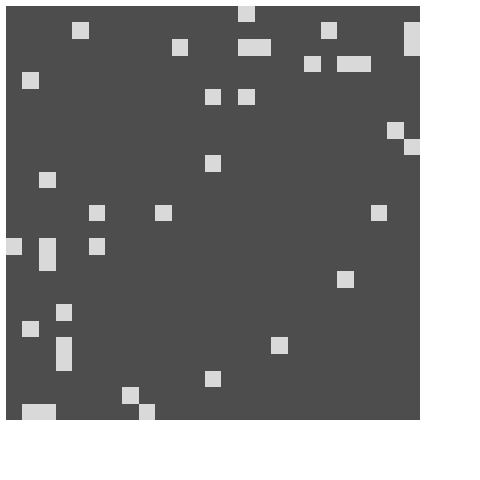}
        \includegraphics[width=0.09\textwidth]{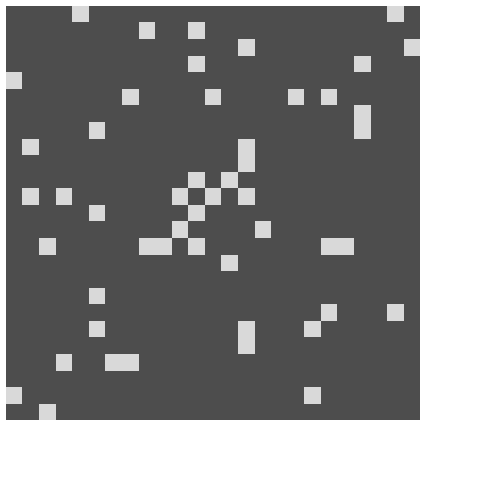}
        \includegraphics[width=0.09\textwidth]{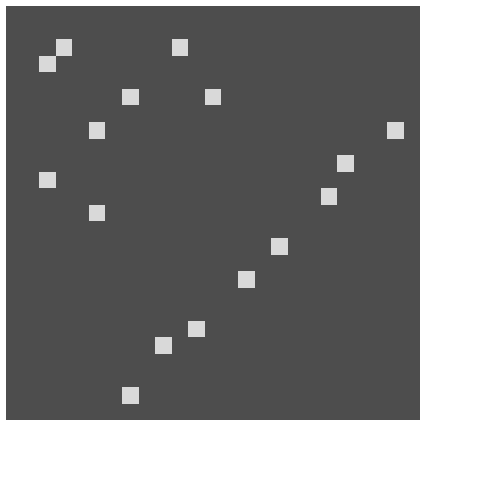}
        \includegraphics[width=0.09\textwidth]{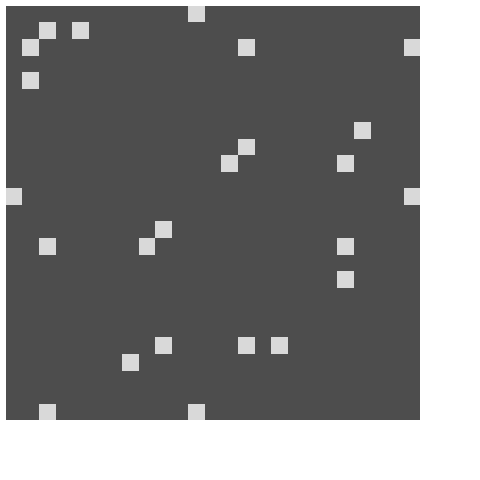}
        \includegraphics[width=0.09\textwidth]{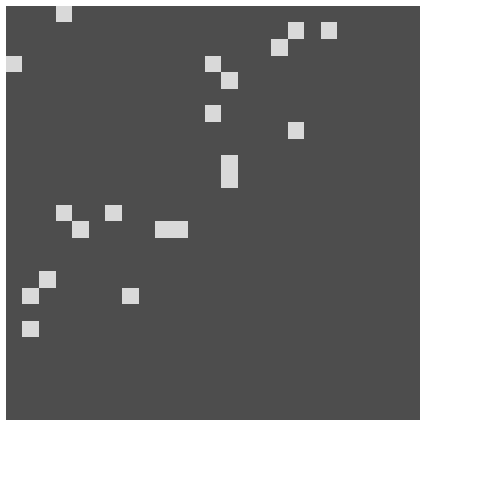}
        \includegraphics[width=0.09\textwidth]{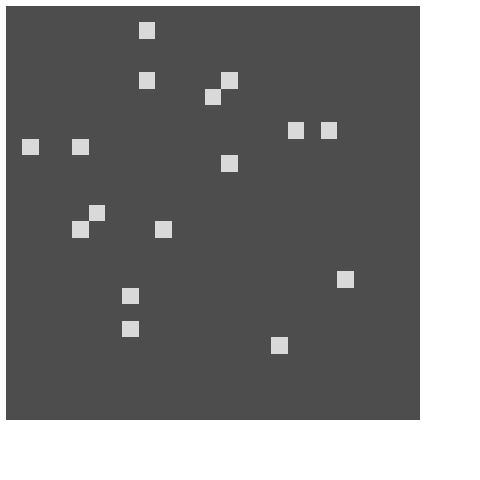}
        \includegraphics[width=0.09\textwidth]{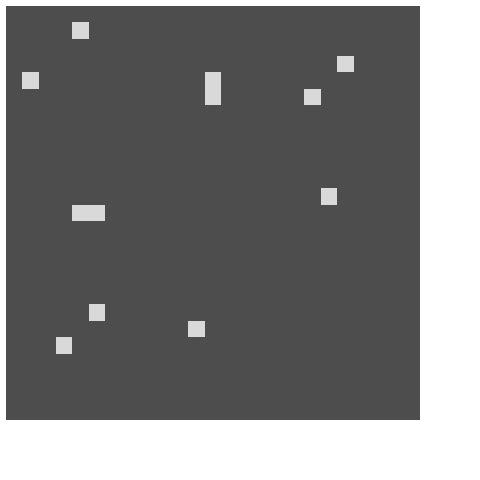}
        \includegraphics[width=0.09\textwidth]{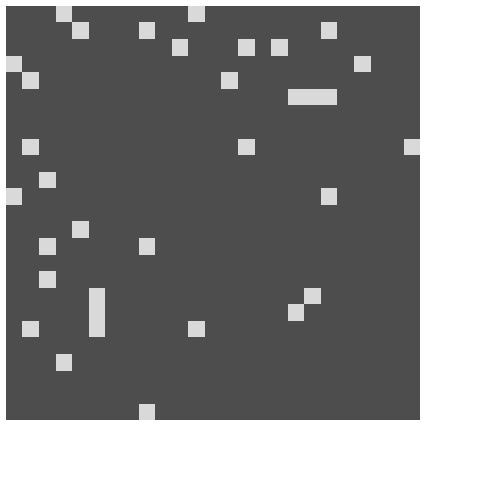}
    \end{subfigure}
    \begin{subfigure}[b]{1\textwidth}
        \centering
        \includegraphics[width=0.09\textwidth]{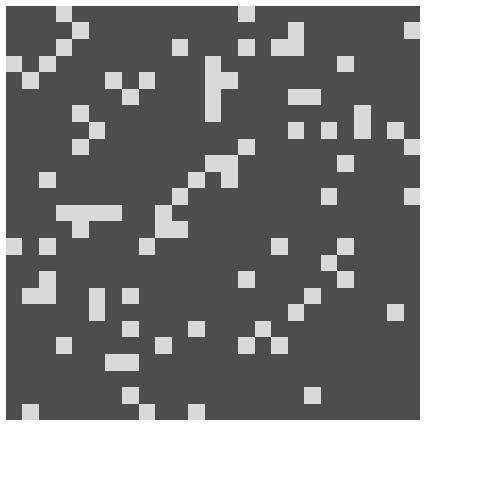}
        \includegraphics[width=0.09\textwidth]{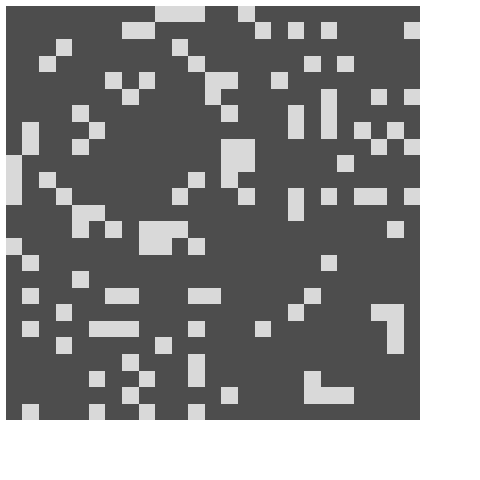}
        \includegraphics[width=0.09\textwidth]{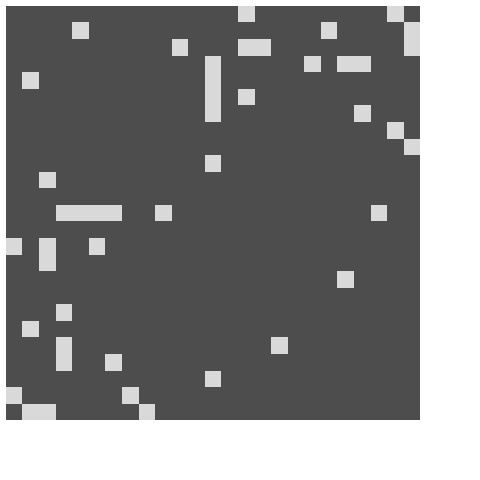}
        \includegraphics[width=0.09\textwidth]{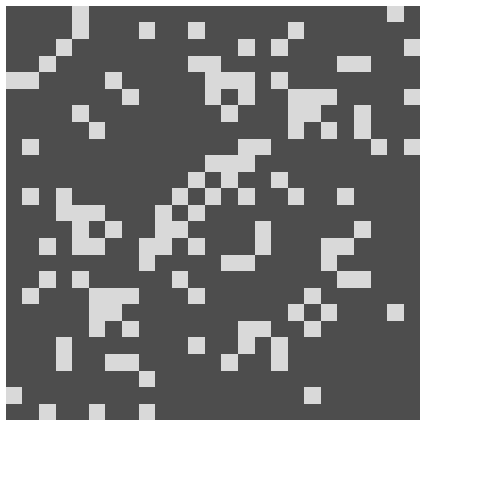}
        \includegraphics[width=0.09\textwidth]{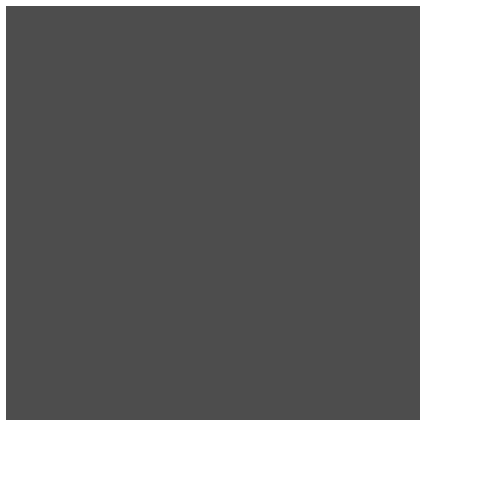}
        \includegraphics[width=0.09\textwidth]{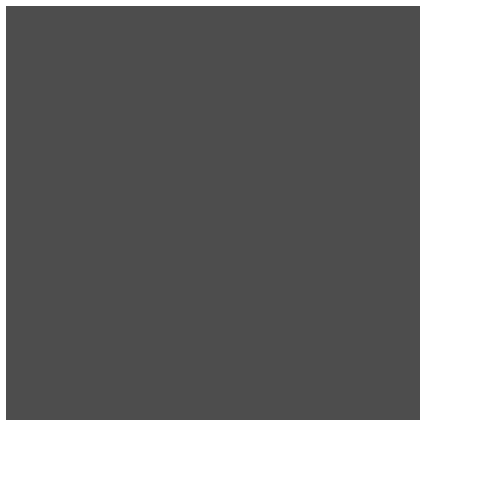}
        \includegraphics[width=0.09\textwidth]{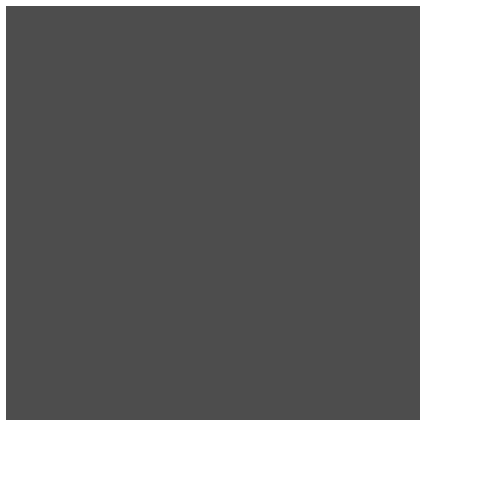}
        \includegraphics[width=0.09\textwidth]{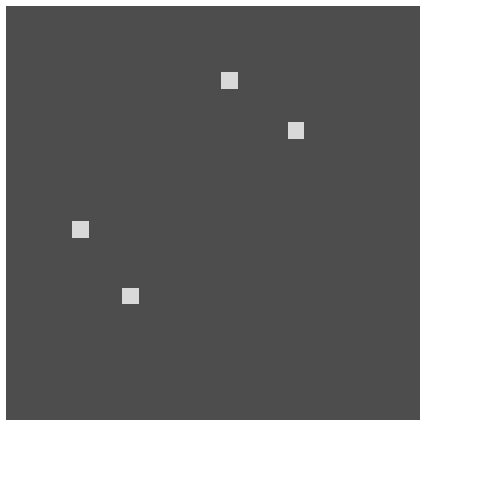}
        \includegraphics[width=0.09\textwidth]{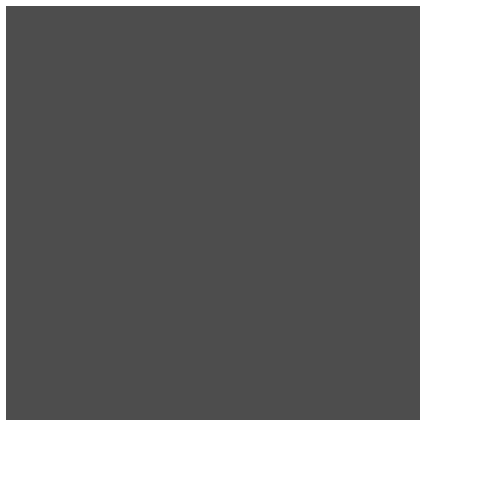}
        \includegraphics[width=0.09\textwidth]{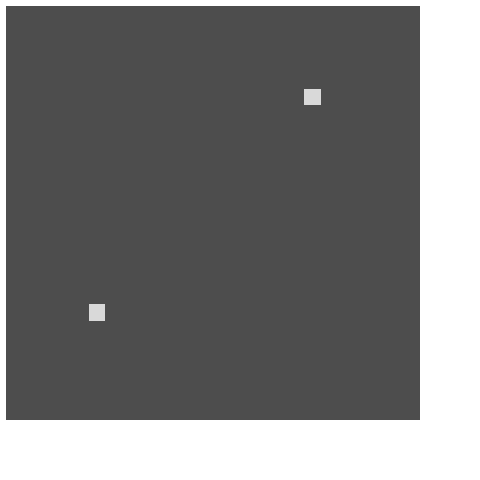}
    \end{subfigure}
     \caption{$N = 500$
     \label{fig:sim.pcoef_500}}
         \end{subfigure}
   \caption{Adjacency matrices corresponding to the precision coefficients $\bB_{1:10}$ (left to right), from one simulated dataset with sample size $N = 200$ (top) and $N = 500$ (bottom). In each plot, the five rows are the true matrices (1st row), and estimated matrices by Lasso (2nd row), GMMReg (3rd row), BSGSSS (4th row) and DGSS (5th row).}
    \label{fig:sim.pcoef_both}
\end{figure}

\subsection{Results}
Figure~\ref{fig:sim.pcoef_both} shows examples of the adjacency matrices corresponding to the precision coefficients $\bB_k, ~k=1,\ldots,q$, for one simulated dataset with $N=200,500$. For both sample sizes, we observe that Lasso and BSGSSS do not penalize the coefficients sufficiently, failing to eliminate covariates $X_{5-10}$. On the contrary, GMMReg penalizes the coefficients excessively, selecting too few edges. This phenomenon may be caused by their sparsity assumptions. Lasso and BSGSSS totally ignore the covariate-level sparsity by treating the problem as a high-dimensional linear regression at each node, ultimately failing to exclude those covariates with no impact on edges. On the other hand, GMMReg assumes a dense intercept and sparse covariates, and tunes the parameters via cross-validation, which fails to find an optimal penalty for both dense intercept and covariates under their assumption, ultimately limiting the number of edges selected. 
Compared to the three existing methods, the proposed DGSS achieves relatively good sparse estimates, identifying the important covariates with a reasonable sparsity level.  In unreported results, selecting prior parameters with greater sparsity, as formulated in Eq. \eqref{eq:prior.sparsity}, with $(a_k, b_k) = (1,3)$ instead of $(1,1)$, results in sparser graph estimates for $\bB_{1:4}$ in Figure~\ref{fig:sim.pcoef_both} and improved FPRs and other metrics in Table~\ref{tab:sim.all}.

\renewcommand{\arraystretch}{1} 
\begin{table}[!hbt]
\caption{Performance for edge and covariate detection. Results are averaged over 50 repeated simulations, with standard errors reported in parentheses.} \label{tab:sim.all}
\begin{subtable}[t]{1\textwidth}
\caption{\footnotesize Covariate-dependent edge detection}
\label{tab:sim.reg_cov_re}
\centering
\begin{adjustbox}{max width = 1\textwidth}
\begin{tabular}{cccccccccccc}
  \hline
  \hline
  & &\multicolumn{4}{c}{sample size $N=200$} & & & \multicolumn{4}{c}{sample size $N=500$}  \\
  \hline
   & & \texttt{Lasso} & \texttt{GMMReg} & \texttt{BSGSSS}  & \texttt{DGSS}  & & & \texttt{Lasso} & \texttt{GMMReg} & \texttt{BSGSSS} &  \texttt{DGSS}\\ 
    \hdashline 
        \multirow{2}{*}{TPR}    && 0.146 & 0.228 & 0.173 & 0.271    && &    0.429 & 0.303 & 0.508 & 0.679 \\ 
      && (0.009) & (0.007) & (0.009) & (0.013)   &&&    (0.016) & (0.005) & (0.012) & (0.012) \\ 
    \hdashline 
    \multirow{2}{*}{FPR}    && 0.023 & 0.025 & 0.017 & 0.025   &&&  0.043 & 0.028 & 0.062 & 0.066 \\ 
       && (0.001) & (0.001) & (0.001) & (0.001)   &&&   (0.001) & (0.001) & (0.002) & (0.002) \\ 
    \hdashline 
    \multirow{2}{*}{F1}    && 0.164 & 0.246 & 0.211 & 0.283   &&&  0.341 & 0.307 & 0.337 & 0.414 \\
       && (0.007) & (0.006) & (0.007) & (0.009)   &&  &   (0.007) & (0.005) & (0.004) & (0.004) \\ 
    \hdashline 
    \multirow{2}{*}{MCC} && 0.142 & 0.220 & 0.197 & 0.260   &&&    0.318 & 0.279 & 0.322 & 0.417 \\
    && (0.007) & (0.007) & (0.006) & (0.009)   &&&  (0.009) & (0.005) & (0.005) & (0.005) \\ 
  \hline
  \hline
\end{tabular}
\end{adjustbox}
\end{subtable} 

\begin{subtable}[t]{1\textwidth}
\caption{\footnotesize Group edge detection in the overall graph}
\label{tab:sim.reg_cov_re_graph} 
\centering
\begin{adjustbox}{max width = 1\textwidth}
\begin{tabular}{cccccccccccc}
  \hline
  \hline
  & &\multicolumn{4}{c}{sample size $N=200$} & & & \multicolumn{4}{c}{sample size $N=500$}  \\
  \hline
   & & \texttt{Lasso} & \texttt{GMMReg} & \texttt{BSGSSS}  & \texttt{DGSS}  & & & \texttt{Lasso} & \texttt{GMMReg} & \texttt{BSGSSS} &  \texttt{DGSS}\\ 
    \hdashline 
     \multirow{2}{*}{TPR}     && 0.345 & 0.552 & 0.266 & 0.412  &&& 0.658 & 0.748 & 0.648 & 0.780 \\ 
        && (0.016) & (0.011) & (0.011) & (0.013)  &&& (0.014) & (0.006) & (0.011) & (0.010) \\ 
    \hdashline 
    \multirow{2}{*}{FPR}    && 0.054 & 0.160 & 0.018 & 0.060  &&& 0.074 & 0.123 & 0.056 & 0.132 \\ 
       && (0.004) & (0.005) & (0.002) & (0.005)  &&& (0.005) & (0.007) & (0.003) & (0.005) \\ 
    \hdashline 
    \multirow{2}{*}{F1}  && 0.470 & 0.605 & 0.404 & 0.538  &&& 0.735 & 0.771 & 0.743 & 0.782 \\ 
      && (0.015) & (0.008) & (0.013) & (0.011)  &&& (0.009) & (0.004) & (0.008) & (0.005) \\ 
    \hdashline 
    \multirow{2}{*}{MCC} && 0.382 & 0.412 & 0.380 & 0.433  &&& 0.622 & 0.636 & 0.639 & 0.650 \\ 
        && (0.012) & (0.011) & (0.010) & (0.010)  &&& (0.010) & (0.009) & (0.010) & (0.008) \\ 
  \hline
  \hline
\end{tabular}
\end{adjustbox}
\end{subtable} 

\begin{subtable}[t]{1\textwidth}
\caption{\footnotesize Covariate selection}
\label{tab:sim.reg_cov_re_sel}
\centering
\begin{adjustbox}{max width = 1\textwidth}
\begin{tabular}{cccccccccccc}
  \hline
  \hline
  & &\multicolumn{4}{c}{sample size $N=200$} & & & \multicolumn{4}{c}{sample size $N=500$}  \\
  \hline
   & & \texttt{Lasso} & \texttt{GMMReg} & \texttt{BSGSSS}  & \texttt{DGSS}  & & & \texttt{Lasso} & \texttt{GMMReg} & \texttt{BSGSSS} &  \texttt{DGSS}\\ 
    \hdashline 
        \multirow{2}{*}{TPR}  && 1 & 0.845 & 1 & 0.925  &&&  1 & 0.895 & 1 & 1 \\ 
       && (0) & (0.030) & (0) & (0.019)  &&&  (0) & (0.022) & (0) & (0) \\ 
    \hdashline 
    \multirow{2}{*}{FPR} && 0.993 & 0.463 & 0.993 & 0.327  &&& 1 & 0.303 & 1 & 0.327 \\ 
      && (0.005) & (0.036) & (0.005) & (0.031)  &&& (0) & (0.039) & (0) & (0.029) \\ 
    \hdashline 
    \multirow{2}{*}{F1}  && 0.573 & 0.663 & 0.573 & 0.776  &&&  0.571 & 0.775 & 0.571 & 0.816 \\ 
     && (0.001) & (0.018) & (0.001) & (0.020)  &&& (0) & (0.018) & (0) & (0.015) \\ 
    \hdashline 
    \multirow{2}{*}{MCC} && 0.272 & 0.419 & 0.272 & 0.605  &&&  - & 0.637 & - & 0.686 \\ 
     && (0) & (0.035) & (0) & (0.038)  &&& - & (0.031) & - & (0.026) \\ 
  \hline
  \hline
\end{tabular}
\end{adjustbox}
\end{subtable}
\end{table}

We now proceed to assess each method using the edge and covariate selection metrics introduced above.  We consider TPR and FPR as a trade-off, as finding more true positives often comes at the cost of increasing false positives, and report them for completeness. For model comparison, we focus on F1 and MCC.
All results are reported in Table~\ref{tab:sim.all}, averaged across $50$ replicates. 
For covariate-dependent edge detection in Table~\ref{tab:sim.reg_cov_re}, performance metrics tend to be low for all methods, even with the relatively large sample size $N = 500$. In this scenario, Lasso only considers the local-level penalty, while the other methods, GMMReg, BSGSSS and DGSS include at least two level of selection/penalties. With sample size $N = 200$, GMMReg and BSGSSS outperform Lasso in terms of F1 and MCC, because their second level selection/penalty can efficiently exclude the empty coefficients. On the contrary, with the relatively large sample size, $N = 500$, their performances become comparable to the Lasso in terms of those two scores as their second level of selection/penalty does not fit the sparsity pattern in the data-generation process. Meanwhile the proposed DGSS method takes advantage of the multi-level selection and outperforms the other methods in terms of F1 and MCC.

Next, we evaluate the models' performance in edge detection for the \textit{overall graph}, which we define as the graph where an individual edge is present if affected by any of the covariates $x^k$. This corresponds to group edge selection at the node level. There are $p-1 = 24$ groups of edges (at the node group level) for each node, and on average $(p-1)\times 0.4 = 9.6$ of them are signals, yielding $(p - 1) p = 600$ groups across all nodes. 
Results in Table~\ref{tab:sim.reg_cov_re_graph} show comparable performance across all methods. Although the GMMReg seems to have a higher  F1 score than other methods when the sample size is $N=200$, this appears to be due to higher TPR score at the cost of higher FPR score. As evidence, its MCC scores are close to those of other methods, and hence, we do not conclude that there is a significant outperformance.

Finally, we look at the task of covariate selection based on the precision coefficient estimates $\hat{\bB}_k$. Four out of ten covariates are signals. We select $X_k$ if $\hat{\bB}_k \ne \bzeros$. From Table~\ref{tab:sim.reg_cov_re_sel} we observe the failure of the Lasso and BSGSSS methods in this task. Without sufficient selection/penalties, these two methods include all covariates in almost all replicates, as also illustrated in Figures~\ref{fig:sim.pcoef_200} and~\ref{fig:sim.pcoef_500}, with the exception of 2 out of 100 cases. Although GMMReg tends to favor a dense precision coefficient for the intercept and sparse precision coefficients for the other covariates, its performance is similar to that of DGSS. The proposed DGSS, on the other hand, outperforms GMMReg in terms of all averaged metrics.

\section{Application to Microbiome Data}
\label{sec:app}
We demonstrate the proposed method with data from the Multi-Omic Microbiome Study: Pregnancy Initiative (MOMS-PI), a study funded by the NIH Roadmap Human Microbiome Project to understand the impact of the vaginal microbiome on pregnancy and the fetal microbiome. This study contains samples from multiple body sites, including mouth, skin, vagina and rectum, of 596 subjects throughout pregnancy and for a short term after childbirth. 
%Previous research found the vaginal microbiome can change early in pregnancy and be predictive of pregnancy outcomes \citep{Serrano2019,Fettweis2019}.

\subsection{Data}
Data from the MOMS-PI study is publicly available and can be found in the R package \texttt{HMP2Data}.
Following \cite{Osborne2022}, we focus on the interplay between microbial abundances and vaginal cytokines, a mechanism by which the host regulates the composition of the vaginal microbiome. We use the first baseline visit data of the $N=225$ subjects whose microbiome and cytokine profiling of the vagina are available, ensuring a complete dataset for analyzing their associations. Furthermore, we consider $p=90$ OTUs whose absolute abundance is greater than 1 in at least $10\%$ of the subjects and use all the 29 available cytokines as covariates, adding an intercept term, which implies $q = 30$. We apply the centered log ratio transformation to normalize the abundance counts \citep{Aitchison1982,Gloor2017,Lin2020}, as commonly done in Gaussian graphical modeling for microbiome data to satisfy the Gaussian assumption \citep{Kurtz2015,Wilms2022}. After transformation, we center the data such that each OTU has zero mean. For the covariates, we transform the data to the log scale and use the min-max normalization, so that values fall within the $[0,1]$ interval. 

\subsection{Results}
Given the results obtained in the simulation study, we restrict comparisons to the GMMReg and the proposed DGSS methods.
For DGSS, we use the same non-informative prior specifications and MCMC settings as in the simulation study. 
On a server, with two 20-core 2.4 GHz Intel(R) Xeon CPUs, running the MCMC algorithm of our DGSS method, coded in Rcpp, took about 15 seconds per iteration. 

\begin{figure}[!thb]
    \centering
    \begin{subfigure}[b]{0.24\textwidth}
        \centering
        \includegraphics[width=1\textwidth]{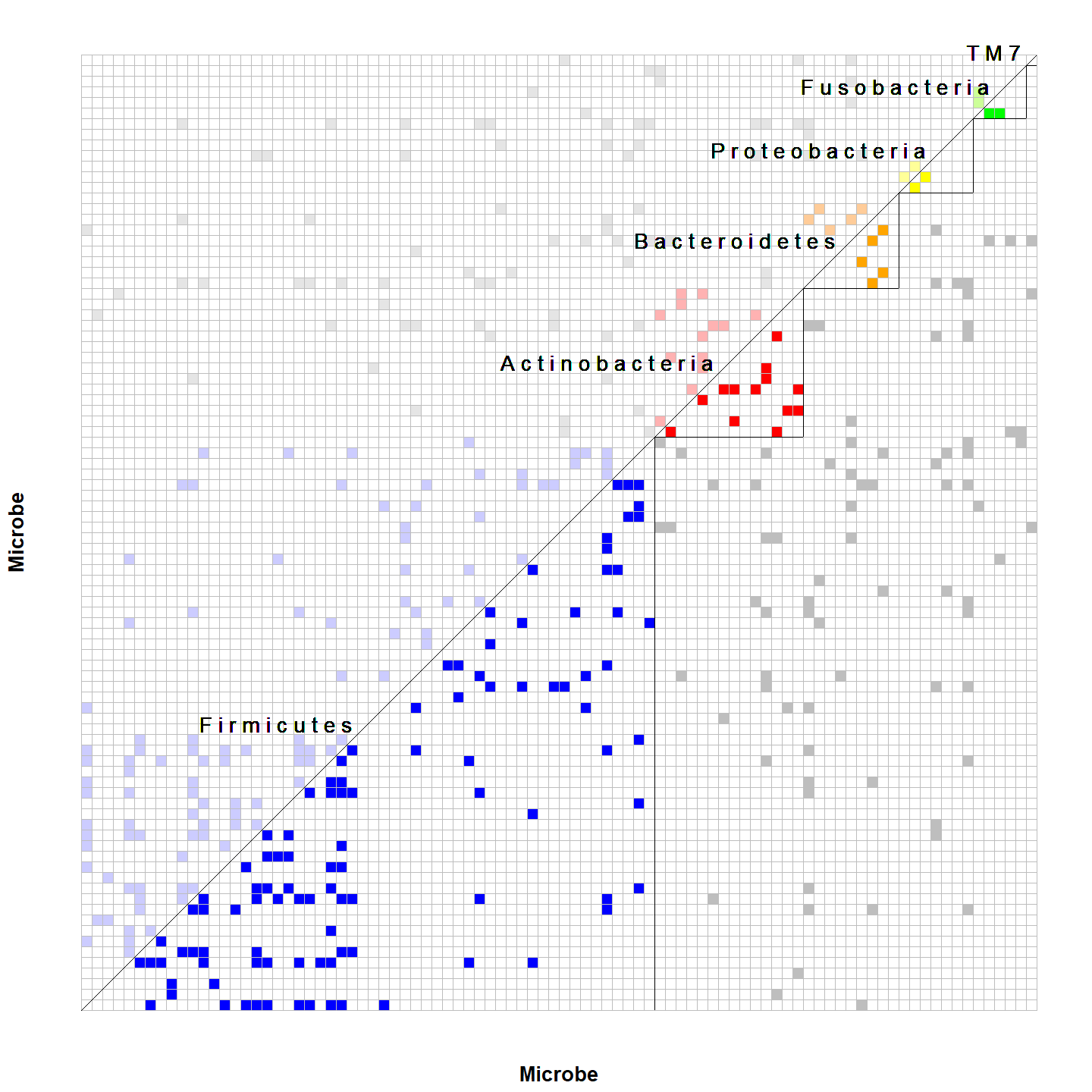}
    \caption*{ IL-1ra (204)}
    \end{subfigure}
    \begin{subfigure}[b]{0.24\textwidth}
        \centering
        \includegraphics[width=1\textwidth]{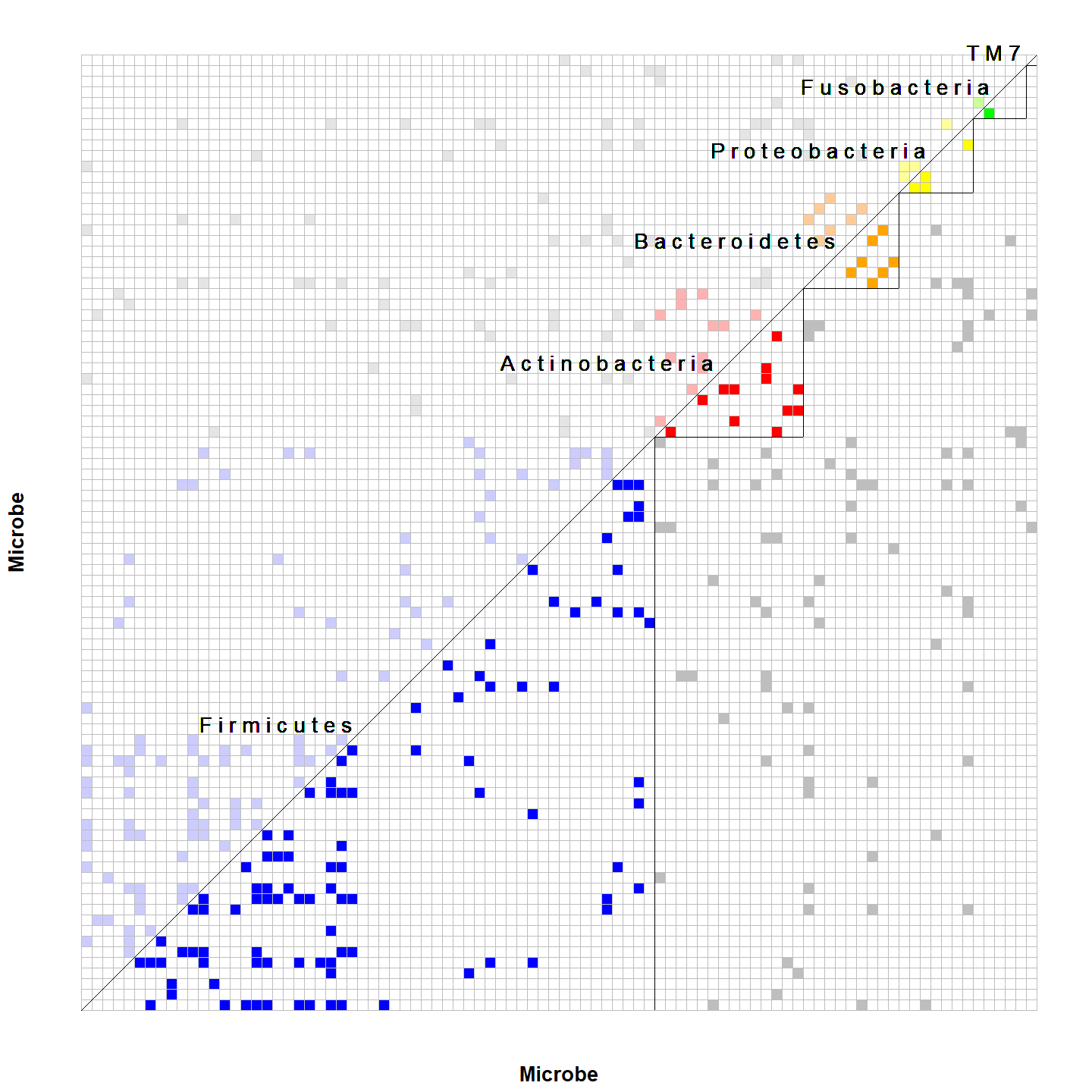}
    \caption*{ Baseline (201)}
    \end{subfigure}
    \begin{subfigure}[b]{0.24\textwidth}
        \centering
        \includegraphics[width=1\textwidth]{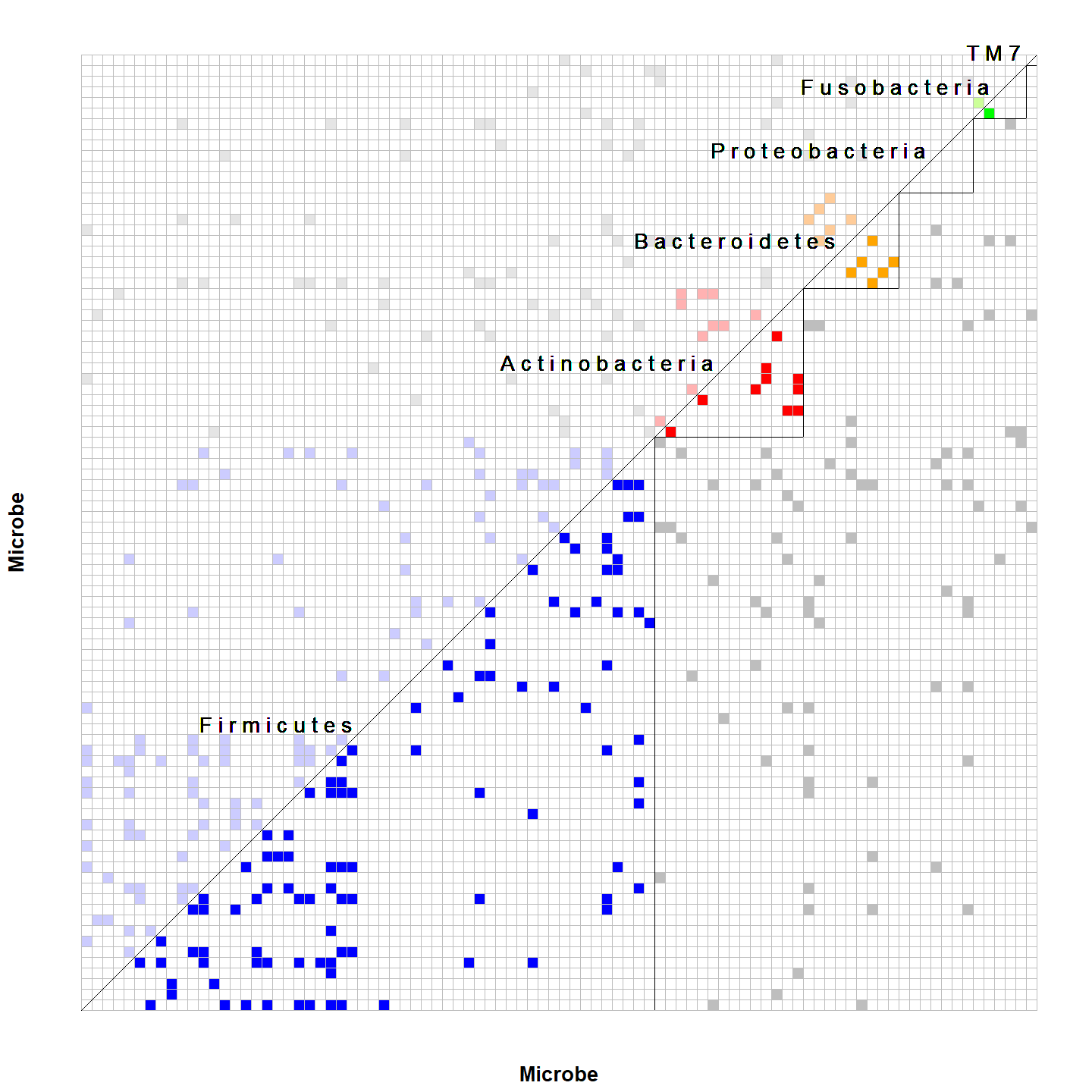}
    \caption*{ IL-2 (195)}
    \end{subfigure}
    \begin{subfigure}[b]{0.24\textwidth}
        \centering
        \includegraphics[width=1\textwidth]{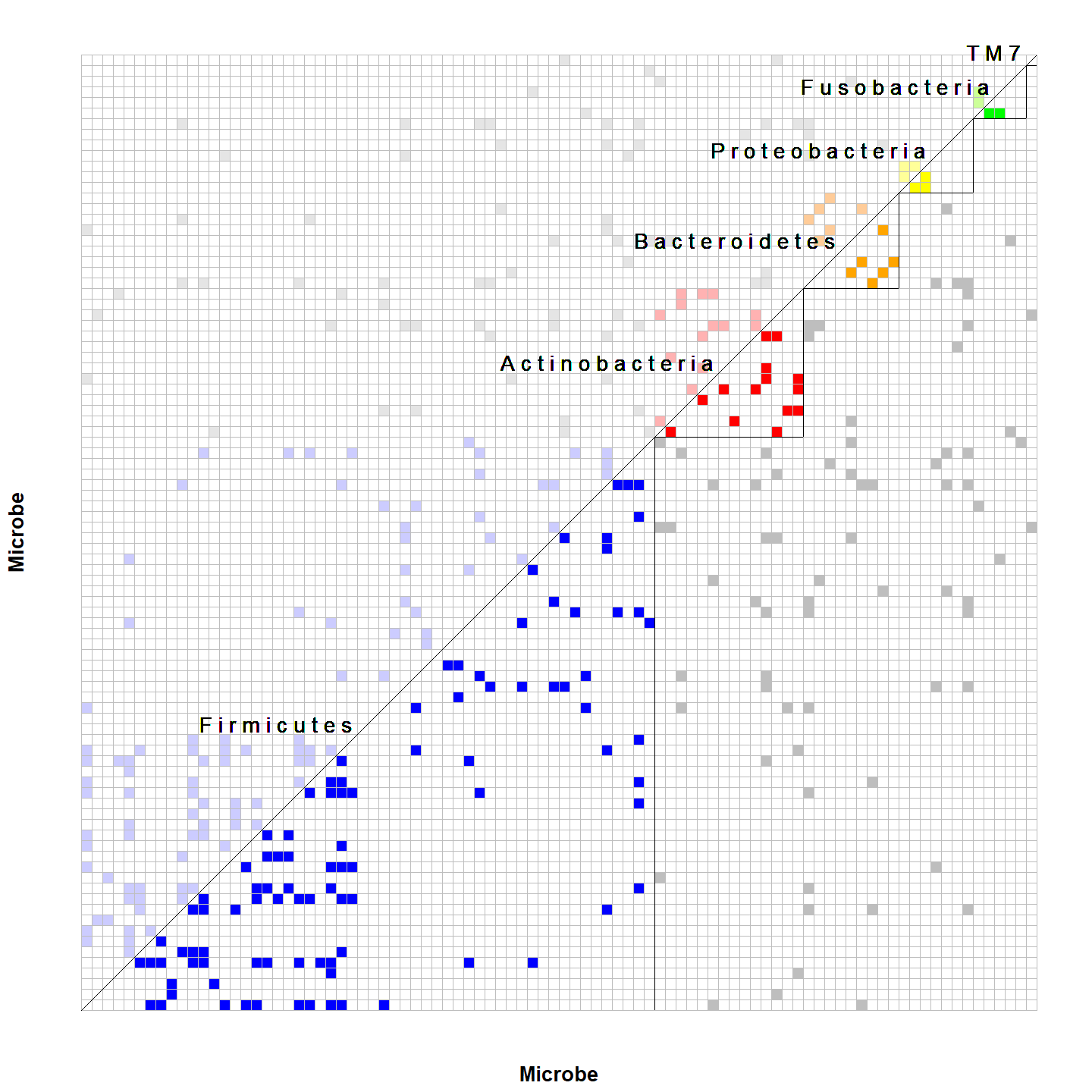}
    \caption*{ MCP-1(MCAF) (192)}
    \end{subfigure}
    \begin{subfigure}[b]{0.24\textwidth}
        \centering
        \includegraphics[width=1\textwidth]{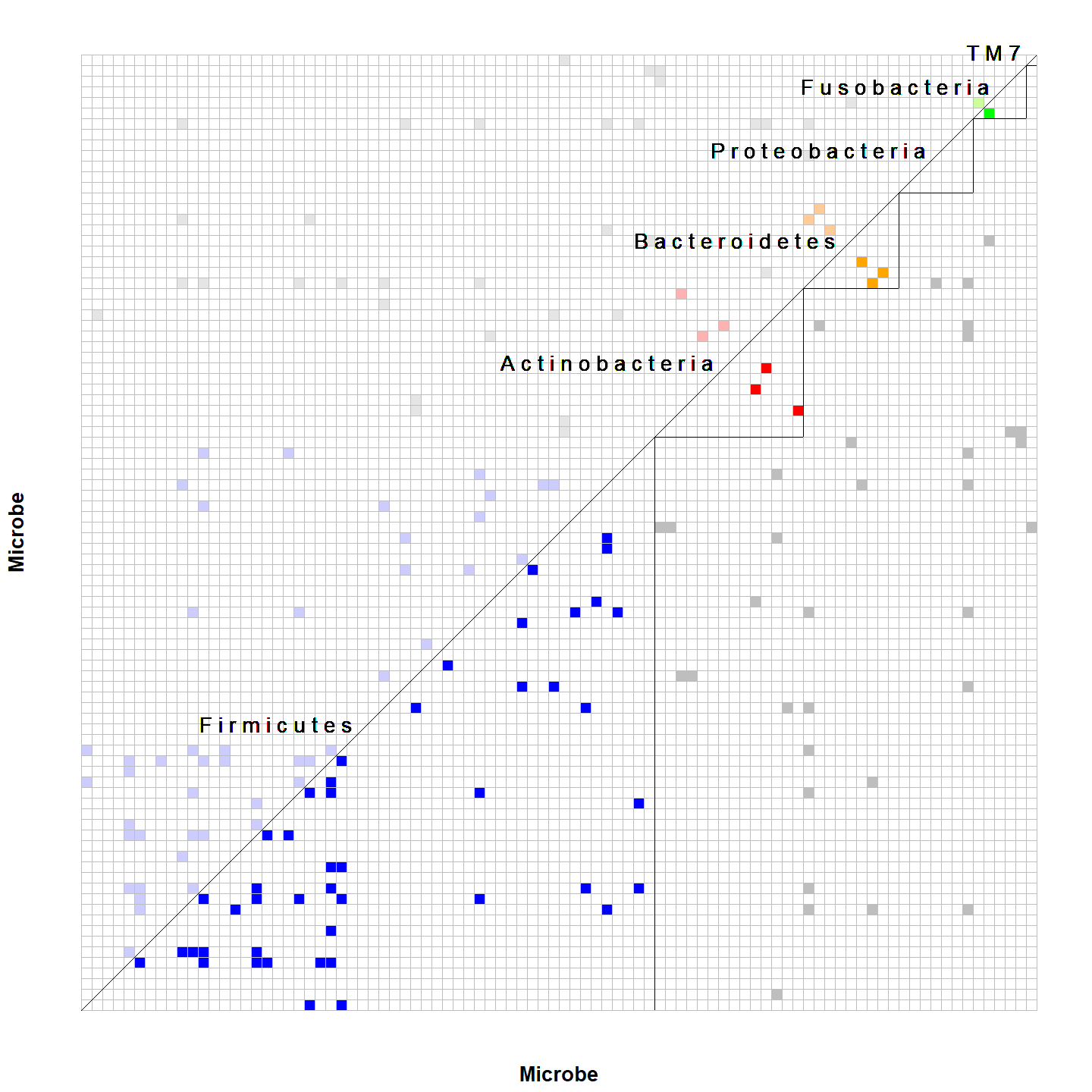}
    \caption*{ IL-4 (87)}
    \end{subfigure}
    \begin{subfigure}[b]{0.24\textwidth}
        \centering
        \includegraphics[width=1\textwidth]{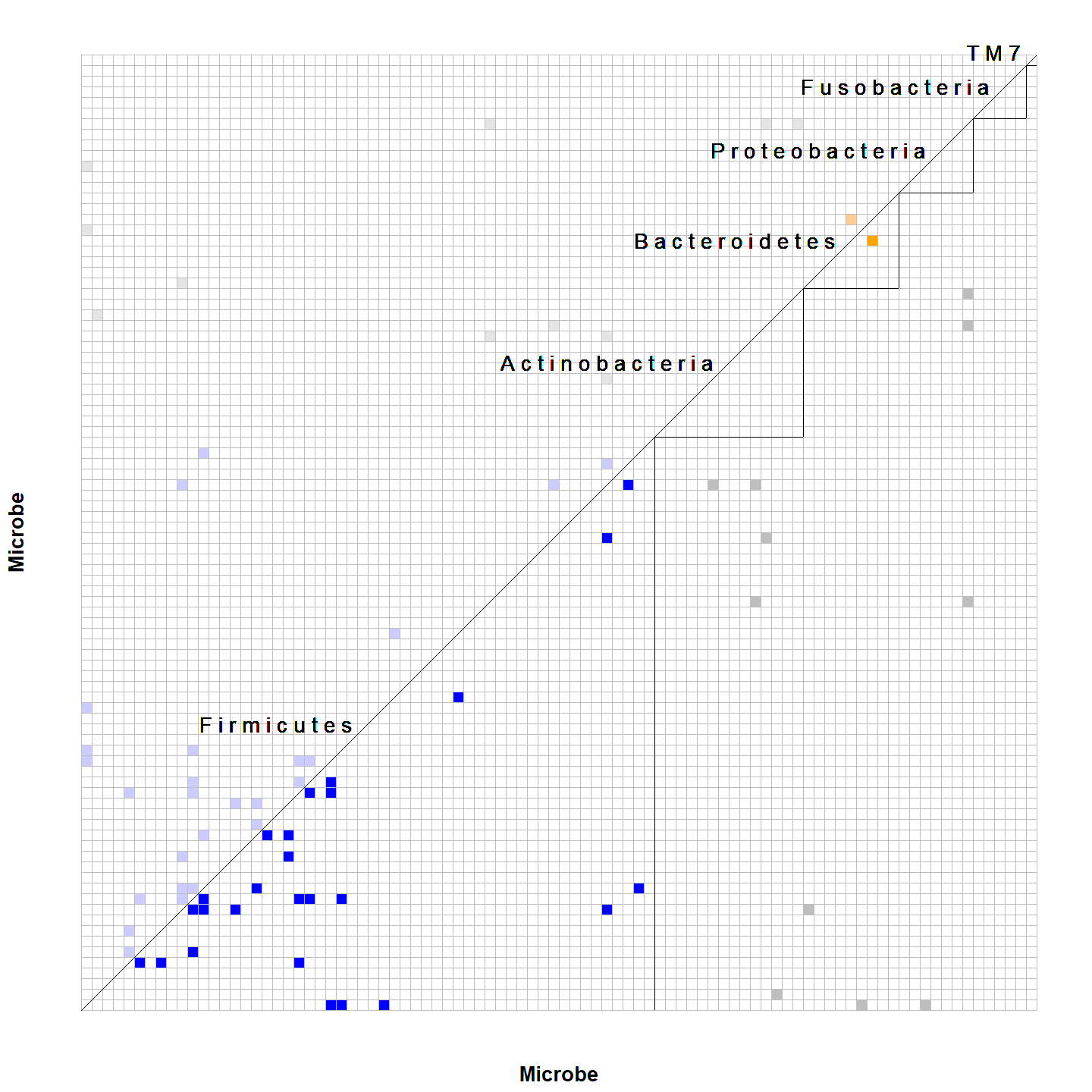}
    \caption*{ FGF basic (38)}
    \end{subfigure}
    \begin{subfigure}[b]{0.24\textwidth}
        \centering
        \includegraphics[width=1\textwidth]{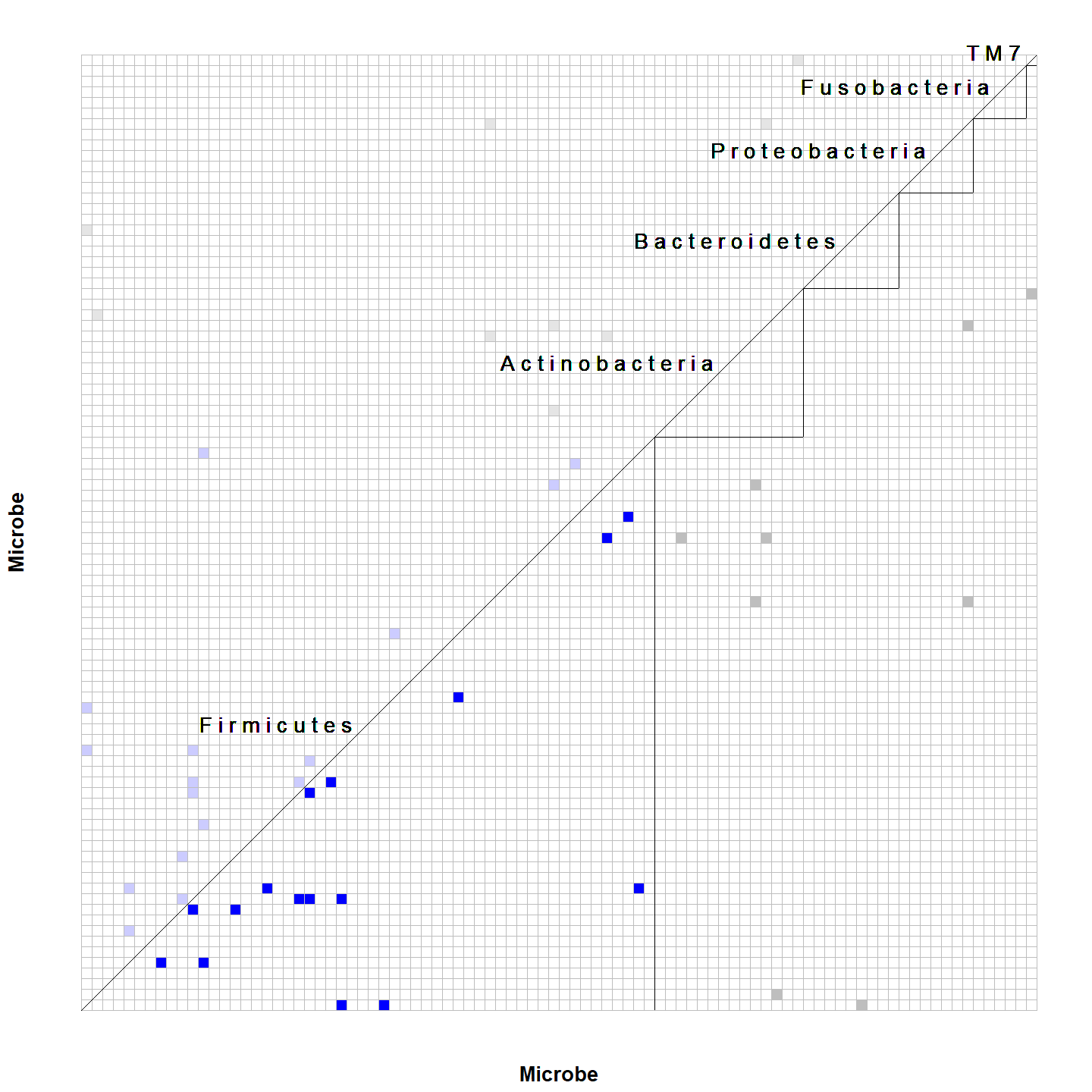}
    \caption*{ IL-17 (25)}
    \end{subfigure}
    \begin{subfigure}[b]{0.24\textwidth}
        \centering
        \includegraphics[width=1\textwidth]{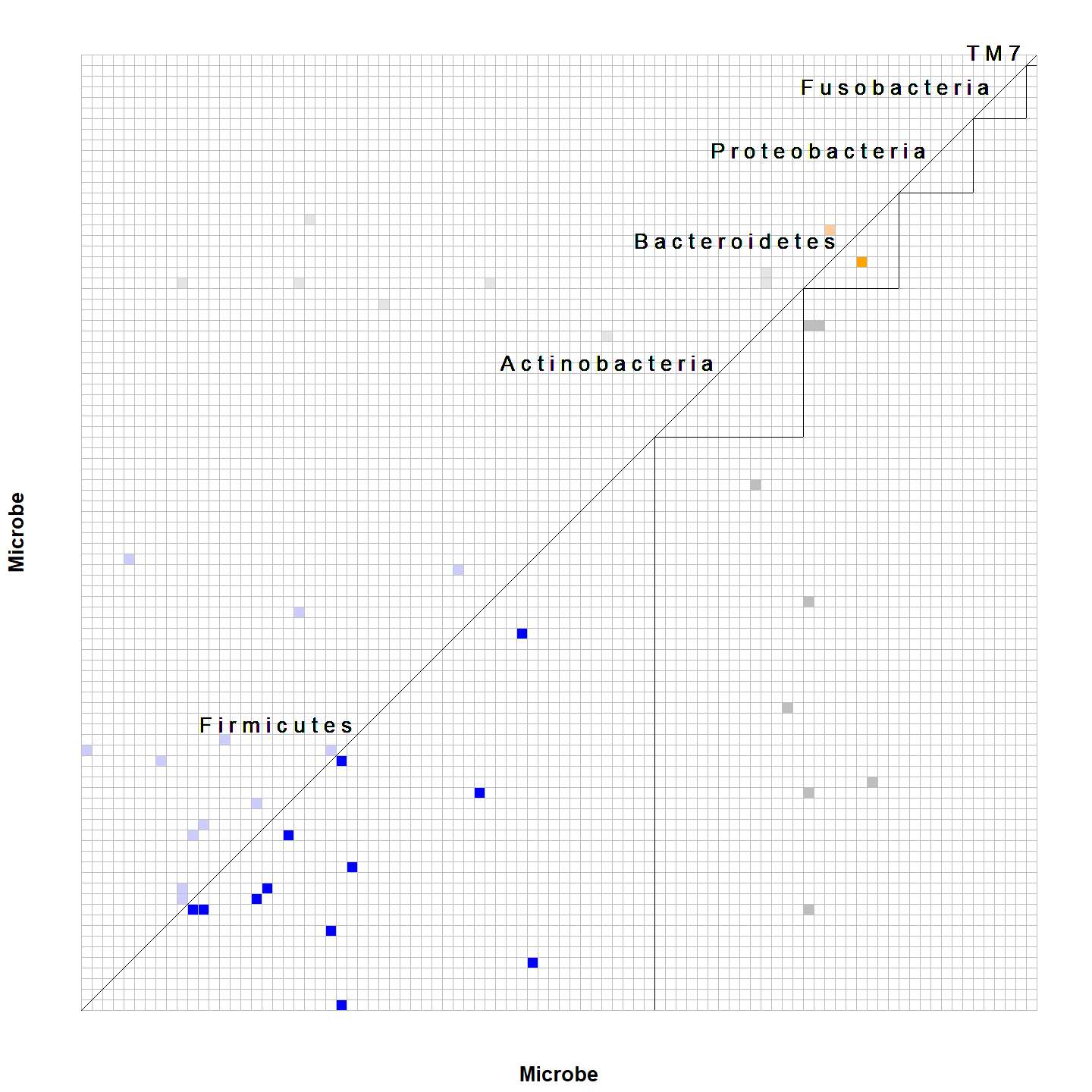}
    \caption*{ TNF-a (21)}
    \end{subfigure}
    \caption{Adjacency matrices corresponding to the precision coefficients $\bB_k$ of the four covariates with the most covariate-dependent edges (first row) and of the four covariates with the least covariate-dependent edges (second row), labeled by each covariate's name, with the number of edges indicated in parentheses.}
    \label{fig:app.NGLSS.coef_network}
\end{figure}

Table~\ref{tab:app.bio.edges.regression.} reports the number of covariate-dependent edges selected by DGSS and GMMReg for each covariate. Similar to the simulation study, we observe that GMMReg tends to select zero edges for almost all covariates, and a few more for the intercept (baseline). DGSS selects significantly more edges than GMMReg, with sparsity ranging from $21/4005 \approx 0.52\%$ for the sub-network influenced by TNF(a) to $204/4005 \approx 5.09\%$ for the sub-network influenced by IL-1ra. Figure~\ref{fig:app.NGLSS.coef_network} shows the adjacency matrices of the graphs corresponding to the precision coefficients $\bB_k$ of the four covariates with the most covariate-dependent edges and the four covariates with the least covariate-dependent edges. We observe that a large number of edges are simultaneously influenced by multiple cytokines. Additionally, it appears that some edges remain within a block of the OTU 1-26 across covariates, which aligns with a finding from \cite{Osborne2022}, as discussed next.

\begin{table}[!hbt]
\caption{Number of selected covariate-dependent edges}
\label{tab:app.bio.edges.regression.}
\centering
\begin{adjustbox}{max width = 1\textwidth}
\begin{tabular}{ccccccccccccccccc}
  \hline
  \hline
 &  \multirow{2}{*}{Baseline} & \multirow{2}{*}{Eotaxin}  & \multirow{2}{*}{FGF}  & \multirow{2}{*}{G-CSF}  & \multirow{2}{*}{GM-CSF}  & \multirow{2}{*}{IFN-g}  & \multirow{2}{*}{IL-10}& \multirow{2}{*}{  \shortstack[c]{IL-12\\(p70)}  }  & \multirow{2}{*}{IL-13}  & \multirow{2}{*}{IL-15}  & \multirow{2}{*}{IL-17A}  & \multirow{2}{*}{IL-1b}  & \multirow{2}{*}{IL-1ra}  & \multirow{2}{*}{IL-2} & \multirow{2}{*}{IL-4}  \\ 
 &   &  &   &   &    &    &  &   &   &    &    &    &    &  &   \\ 
    \hline 
GMMReg  & 97 & 0 & 0 & 0 & 0 & 0 & 0 & 0 & 0 & 0 & 0 & 0 & 1 & 0 & 0\\ 
DGSS & 201 & 104 & 105 & 190 & 167 & 163 & 172 & 104 & 189 & 132 & 146 & 123 & 204 & 195 & 87\\ 
\hline
\hline
 & \multirow{2}{*}{IL-5} & \multirow{2}{*}{IL-6} & \multirow{2}{*}{IL-7} & \multirow{2}{*}{IL-8} & \multirow{2}{*}{IL-9} & \multirow{2}{*}{IP-10} & \multirow{2}{*}{ \shortstack[c]{MCP-1 \\(MCAF)} } & \multirow{2}{*}{ \shortstack[c]{MIP \\ (1a)} } & \multirow{2}{*}{ \shortstack[c]{MIP \\ (1b)} } & \multirow{2}{*}{ \shortstack[c]{PDGF \\ (bb)} } & \multirow{2}{*}{ \shortstack[c]{RAN- \\ TES} } & \multirow{2}{*}{ \shortstack[c]{TNF \\ (a)} }  & \multirow{2}{*}{VEGF} & \multirow{2}{*}{ \shortstack[c]{FGF \\ basic}  } & \multirow{2}{*}{IL-17}\\ 
  &   &  &   &   &    &    &  &   &   &    &    &    &    &  & \\
    \hline 
GMMReg  & 1 & 0 & 0 & 0 & 0 & 0 & 0 & 0 & 0 & 0 & 0 & 0 & 0 & 0 & 0 \\ 
DGSS& 93 & 154 & 165 & 130 & 156 & 170 & 192 & 186 & 100 & 147 & 167 & 21 & 127 & 38 & 25 \\ 
\hline
\hline
\end{tabular}
\end{adjustbox}
\end{table} 

\begin{figure}[!thb]
    \centering
    \begin{subfigure}[b]{0.49\textwidth}
        \centering
        \includegraphics[width=1\textwidth]{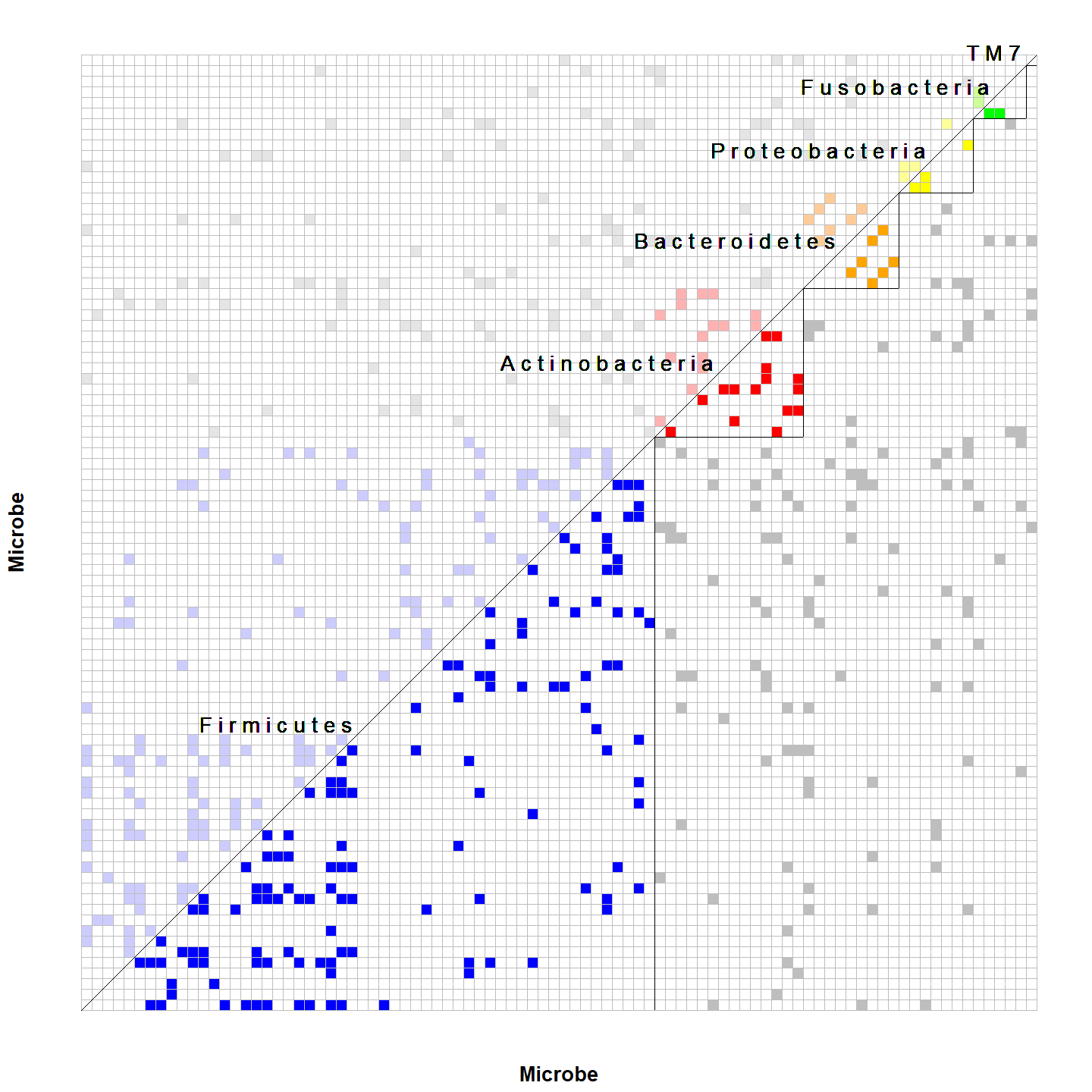}
    \caption*{ (a) DGSS}
    \end{subfigure}
    \begin{subfigure}[b]{0.49\textwidth}
        \centering
        \includegraphics[width=1\textwidth]{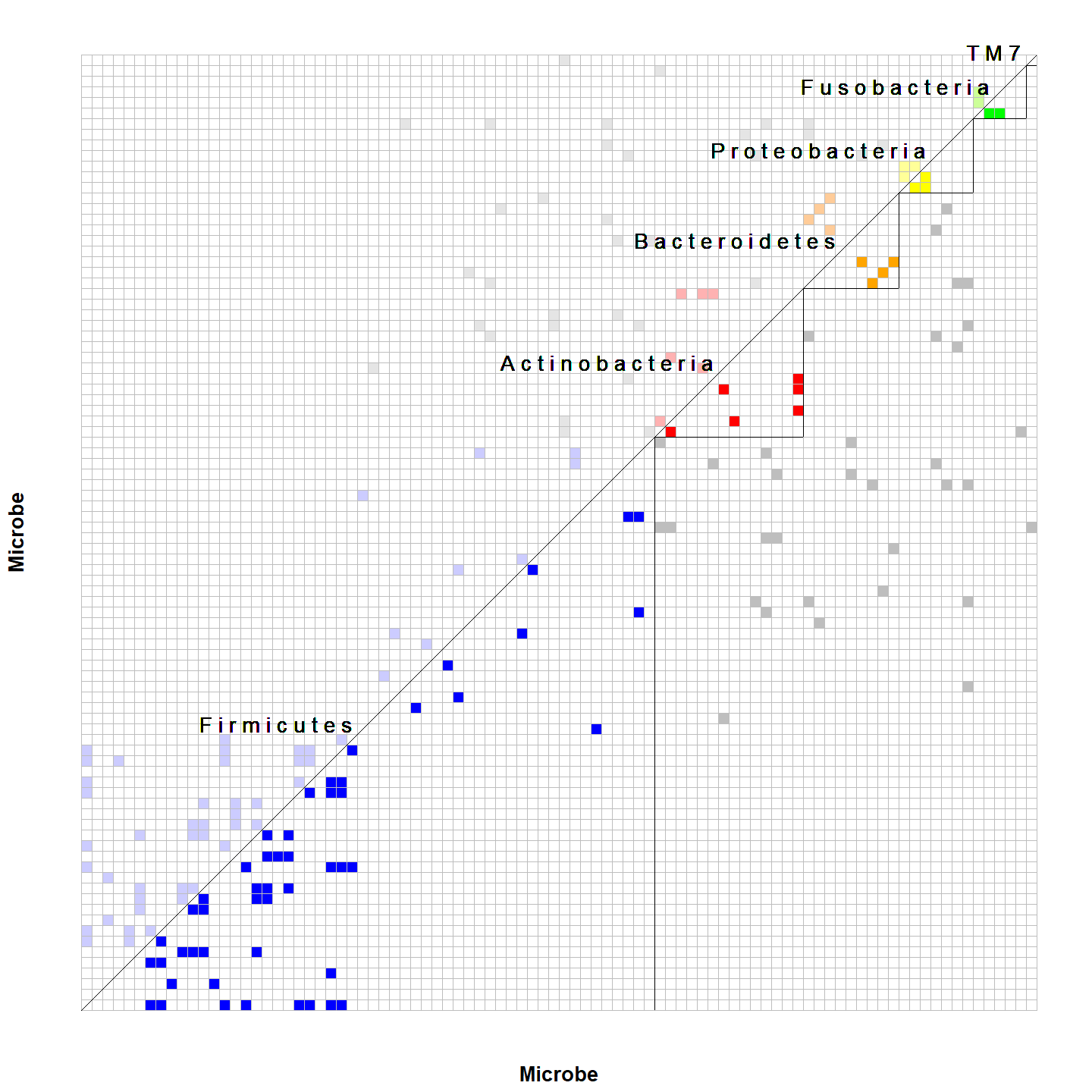}
    \caption*{ (b) Common edges }
    \end{subfigure}
\caption{Adjacency matrix of the overall graph selected by DGSS  (a) and the common edges selected by both DGSS and \protect\cite{Osborne2022} (b).}
    \label{fig:app.NGLSS.network}
\end{figure}

Figure~\ref{fig:app.NGLSS.network} shows the adjacency matrix of the overall graph selected by DGSS, together with a plot showing the commonly selected edges with the graph selected in \cite{Osborne2022}. OTUs are grouped based on their phylum (Firmicutes, Actinobacteria, Bacteroidetes, Proteobacteria, Fusobacteria, and TM7). Interestingly, even though \cite{Osborne2022} used a different Bayesian approach from our method, based on a latent Gaussian graphical model with separate variable selection priors for both covariate-dependent mean and covariate-independent precision,  we found substantial agreement in the selected edges. The overall graph sparsity is similar between the two approaches, with \cite{Osborne2022} reporting a sparsity level of $294/4005 \approx 7.34\%$ and our method yielding $271/4005 \approx 6.77\%$. Moreoever, a large number of edges in the overall graph selected by DGSS were also detected by their method (98 out of 271 edges). In particular, the common edges highlight the subnetwork formed by OTUs 1 - 26 within \textit{Firmicutes}, which appears to be the area consistently detected as having covariate-dependent edges for various cytokines. Jointly, these findings may imply the existence of a subnetwork within the \textit{Firmicutes} that is widely affected by cytokines. Additionally, the proposed method selects more inter-phylum edges compared to the model from \cite{Osborne2022}, which capture correlations among different OTUs across phyla, suggesting more complex latent effects of microbiome during pregnancy.

In the Supplementary Materials, we provide additional details on the cytokines impacting edges selection and a sensitivity analysis of the covariate-level selection threshold $d_k$ for the DGSS method. The code to reproduce the results, along with a table documenting the cytokines impacting edges, is available on GitHub. %This allows to perform a more detailed examination of covariate-influential edges.}

\section{Concluding remarks}
\label{sec:final}
We have considered the framework of covariate-dependent Gaussian graphical modeling for learning heterogeneous graphs and proposed a dual group spike-and-slab prior that achieves simultaneous local sparsity and bi-directional group sparsity.  The proposed prior accomplishes covariate-level selection, inferred by the local-level selection, on grouped precision coefficients sliced in one direction and the node-level selection on grouped coefficients sliced in another direction. 
Our approach has led to a parsimonious model for covariate-dependent precision matrices with improved interpretability. For posterior inference, we have designed a Gibbs sampler to automatically tune the hyper-parameters while incorporating their uncertainty, leads to interpretable and flexible selection results. Through simulation studies, we have demonstrated that the proposed model outperforms existing methods in its accuracy of graph recovery. We have applied our model to microbiome data to estimate the interaction between microbes in the vagina, as well as the interplay between vaginal cytokines and microbial abundances, providing insight into mechanisms of host-microbial interaction during pregnancy. 

There are several interesting future directions to extend our model. 
First, the model can be expanded to incorporate a covariate-adjusted mean. A potential challenge here is the increased computational complexity due to a larger parameter space.  Secondly, although our focus is on Gaussian graphical models, the structured sparsity we consider can be useful for other models with ultra  high-dimensional parameter spaces, such as arrays, that exhibit various grouping directions. Finally, approximation methods such as variational inference may merit investigation, as they improve scalability and enable application to larger datasets; the main challenges involve deriving a variational algorithm that accounts for dependencies across multi-level selection indicators and ensuring proper uncertainty quantification.

\section*{Supplementary Materials}
Web Appendices, Tables, and Figures, and data and code referenced in Sections~\ref{sec:sampler},~\ref{sec:sim} and~\ref{sec:app} are available with this paper at the Biometrics website on Oxford Academic and on GitHub at \href{https://github.com/ZijianZeng/BCDR_DGSS}{https://github.com/ZijianZeng/BCDR\_DGSS}.

\section*{Data Availability}
Both simulation and application data are available on GitHub: \\
\href{https://github.com/ZijianZeng/BCDR_DGSS}{https://github.com/ZijianZeng/BCDR\_DGSS}

% The supplementary material includes detailed derivations of the Gibbs sampler in Section~\ref{sec:sampler}. R code and scripts to reproduce the results from the simulation study and the real data applications, with main functions coded in Rcpp, will be made available on Github upon acceptance of the paper.

\bibliographystyle{apalike}
\bibliography{reference}

\newpage
\begin{center}
    \Large\bfseries Supplementary Materials for ``Bayesian Covariate-Dependent Graph Learning with a Dual Group Spike-and-Slab Prior"
\end{center}

\section*{S1. Markov Chain Monte Carlo Sampling (MCMC)}
We provide detailed derivations and the pseudocode for the Gibbs sampler used in the main paper. 

\begin{itemize}
	\item \textbf{Update the covariate-level selection parameters $\left\{ \tau^{ij}_k, \tilde{\tau}^{ij}_k, \gamma^{ij}_k, \pi_k\right\}$} \\ 
    Rewriting  the likelihood of $y^i_n$ from the covariate-level group perspective, we have that the mean part is
	$$
		\underbrace{ \sum_{s \ne k }  \sum_{l\ne i} \beta^{il}_s   y^l_n x^s_n}_{  \substack{  \text{ conditional on $s$} \\ \text{denoted as } c^{1, ijk}_{n} }} +  \sum_{j \ne i } \beta^{ij}_k  y^j_n x^k_n, 
    $$
    leading to the conditional distribution 
    $$
    (y^i_n - c^{1, ijk}_{n}) | - \sim N \left(  \sum_{j\ne i} \beta^{ij}_k y^j_n x^k_n,  \sigma^2_i \right).
    $$
    Similarly, with $\beta^{ij}_k = \tau^{ij}_kb^{ij}_k$, we have
    \begin{eqnarray*}
    (y^i_n - c^{1, ijk}_{n}) | - \sim N \left(  \underbrace{\sum_{j' \notin \{i, j\} } \beta^{ij'}_k y^{j'}_n x^k_n }_{  \substack{   \text{ conditional on } j' \notin \{i,j\} \\ \text{denoted as } c^{2,ijk}_n }} + \tau^{ij}_kb^{ij}_k  y^j_n x^k_n,\sigma^2_i\right).
    \end{eqnarray*}
    Denoting $y^{ijk}_n = y^i_n - c^{1,ijk}_n - c^{2,ijk}_n$,
    we have that the distribution of the latent coefficients conditional upon the indicators is
    \begin{eqnarray*}
    y^{ijk}_n | \tilde{\tau}^{ij}_{k}, \gamma^{ij}_k = 1, - & \sim & N\left( \tilde{\tau}^{ij}_{k}  b^{ij}_k y^j_n x^k_n ,  \sigma^2_i  \right) \\
    y^{ijk}_n |  \gamma^{ij}_k = 0, - & \sim & N\left( 0,  \sigma^2_i  \right).
    \end{eqnarray*}
    Following \cite{Zeng2024}, we integrate out the latent coefficients, obtaining
    \begin{equation*}
    \begin{aligned}
    & p\left( \gamma^{ij}_k  = 1 \left|  - \right. \right)   \\ 
    & = \frac{  \int p\left(  \gamma^{ij}_k = 1, \tilde{\tau}^{ij}_k | - \right)d\tilde{\tau}^{ij}_k }{  p\left(  \gamma^{ij}_k = 0, \tilde{\tau}^{ij}_k =0 |- \right)  + \int p\left(  \gamma^{ij}_k = 1, \tilde{\tau}^{ij}_k | - \right)d\tilde{\tau}^{ij}_k } \\
    & = \frac{1}{1 + \theta^{ij}_k },
    \end{aligned}
    \end{equation*}
    where the posterior probability ratio is
    \begin{equation*}
    \begin{aligned}
    & \theta^{ij}_k   = \frac{ p\left(  \gamma^{ij}_k = 0, \tilde{\tau}^{ij}_k =0 |- \right)  }{ \int p\left(  \gamma^{ij}_k = 1, \tilde{\tau}^{ij}_k | - \right)d\tilde{\tau}^{ij}_k}  \\
    & = \frac{ \frac{1}{p\left( y^{ijk}_{\cdot}\right)} p\left(  y^{ijk}_{\cdot} | \gamma^{ij}_k = 0, \tilde{\tau}^{ij}_k = 0 \right) \times \left( 1 - \pi_k \right) }{  \frac{1}{ p\left( y^{ijk}_{\cdot} \right)  }  \int p\left( y^{ijk}_\cdot | \gamma^{ij}_k = 1, \tilde{\tau}^{ij}_k \right)p\left( \tilde{\tau}^{ij}_k \right)d \tilde{\tau}^{ij}_k \times \pi_k }  \\
    & = \frac{  p\left(  y^{ijk}_{\cdot} | \gamma^{ij}_k = 0, \tilde{\tau}^{ij}_k = 0 \right) \times \left( 1 - \pi_k \right) }{   \int p\left( y^{ijk}_\cdot | \gamma^{ij}_k = 1, \tilde{\tau}^{ij}_k \right)p\left( \tilde{\tau}^{ij}_k \right)d \tilde{\tau}^{ij}_k \times \pi_k } \\
    & = \left(  \frac{\int p\left( y^{ijk}_\cdot | \gamma^{ij}_k = 1, \tilde{\tau}^{ij}_k \right)p\left( \tilde{\tau}^{ij}_k \right)d \tilde{\tau}^{ij}_k  }{ p\left(  y^{ijk}_{\cdot} | \gamma^{ij}_k = 0, \tilde{\tau}^{ij}_k = 0 \right) } \right)^{-1} \times \frac{1-\pi_k}{\pi_k}.
    \end{aligned}
    \end{equation*}

    Next, we estimate $\theta^{ij}_k $. First we note that 
    $$
        p\left( y^{ijk}_\cdot | \gamma^{ij}_k = 0, \tilde{\tau}^{ij}_k = 0\right) = \left( 2\pi \sigma^2_i\right)^{-\frac{N}{2}}  \exp\left\{ -\frac{1}{2}  \sum^N_{n=1} \left( y^{ijk}_n\right)^2 / \sigma^2_i\right\} .
    $$
    Therefore, 
    \begin{equation*}
    \begin{aligned}
        & \int p\left( y^{ijk}_\cdot | \gamma^{ij}_k = 1, \tilde{\tau}^{ij}_k \right)p\left( \tilde{\tau}^{ij}_k \right)d \tilde{\tau}^{ij}_k  \\
    = &  \int \left( 2\pi \sigma^2_i \right)^{-\frac{N}{2}}\exp \left\{ - \frac{1}{2} \sum^N_{n=1} \left( y^{ijk}_n - y^j_n x^k_nb^{ij}_k\tilde{\tau}^{ij}_k \right)^2  / \sigma^2_i \right\} \\
    & \times 2\left( 2 \pi s^2_k \right)^{-\frac{1}{2}} \exp \left\{ - \frac{1}{2} \left( \tilde{\tau}^{ij}_k \right)^2 / s^2_k\right\} \mathbbm{1}\left( \tilde{\tau}^{ij}_k \ge 0 \right) d \tilde{\tau}^{ij}_k \\
    = & 2 \left( 2 \pi s^2_k \right)^{-\frac{1}{2}} \underbrace{ \left( 2 \pi \sigma^2_i \right)^{-\frac{N}{2}}  \exp\left\{  -\frac{1}{2} \sum^N_{n=1}\left( y^{ijk}_n \right)^2/ \sigma^2_i\right\}}_{ = p\left( y^{ijk}_\cdot | \gamma^{ij}_k = 0, \tilde{\tau}^{ij}_k = 0\right) }\\
    & \times \int  \exp\left\{ -\frac{1}{2} \left[ \left(   \sum^N_{n=1} \left(y^j_n x^k_n \right)^2 \left(b^{ij}_k\right)^2 / \sigma^2_i   + 1/ s^2_k\right) \left( \tilde{\tau}^{ij}_k \right)^2-2\left(  b^{ij}_k \sum^N_{n=1} y^j_n x^k_n y^{ijk}_n / \sigma^2_i  \right) \tilde{\tau}^{ij}_k \right] \right\} \\ 
    & \mathbbm{1} \left(\tilde{\tau}^{ij}_k \ge 0\right) d \tilde{\tau}^{ij}_k.
    \end{aligned}
    \end{equation*}
    Letting $\tilde{\nu}^2_{ijk} = \left(  \sum^N_{n=1} \left(y^j_n x^k_n \right)^2 \left(b^{ij}_k\right)^2 / \sigma^2_i   + 1/ s^2_k\right)^{-1} $ and  $\tilde{m}_{ijk} = \tilde{\nu}^{2}_{ijk}  b^{ij}_k \sum^N_{n=1} y^j_n x^k_n y^{ijk}_n / \sigma^2_i$, we obtain the ratio
    \begin{equation*}
    \begin{aligned}
    & \frac{\int p\left( y^{ijk}_\cdot | \gamma^{ij}_k = 1, \tilde{\tau}^{ij}_k \right)p\left( \tilde{\tau}^{ij}_k \right)d \tilde{\tau}^{ij}_k  }{ p\left(  y^{ijk}_{\cdot} | \gamma^{ij}_k = 0, \tilde{\tau}^{ij}_k = 0 \right) }  \\
        = & 2 \left( 2 \pi s^2_k \right)^{-\frac{1}{2}}  \left( 2\pi \tilde{\nu}^{2}_{ijk}\right)^{\frac{1}{2}} \exp\left( \frac{1}{2} \frac{ \tilde{m}^2_{ijk}}{ \tilde{\nu}^2_{ijk} } \right) \\
        & \times  \int \left( 2\pi \tilde{\nu}^2_{ijk}\right)^{-\frac{1}{2}} \exp \left\{ - \frac{1}{2} \left[ \left( \tilde{\tau}^{ij}_k \right)^2  - 2 \tilde{\tau}^{ij}_k  \tilde{m}_{ijk} + \left(  \tilde{m}_{ijk}\right)^2 \right] / \tilde{\nu}^2_{ijk}\right\} \mathbbm{1}\left(\tilde{\tau}^{ij}_k \ge 0 \right) d\tilde{\tau}^{ij}_k \\
        = &   2 \left( s^2_k \right)^{-\frac{1}{2}}  \times \left( \tilde{\nu}^2_{ijk} \right)^{\frac{1}{2}} \exp\left\{  \frac{1}{2}  \frac{ \tilde{m}^2_{ijk} }{ \tilde{\nu}^2_{ijk}} \right\}  \times  { \Phi\left( \frac{ \tilde{m}_{ijk}}{\tilde{\nu}_{ijk}}\right)}, \label{eq:ratio.simplified}
        \end{aligned}
    \end{equation*}
    where the last line follows from the fact that the integral to be evaluated is associated with the truncated normal kernel, $N^{+}(\tilde{m}_{ijk}, \tilde{\nu}^2_{ijk})$, which leads to the result $\Phi(  \tilde{m}_{ijk}/{\tilde{\nu}_{ijk}}).$

    Substituting the ratio above yields 
    \begin{equation*}
    \begin{aligned}
     \theta^{ij}_k   & = \left(  \frac{\int p\left( y^{ijk}_\cdot | \gamma^{ij}_k = 1, \tilde{\tau}^{ij}_k \right)p\left( \tilde{\tau}^{ij}_k \right)d \tilde{\tau}^{ij}_k  }{ p\left(  y^{ijk}_{\cdot} | \gamma^{ij}_k = 0, \tilde{\tau}^{ij}_k = 0 \right) } \right)^{-1} \times \frac{1-\pi_k}{\pi_k} \\
     & = \frac{  1 - \pi_k }{ 2 \left(  s^2_k \right)^{-\frac{1}{2}}\times \left( \tilde{\nu}^2_{ijk} \right)^{\frac{1}{2}} \exp\left\{  \frac{1}{2}  \frac{ \tilde{m}^2_{ijk} }{ \tilde{\nu}^2_{ijk}} \right\}  \Phi\left( \frac{ \tilde{m}_{ijk}}{\tilde{\nu}_{ijk}}\right) \times \pi_k}.
    \end{aligned}
    \end{equation*}

    Hence, we sample each $\gamma^{ij}_k$ as 
    $$
    \gamma^{ij}_k | - \sim \text{Bernoulli}\left( \frac{1}{1+ \theta^{ij}_k} \right).
    $$
    Then, if $\gamma^{ij}_k = 1$, we update $\tilde{\tau}^{ij}_{k} | - \sim N^+\left(  \tilde{m}_{ijk}, \tilde{\nu}^2_{ijk}\right)$; else, if $\gamma^{ij}_k = 0$, we set $\tilde{\tau}^{ij}_{k}  = 0$. 
    After updating all indicators for covariate $x^k$, we update
    $$
    \pi_k| - \sim \text{Beta}\left(  a_k + \sum_{ 1 < i \ne j \le p } \gamma^{ij}_{k}, b_k + p(p-1) - \sum_{ 1 < i \ne j \le p } \gamma^{ij}_{k}\right),
    $$
leading to $\tau^{ij}_k = \tilde{\tau}^{ij}_k \delta_k = \tilde{\tau}^{ij}_k I\left( \pi_k \ge d\right).$

    \item \textbf{Update the node-level selection parameters $\left\{ \bm{b}^{ij}, \delta^{ij}, \pi^i\right\}$ together with $\{\beta^{ij}_k\}$} \\
    Rewriting  the likelihood of $y^i_n$ from the node-level group perspective, we have the mean part
    $$
         \sum_{ j' \notin \left\{ i,j \right\} } \sum_{k} \beta^{ij'}_k y^{j'}_n x^k_n  + \sum_{k} \beta^{ij}_k y^j_n x^k_n.
    $$
    With $\beta^{ij}_k = \tau^{ij}_k b^{ij}_k$, we denote $z^{ij}_n = y^i_n -  \sum_{ j' \notin \left\{ i,j \right\} } \sum_{k} \beta^{ij'}_k y^{j'}_n x^k_n $, leading to 
    $$
    z^{ij}_n| - \sim N\left(  \sum_{k} b^{ij}_k \tau^{ij}_k y^j_n x^k_n, \sigma^2_i \right)
    $$
    and
    \begin{eqnarray*}
    		 && z^{ij}_n  | \delta^{ij} = 1, - \sim N\left(  \left(\bm{X}^{ij}_n\right)^T \bm{V}^{ij} \bm{b}^{ij}, \sigma^2_i \right) \\
    		&& z^{ij}_n | \delta^{ij} = 0, -   \sim N\left( 0, \sigma^2_i \right),
    \end{eqnarray*}
    where $\bm{X}^{ij}_n =  \left( y^j_n x^1_n , ..., y^j_n x^q_n\right)^T$ and $\bm{V}^{ij} = \text{diag}\left( \tau^{ij}_1, \ldots, \tau^{ij}_q\right)$. In addition, we denote $\bm{Z}^{ij} = \left( z^{ij}_1, \ldots, z^{ij}_n\right)^T$ and $\bX^{ij} =\left( \bX^{ij}_1,..., \bX^{ij}_n \right)^T$, leading to the vector form formulations
    \begin{eqnarray*}
    		 && \bm{Z}^{ij} | \delta^{ij} = 1, - \sim MVN\left(  \bm{X}^{ij} \bm{V}^{ij} \bm{b}^{ij}, \sigma^2_i \bm{I}_n \right) \\
    		&& \bm{Z}^{ij} | \delta^{ij} = 0, - \sim MVN\left(  \bzeros_n, \sigma^2_i \bm{I}_n \right) .
    \end{eqnarray*}

    Similarly, we integrate out $\bb^{ij}$
	 \begin{equation*}
    \begin{aligned}
    & p\left( \delta^{ij}  = 1 \left|  - \right. \right)   \\ 
    & = \frac{  \int p\left(  \delta^{ij} = 1, \bm{b}^{ij} | - \right)d\bm{b}^{ij} }{  p\left(  \delta^{ij} = 0, \bm{b}^{ij} =\bzeros |- \right)  + \int p\left(  \delta^{ij} = 1, \bm{b}^{ij} | - \right)d\bm{b}^{ij} } \\
    & = \frac{1}{1 + \theta^{ij} },
    \end{aligned}
    \end{equation*}
    where the posterior probability ratio is
    \begin{equation*}
    \begin{aligned}
    & \theta^{ij}   = \frac{ p\left(  \delta^{ij} = 0, \bm{b}^{ij} =  \bm{0} |- \right)  }{ \int p\left(  \delta^{ij} = 1, \bm{b}^{ij} | - \right)d\bm{b}^{ij}}  \\
    & = \frac{ \frac{1}{p\left( \bm{Z}^{ij}\right)} p\left(  \bm{Z}^{ij} | \delta^{ij} = 0, \bm{b}^{ij} =  \bm{0} \right) \times \left( 1 - \pi^{i} \right) }{  \frac{1}{ p\left( \bm{Z}^{ij} \right)  }  \int p\left( \bm{Z}^{ij}| \delta^{ij} = 1, \bm{b}^{ij} \right)p\left( \bm{b}^{ij} \right)d \bm{b}^{ij} \times \pi^{i} }  \\
    & = \frac{  p\left(  \bm{Z}^{ij} | \delta^{ij} = 0, \bm{b}^{ij} =  \bm{0} \right) \times \left( 1 - \pi^{i} \right) }{   \int p\left( \bm{Z}^{ij}| \delta^{ij} = 1, \bm{b}^{ij} \right)p\left( \bm{b}^{ij} \right)d \bm{b}^{ij} \times \pi^{i} } \\
    & = \left( \frac{   \int p\left( \bm{Z}^{ij}| \delta^{ij} = 1, \bm{b}^{ij} \right)p\left( \bm{b}^{ij} \right)d \bm{b}^{ij}  }{  p\left(  \bm{Z}^{ij} | \delta^{ij} = 0, \bm{b}^{ij} =  \bm{0} \right) } \right)^{-1} \times \frac{\left( 1 - \pi^{i} \right) }{ \pi^{i}}.
    \end{aligned}
    \end{equation*}
    We next estimate $\theta^{ij}$. First, we note that
    $$
    p\left(  \bm{Z}^{ij}  | \delta^{ij} = 0, \bm{b}^{ij} = \bm{0} \right)  =  \left( 2\pi \sigma^2\right)^{-\frac{N}{2}} \exp\left\{ -\frac{1}{2\sigma^2_i} \left(\bm{Z}^{ij}\right)^T \bm{Z}^{ij}\right\} .
    $$
    Therefore,
    \begin{equation*}
        \begin{aligned}
        & \int p\left( \bm{Z}^{ij} | \delta^{ij} = 1, \bm{b}^{ij} \right)p\left( \bm{b}^{ij} \right)d \bm{b}^{ij} \\
        = & \int  \left( 2\pi \sigma^2_i \right)^{-\frac{N}{2}}\exp \left\{ - \frac{1}{2\sigma^2_i}  \left( \bm{Z}^{ij}  - \bm{X}^{ij}\bm{V}^{ij} \bm{b}^{ij}\right)^T\left( \bm{Z}^{ij}  - \bm{X}^{ij}\bm{V}^{ij}\bm{b}^{ij}\right)\right\}  \\
        & \times \left( 2\pi \right)^{-\frac{q}{2}} \exp \left\{ - \frac{1}{2} \left(\bm{b}^{ij}\right)^T \bm{b}^{ij} \right\} d \bm{b}^{ij} \\
        = & \underbrace{ \left( 2\pi \sigma^2_i\right)^{ -\frac{N}{2}} \exp\left\{ - \frac{1}{2\sigma^2_i}   \left(\bm{Z}^{ij}\right)^T \bm{Z}^{ij}\right\} }_{= p\left(  \bm{Z}^{ij}  | \delta^{ij} = 0, \bm{b}^{ij} = \bm{0} \right) } \left( 2 \pi \right)^{-\frac{q}{2}}  \left( 2 \pi \right)^{\frac{q}{2}} \left| \tilde{\bm{\Sigma}}^{ij} \right|^{\frac{1}{2}} \exp \left\{ \frac{1}{2} \left( \tilde{\bm{\mu}}^{ij}\right)^T\left( \tilde{\bm{\Sigma}}^{ij}  \right)^{-1} \tilde{\bm{\mu}}^{ij} \right\} \\
        & \times \int  \left( 2 \pi \right)^{-\frac{q}{2}} \left| \tilde{\bm{\Sigma}}^{ij} \right|^{-\frac{1}{2}} \exp \left\{ -\frac{1}{2} \left[  \left( \bm{b}^{ij}\right)^T  \left(\frac{1}{\sigma^2_i}\left( \bm{X}^{ij}\bm{V}^{ij} \right)^T\left(\bm{X}^{ij}\bm{V}^{ij}\right) + \bm{I}_q\right)  \bm{b}^{ij} \right. \right.
        \end{aligned}
    \end{equation*}
    \begin{equation*}
        \begin{aligned}
            &   \left. \left. - 2       \frac{1}{\sigma^2_i} \left(\bm{Z}^{ij}\right)^T \bm{X}^{ij} \bm{V}^{ij} \tilde{\bm{\Sigma}}^{ij}  \left( \tilde{\bm{\Sigma}}^{ij}  \right)^{-1} \bm{b}^{ij} + \left( \tilde{\bm{\mu}}^{ij}\right)^T\left( \tilde{\bm{\Sigma}}^{ij}  \right)^{-1} \tilde{\bm{\mu}}^{ij}  \right] \right\}.
        \end{aligned}
    \end{equation*}
    Letting $\tilde{\bm{\Sigma}}^{ij}   = \left( \frac{1}{\sigma^2_i}\left( \bm{X}^{ij}\bm{V}^{ij} \right)^T\left(\bm{X}^{ij}\bm{V}^{ij}\right) + \bm{I}_q \right)^{-1}$ and $\tilde{\bm{\mu}}^{ij}  =  \left( \frac{1}{\sigma^2_i} \left(\bm{Z}^{ij}\right)^T \bm{X}^{ij} \bm{V}^{ij} \tilde{\bm{\Sigma}}^{ij} \right)^T$, the integral to be evaluated above is 1 because the integrand is the density function of an $MVN\left(\tilde{\bm{\mu}}^{ij}, \tilde{\bm{\Sigma}}^{ij} \right)$ random vector. Therefore, we obtain that
    \begin{equation*}
        \begin{aligned}
        &     \frac{   \int p\left( \bm{Z}^{ij}| \delta^{ij} = 1, \bm{b}^{ij} \right)p\left( \bm{b}^{ij} \right)d \bm{b}^{ij}  }{  p\left(  \bm{Z}^{ij} | \delta^{ij} = 0, \bm{b}^{ij} =  \bm{0} \right) } \\
            = &  \left| \tilde{\bm{\Sigma}}^{ij} \right|^{\frac{1}{2}} \exp \left\{ \frac{1}{2} \left( \tilde{\bm{\mu}}^{ij}\right)^T\left( \tilde{\bm{\Sigma}}^{ij}  \right)^{-1} \tilde{\bm{\mu}}^{ij} \right\},
        \end{aligned}
    \end{equation*}
    and consequently,    
\begin{equation*}
\begin{aligned}
 \theta^{ij}   & = \left( \frac{   \int p\left( \bm{Z}^{ij}| \delta^{ij} = 1, \bm{b}^{ij} \right)p\left( \bm{b}^{ij} \right)d \bm{b}^{ij}  }{  p\left(  \bm{Z}^{ij} | \delta^{ij} = 0, \bm{b}^{ij} =  \bm{0} \right) } \right)^{-1} \times \frac{\left( 1 - \pi^{i} \right) }{ \pi^{i}} \\
 & = \frac{  1 - \pi^{i} }{\left| \tilde{\bm{\Sigma}}^{ij} \right|^{\frac{1}{2}} \exp \left\{ \frac{1}{2} \left( \tilde{\bm{\mu}}^{ij}\right)^T\left( \tilde{\bm{\Sigma}}^{ij}  \right)^{-1} \tilde{\bm{\mu}}^{ij} \right\} \times \pi^{i}}.
\end{aligned}
\end{equation*}
Hence, we first sample each $\delta^{ij}$  by
$$
\delta^{ij} | - \sim \text{Bernoulli}\left( \frac{1}{1+ \theta^{ij}} \right).
$$
Then, if $\delta^{ij} = 1$, we update $\bm{b}^{ij} | - \sim N\left(  \tilde{\bm{\mu}}^{ij}, \tilde{\bm{\Sigma}}^{ij} \right)$; else, if $\delta^{ij} = 0$, we set $\bm{b}^{ij} = \bm{0}_q$.  \\
After updating all indicators for node $i$, we update
$$
    \pi^{i}| - \sim \text{Beta}\left( a^{i} + \sum_{ j\ne i } \delta^{ij}, b^i + (p-1) - \sum_{ j\ne i } \delta^{ij}\right),
$$
and
$$
        (\beta^{ij}_k)_{1\le k \le q} = \bm{B}^{ij} = \bm{V}^{ij}\bb^{ij}.
$$ 

\item \textbf{Update the variances $\left\{ \sigma^2_i \right\}$} \\
The posterior of $\sigma^2_i$ is:
\begin{eqnarray*}
	p\left( \sigma^2_i | - \right) & \propto &  \left( \sigma^2_i \right)^{-\frac{N}{2}} \exp\left\{ - \frac{1}{2} \frac{1}{\sigma^2_i} \sum^N_{n=1} \left( y^i_n - \sum_{j\ne i }\sum_{k = 1}^q  \beta^{ij}_k y^j_n x^k_n\right)^2 \right\} \\
	& & \times \left( \sigma^{-2}_i \right)^{a^i_\sigma + 1} \exp\left\{ -\sigma^{-2}_i b^i_{\sigma}\right\} \\
	& \propto & \left( \sigma^{-2}_i \right)^{ \frac{N}{2} + a^i_\sigma + 1 } \exp\left\{ - \sigma^{-2}_i \left[ \frac{1}{2} \sum^N_{n=1} \left( y^i_n - \sum_{ j\ne i }\sum_{k = 1}^q  \beta^{ij}_k y^j_n x^k_n\right)^2 + b^i_\sigma \right]\right\},
\end{eqnarray*}
which is an Inverse-Gamma distribution
$$
		\sigma^2_i | - \sim \text{InvGamma}\left(  \frac{N}{2} + a^i_{\sigma},  \frac{1}{2} \sum^N_{n=1} \left( y^i_n - \sum_{ j\ne i}\sum_{k = 1}^q \beta^{ij}_k y^j_n x^k_n\right)^2 + b^i_\sigma\right).
$$
\item \textbf{Update $\left\{ s^2_k \right\}$ and $t$} \\
    We have conjugate updates:
    \begin{eqnarray*}
	p\left( s^2_k | - \right) & = & \prod_{ \substack{1\le i\ne j\le p\\ \gamma^{ij}_k = 1}  } 2\left( 2\pi s^2_k \right)^{-\frac{1}{2}} \exp\left\{ - \frac{1}{2} \frac{\left( \tilde{\tau}^{ij}_k \right)^2}{s^2_k} \right\} \mathbbm{1}\left( \tilde{\tau}^{ij}_k  \ge 0 \right)\\
	& & \times \frac{t^1}{\Gamma(1)}\left( s^2_k\right)^{- 2} \exp\left\{ -\frac{t}{s^2_k}\right\} \\
	& \propto & \left( s^{2}_k \right)^{ - \left(\frac{1}{2} \sum_{ 1\le i\ne j\le p } \gamma^{ij}_k +1  + 1 \right)} \exp\left\{ - s^{-2}_k \left[  \frac{1}{2}  \sum_{ 1\le i\ne j\le p } \left( \tilde{\tau}^{ij}_k \right)^2 + t \right]\right\},
 \end{eqnarray*}
 which is an Inverse-Gamma distribution
 \begin{eqnarray*}
 s^2_k | - & \sim & \text{InvGamma}\left( 1 +  \frac{1}{2} \sum_{ 1\le i\ne j\le p } \gamma^{ij}_k, t + \frac{1}{2}  \sum_{ 1\le i\ne j\le p } \left( \tilde{\tau}^{ij}_k \right)^2 \right).
\end{eqnarray*}
The posterior of $t$ is:
 \begin{eqnarray*}
 p\left( t | - \right) & = & \prod_{k = 1}^q \frac{t^1}{\Gamma(1)}\left( s^2_k\right)^{- 2} \exp\left\{ -\frac{t}{s^2_k}\right\} \\
    & \propto & t^q  \exp\left( - t \sum^q_{k=1} \frac{1}{s^2_k}\right),
 \end{eqnarray*}
which is a Gamma distribution
\begin{eqnarray*}
 t | -  & \sim & \text{Gamma}\left( q + 1,  \sum^q_{k=1} \frac{1}{s^2_k}\right).
    \end{eqnarray*}
\end{itemize}

\begin{algorithm}
\caption{Full Gibbs Sampler for the DGSS}
{\footnotesize
\setstretch{1}
\begin{algorithmic}[1]
    \For {Each MCMC Iteration}
        \State Update the covariate-level selection parameters
            \For {Each Covariate $k$}
                \For {Each Response Node $i$}
                    \For {Each Explanatory Node $j$}
                        \State Update the local-level Spike-and-Slab indicator from its posterior distribution 
                        \State $\gamma^{ij}_k|- \sim \text{Bernoulli}\left(\frac{1}{1+\theta^{ij}_k}\right)$
                        \If { $\gamma^{ij}_k = 1$}
                            \State Update $\tilde{\tau}^{ij}_k$ from its posterior distribution $\tilde{\tau}^{ij}_k | - \sim N^+\left(\tilde{m}_{ijk}, \tilde{\nu}^2_{ijk} \right)$
                        \Else { $\gamma^{ij}_k = 0$}
                            \State Update $\tilde{\tau}^{ij}_k = 0$
                        \EndIf
                    \EndFor
                \EndFor
                \State Update the participation rate from its posterior distribution 
                \State $\pi_k | - \sim \text{Beta}\left(  a_k + \sum_{ 1 < i \ne j \le p } \gamma^{ij}_{k}, b_k + p(p-1) - \sum_{ 1 < i \ne j \le p } \gamma^{ij}_{k}\right)$
                \If { $\pi_k \ge d_k$}
                    \State $\tau^{ij}_k = \tilde{\tau}^{ij}_k$
                \Else { $\pi_k < d_k$}
                    \State $\tau^{ij}_k = 0$
                \EndIf
            \EndFor
        \State Update the node-level selection parameters
        \For {Every Response Node $i$}
            \For {Every Response Node $j$}
                \State Update the node-level Spike-and-Slab indicator from its posterior distribution
                \State $\delta^{ij} | - \sim \text{Bernoulli}\left( \frac{1}{1+ \theta^{ij}}\right)$
                \If { $\delta^{ij} = 1$}
                    \State Update $\bb^{ij}$ from its posterior distribution $\bb^{ij}|- \sim \text{Bernoulli}\left(\tilde{\bm{\mu}}^{ij}, \tilde{\bSigma}^{ij}\right)$
                \Else { $\delta^{ij} = 0$}
                    \State Update $\bb^{ij} = \bm{0}_q$
                \EndIf
                \State Update coefficients vector $        (\beta^{ij}_k)_{1\le k \le q} = \bm{B}^{ij} = \bm{V}^{ij}\bb^{ij}.$
            \EndFor
            \State Update the node-level sparsity from its posterior distribution
            \State $ \pi^{i}| - \sim \text{Beta}\left( a^{i} + \sum_{ j\ne i } \delta^{ij}, b^i + (p-1) - \sum_{ j\ne i } \delta^{ij}\right) $
        \EndFor 
    \State Update regression parameters
    \For {Each Response Node $i$}
        \State Update regression variance from its posterior distribution 
        \State $\sigma^2_i | - \sim \text{InvGamma}\left(  \frac{N}{2} + a^i_{\sigma},  \frac{1}{2} \sum^N_{n=1} \left( y^i_n - \sum_{ j\ne i}\sum_{k = 1}^q \beta^{ij}_k y^j_n x^k_n\right)^2 + b^i_\sigma\right)$
    \EndFor 
    \State Update the hyper-parameters from the posterior
    \For {Each Covaraite $k$}
        \State $s^2_k | - \sim  \text{InvGamma}\left( 1 +  \frac{1}{2} \sum_{ 1\le i\ne j\le p } \gamma^{ij}_k, t + \frac{1}{2}  \sum_{ 1\le i\ne j\le p } \left( \tilde{\tau}^{ij}_k \right)^2 \right)$
    \EndFor
    \State $t | -  \sim \text{Gamma}\left( q + 1,  \sum^q_{k=1} \frac{1}{s^2_k}\right)$
    \EndFor
\end{algorithmic}
} 
\end{algorithm} 

\section*{S2. Scalability}
We conducted an investigation to evaluate how the computation time of DGSS scales as $N, p$ and $q$ increase. The default setting is $N = 200, p = 25$ and $q=10$. We varied $N, p$ and $q$ individually while keeping the other parameters constant, estimating the runtime in seconds per iteration, averaged over 100 iterations. The test was conducted on a desktop with a 24-core 13th Gen Intel(R) Core i9-13900K CPU. The results are presented in Table~\ref{tab:sup.scalability}. Since the computational bottleneck of the sampler lies in its iterative requirement to update the indicators one-by-one, with a complexity of approximately $O(p^2q)$ for the covariate-level group and $O(pq)$ for the node-level group, the computation time increases most dramatically as $p$ increases, followed by increases in $q$. In contrast, it is not significantly impacted when $N$ increases. Overall, as a sampling-based approach, the computational time appears reasonable given the moderately large model space of dimension $p\times (p-1) \times q$.

\begin{table}[!hbt]
\centering
\caption{Runtime in seconds per iteration, averaged over 100 iterations, as $N, p,$ and $q$ increase. The default setting is $N = 200, p = 25$ and $q=10$.} 
\begin{tabular}{cccccccc}
      \hline
      \hline
      \multicolumn{8}{c}{Computation time when $N$ grows} \\
      \hline 
     \multicolumn{2}{r}{$N=$} & 200 & 500 & 1000 & 1500 & 2000 & \\ 
    \multicolumn{2}{r}{Time} & 0.03 & 0.07 & 0.14 & 0.22 & 0.32 & \\ 
    \hline
    \hline
    \multicolumn{8}{c}{Computation time when $p$ grows} \\
    \hline 
    \multicolumn{2}{r}{$p=$}   & 25 & 50 & 100 & 150 & 200 \\ 
    & Time   & 0.03 & 0.21 & 1.58 & 5.19 & 11.94 & \\ 
      \hline
    \hline
    \multicolumn{8}{c}{Computation time when $q$ grows} \\
    \hline
    \multicolumn{2}{r}{$q=$} & 10 & 50 & 100 & 150 & 200 & \\ 
    & Time   & 0.03 & 0.32 & 1.14 & 2.53 & 4.63 & \\ 
    \hline
    \hline
\end{tabular}
\label{tab:sup.scalability}
\end{table}

\section*{S3. Sensitivity Analyses}
\begin{figure}[H]
    \centering
    \includegraphics[width=0.8\textwidth]{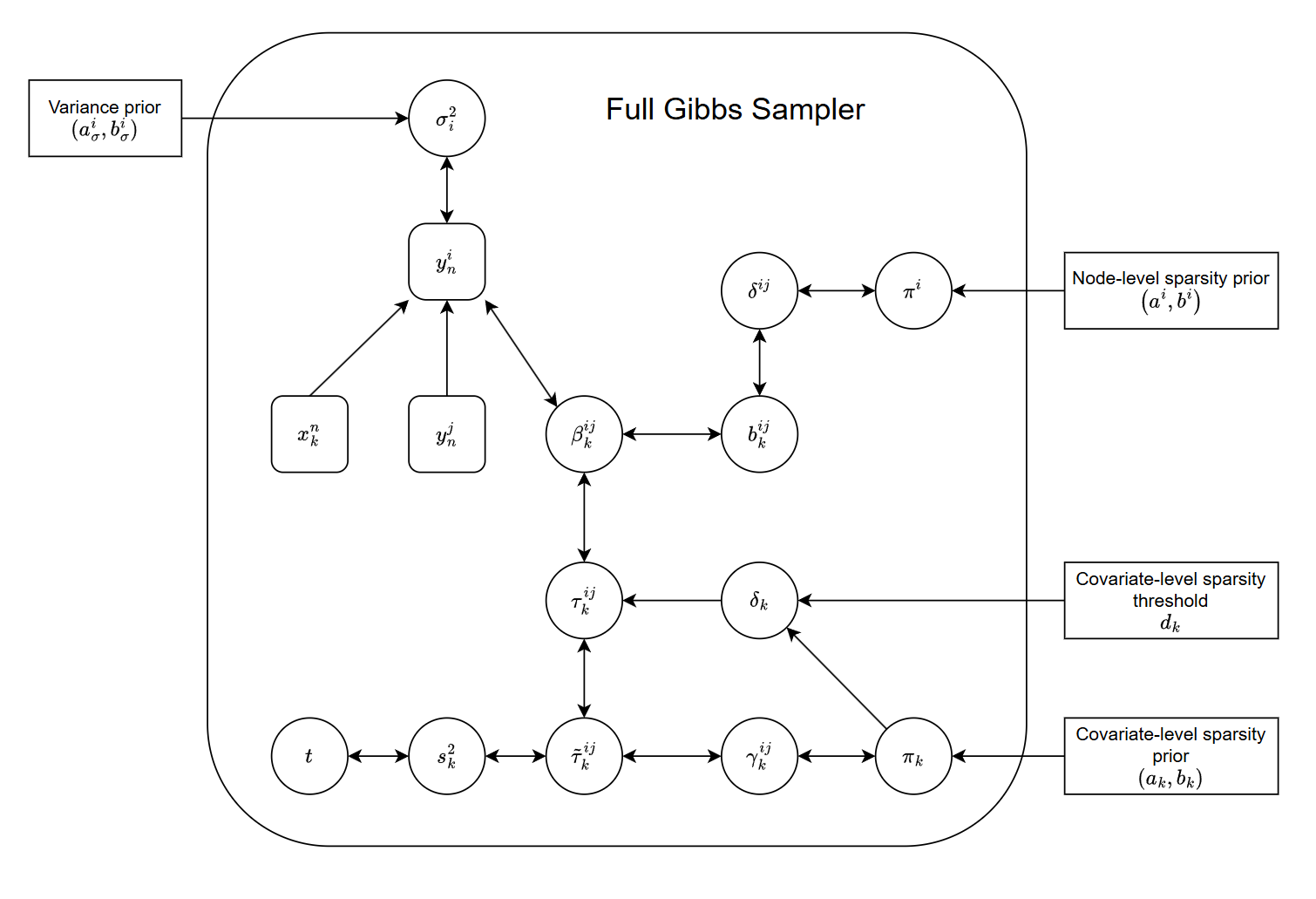}
    \caption{Graphical representation of the model, with the double arrows representing the Gibbs updates.}
    \label{fig:ModelStructure}
\end{figure}

We plot the model structure in Figure~\ref{fig:ModelStructure}, with the double arrows representing the Gibbs updates. By using the model reparameterization, the hyperparameters in the Slab prior are updated entirely by the Gibbs sampler, $\left\{\tau^{ij}_k, b^{ij}_k\right\}^{1\le i < j \le p}_{1\le k \le q}$, leaving the following specifiable hyperparameters in the DGSS:
\begin{itemize}
    \item Node-level prior sparsity $(a^i, b^i)$ from $\pi^i \sim \text{Beta}(a^i,b^i)$;
    \item Local-level prior sparsity $(a_k, b_k)$ from $\pi_k \sim \text{Beta}(a_k, b_k)$;
    \item Covariate-level threshold $d_k$ from $\delta_k = I(\pi_k \ge d_k)$.
\end{itemize}

\subsection*{Simulation Study} 

We follow \cite{Xu2015} to test $(a^i,b^i), (a_k, b_k) \in \left\{(0.5,0.5), (1,1), (1.5,1.5)\right\}$ and \cite{Zeng2024} to test $d_k \in \left\{0.01, 0.05, 0.1\right\}$. We use the data sets generated with $N=500$ cases. We take the results $a^i=b^i=a_k=b_k = 1$ and $d_k = 0.05$ as the baseline. Table~\ref{tab:sensitivity_sim} presents the results.

For the sparsity priors $(a^i,b^i), (a_k,b_k)$, consistent with \cite{Xu2015}, whose results also vary with different sparsity priors, we observe that the results fluctuate as the sparsity priors change, despite all of them being non-informative priors. The most significant changes occur when $(a_k,b_k)=(0.5,0.5)$, where covariate-dependent edge selection deteriorates while covariate selection improves. For the sparsity threshold $d_k$, consistent with \cite{Zeng2024}, we observe that the results for covariate selection vary with changes in $d$. Specifically, with larger values of $d$ empty covariates are more easily excluded from the model, resulting in improved covariate selection under this simulation setting. 

Although we observe variations in results when the sparsity priors and parameters change, these changes do not affect the main conclusions presented in the paper. We can still observe that DGSS, which incorporates both node and covariate grouping directions for sparsity, enhances the results compared to existing methods that consider only the node grouping direction. This improvement is particularly evident when the true sparsity of the linear system arises from the covariate grouping direction.

In the absence of specific evidence from the tests, we adopt the default non-informative Beta prior $(1,1)$ and the commonly used significance threshold $0.05$ as the default settings.

\subsection*{Application} 

Table~\ref{tab:sensitivity_real} reports the edges detected in each covariate-specific subnetwork, along with the ratio of "agreement," where both methods either detect or do not detect an edge. We observe a high agreement ratio across the Cytokines, given that there are a total of $4,005$ locations in each subnetwork. Table~\ref{tab:sensitivity_real} also presents the number of edges identified as being impacted by the Cytokines.

When different thresholds $d$ are used, some Cytokines may be excluded from the model if they impact too few edges to be considered truly influential covariates. In such cases, the impact of an edge may be reassigned to other Cytokines, leading to variations in results across cases. However, based on the ratio of "agreement," we can see that most of the selection results remain consistent across different values of $d$.

\begin{table}
    \centering
    \caption{Sensitivity analysis for DGSS using $N=500$ datasets in Main paper simulation section }
    \label{tab:sensitivity_sim}
    \begin{adjustbox}{max width = 1\textwidth}
    \begin{tabular}{ccccccccccccccc}
      \hline
      \hline
      & \multicolumn{4}{c}{Covariate-dependent edge} && \multicolumn{4}{c}{Group edge } && \multicolumn{4}{c}{Covariate}\\
      \hline
      \hline
      & \multicolumn{4}{c}{Change of $(a^i, b^i)$} && \multicolumn{4}{c}{Change of $(a^i, b^i)$} && \multicolumn{4}{c}{Change of $(a^i, b^i)$}\\
      \hline
       & TPR & FPR & F1 & MCC  && TPR & FPR & F1 & MCC  && TPR & FPR & F1 & MCC  \\ 
      \multirow{2}{*}{$(1.5, 1.5)$} & 0.669 & 0.064 & 0.416 & 0.417 && 0.759 & 0.106 & 0.785 & 0.664 && 1 & 0.303 & 0.828 & 0.708 \\ 
       & (0.013) & (0.002) & (0.004) & (0.005) && (0.010) & (0.004) & (0.005) & (0.008) && (0) & (0.030) & (0.015) & (0.027) \\ 
       
      \multirow{2}{*}{$(1, 1)$} & 0.679 & 0.066 & 0.414 & 0.417 && 0.780 & 0.132 & 0.782 & 0.650 &&  1 & 0.327 & 0.816 & 0.686 \\ 
       & (0.012) & (0.002) & (0.004) & (0.005) && (0.010) & (0.005) & (0.005) & (0.008) && (0) & (0.029) & (0.015) & (0.026) \\
       
      \multirow{2}{*}{$(0.5, 0.5)$} & 0.698 & 0.070 & 0.413 & 0.420 && 0.821 & 0.191 & 0.772 & 0.621 && 1 & 0.273 & 0.842 & 0.732 \\ 
       & (0.013) & (0.002) & (0.005) & (0.006) &&  (0.009) & (0.008) & (0.004) & (0.009) && (0) & (0.028) & (0.014) & (0.025) \\ 
       
      \hline
      \hline
      & \multicolumn{4}{c}{Change of $(a_k, b_k)$} && \multicolumn{4}{c}{Change of $(a_k, b_k)$} && \multicolumn{4}{c}{Change of $(a_k, b_k)$}\\
      \hline
       & TPR & FPR & F1 & MCC && TPR & FPR & F1 & MCC  && TPR & FPR & F1 & MCC\\ 
      \multirow{2}{*}{$(1.5, 1.5)$}  & 0.649 & 0.057 & 0.430 & 0.427 && 0.756 & 0.111 & 0.780 & 0.655 && 1 & 0.377 & 0.792 & 0.643 \\ 
      & (0.013) & (0.002) & (0.005) & (0.006)  && (0.010) & (0.005) & (0.005) & (0.008) && (0) & (0.031) & (0.014) & (0.027) \\ 
     
      \multirow{2}{*}{$(1, 1)$} & 0.679 & 0.066 & 0.414 & 0.417 && 0.780 & 0.132 & 0.782 & 0.650 && 1 & 0.327 & 0.816 & 0.686 \\ 
      & (0.012) & (0.002) & (0.004) & (0.005) && (0.010) & (0.005) & (0.005) & (0.008) && (0) & (0.029) & (0.015) & (0.026) \\ 
      
      \multirow{2}{*}{$(0.5, 0.5)$} & 0.725 & 0.084 & 0.386 & 0.401 && 0.817 & 0.174 & 0.779 & 0.635 && 1 & 0.187 & 0.888 & 0.812 \\ 
      & (0.012) & (0.002) & (0.004) & (0.005)  && (0.009) & (0.007) & (0.004) & (0.008) && (0) & (0.024) & (0.013) & (0.023) \\ 
      
       \hline
      \hline
      & \multicolumn{4}{c}{Change of $d$} && \multicolumn{4}{c}{Change of $d$}  && \multicolumn{4}{c}{Change of $d$}\\
      \hline
      & TPR & FPR & F1 & MCC && TPR & FPR & F1 & MCC && TPR & FPR & F1 & MCC  \\ 
      \multirow{2}{*}{$0.1$} & 0.683 & 0.067 & 0.414 & 0.419 && 0.785 & 0.139 & 0.781 & 0.646 &&  1 & 0.210 & 0.874 & 0.789 \\ 
      & (0.013) & (0.002) & (0.004) & (0.005) && (0.009) & (0.005) & (0.004) & (0.008) && (0) & (0.025) & (0.013) & (0.023) \\ 
      
      \multirow{2}{*}{$0.05$} & 0.679 & 0.066 & 0.414 & 0.417 && 0.780 & 0.132 & 0.782 & 0.650 && 1 & 0.327 & 0.816 & 0.686 \\ 
      & (0.012) & (0.002) & (0.004) & (0.005)  && (0.010) & (0.005) & (0.005) & (0.008) && (0) & (0.029) & (0.015) & (0.026) \\ 
      
      \multirow{2}{*}{$0.01$}  & 0.678 & 0.066 & 0.415 & 0.419  && 0.779 & 0.130 & 0.783 & 0.653 && 1 & 0.353 & 0.801 & 0.660 \\ 
      & (0.013) & (0.002) & (0.004) & (0.005) && (0.009) & (0.005) & (0.005) & (0.008)  && (0) & (0.027) & (0.013) & (0.023) \\ 
      \hline
      \hline
    \end{tabular}
    \end{adjustbox}
\end{table}

\begin{table}[!hbt]
    \caption{Sensitivity analysis for DGSS using Real Data}
    \label{tab:sensitivity_real}
    \centering
    \begin{adjustbox}{max width = 1\textwidth}
    \begin{tabular}{ccccccccccccccccc}
      \hline
      \hline
     &  \multirow{2}{*}{Baseline} & \multirow{2}{*}{Eotaxin}  & \multirow{2}{*}{FGF}  & \multirow{2}{*}{G-CSF}  & \multirow{2}{*}{GM-CSF}  & \multirow{2}{*}{IFN-g}  & \multirow{2}{*}{IL-10}& \multirow{2}{*}{  \shortstack[c]{IL-12\\(p70)}  }  & \multirow{2}{*}{IL-13}  & \multirow{2}{*}{IL-15}  & \multirow{2}{*}{IL-17A}  & \multirow{2}{*}{IL-1b}  & \multirow{2}{*}{IL-1ra}  & \multirow{2}{*}{IL-2} & \multirow{2}{*}{IL-4}  \\ 
     &   &  &   &   &    &    &  &   &   &    &    &    &    &  &   \\ 
        \hline 
    $d =  0.1$  & 197 & 114 & 163 & 142 & 197 & 160 & 160 & 155 & 168 & 145 & 157 & 141 & 195 & 173 & 0 \\ 
    $d = 0.05$ & 201 & 104 & 105 & 190 & 167 & 163 & 172 & 104 & 189 & 132 & 146 & 123 & 204 & 195 & 87 \\ 
    $d = 0.01$ & 201 & 121 & 92 & 169 & 137 & 155 & 166 & 123 & 199 & 146 & 151 & 171 & 201 & 174 & 125 \\ 
    \hline
    \hline
     & \multirow{2}{*}{IL-5} & \multirow{2}{*}{IL-6} & \multirow{2}{*}{IL-7} & \multirow{2}{*}{IL-8} & \multirow{2}{*}{IL-9} & \multirow{2}{*}{IP-10} & \multirow{2}{*}{ \shortstack[c]{MCP-1 \\(MCAF)} } & \multirow{2}{*}{ \shortstack[c]{MIP \\ (1a)} } & \multirow{2}{*}{ \shortstack[c]{MIP \\ (1b)} } & \multirow{2}{*}{ \shortstack[c]{PDGF \\ (bb)} } & \multirow{2}{*}{ \shortstack[c]{RAN- \\ TES} } & \multirow{2}{*}{ \shortstack[c]{TNF \\ (a)} }  & \multirow{2}{*}{VEGF} & \multirow{2}{*}{ \shortstack[c]{FGF \\ basic}  } & \multirow{2}{*}{IL-17}\\ 
      &   &  &   &   &    &    &  &   &   &    &    &    &    &  & \\
        \hline 
    $d =  0.1$ & 75 & 177 & 173 & 153 & 167 & 163 & 207 & 191 & 157 & 198 & 164 & 119 & 132 & 0 & 46 \\ 
    $d = 0.05$ & 93 & 154 & 165 & 130 & 156 & 170 & 192 & 186 & 100 & 147 & 167 & 21 & 127 & 38 & 25 \\ 
    $d = 0.01$ & 88 & 158 & 163 & 154 & 143 & 195 & 191 & 179 & 114 & 127 & 143 & 51 & 108 & 36 & 14 \\ 
    \hline
    \hline
    \multicolumn{16}{c}{Ratio of the same selection results ($d= 0.05$ as baseline, divided by the total number of potential edges $4,005$ )} \\
    \hline
    \hline
     &  \multirow{2}{*}{Baseline} & \multirow{2}{*}{Eotaxin}  & \multirow{2}{*}{FGF}  & \multirow{2}{*}{G-CSF}  & \multirow{2}{*}{GM-CSF}  & \multirow{2}{*}{IFN-g}  & \multirow{2}{*}{IL-10}& \multirow{2}{*}{  \shortstack[c]{IL-12\\(p70)}  }  & \multirow{2}{*}{IL-13}  & \multirow{2}{*}{IL-15}  & \multirow{2}{*}{IL-17A}  & \multirow{2}{*}{IL-1b}  & \multirow{2}{*}{IL-1ra}  & \multirow{2}{*}{IL-2} & \multirow{2}{*}{IL-4}  \\ 
     &   &  &   &   &    &    &  &   &   &    &    &    &    &  &   \\ 
        \hline 
    $d =  0.1$  & 98.4\% & 98.7\% & 97.9\% & 98.0\% & 98.1\% & 98.4\% & 98.4\% & 98.0\% & 98.3\% & 98.8\% & 98.3\% & 98.6\% & 98.5\% & 97.9\% & 97.8\% \\ 
    $d =  0.01$ & 99.0\% & 99.0\% & 99.1\% & 98.6\% & 98.6\% & 99.1\% & 98.9\% & 98.9\% & 99.2\% & 99.0\% & 98.9\% & 98.4\% & 99.0\% & 98.6\% & 98.8\%  \\
    \hline
    \hline
     & \multirow{2}{*}{IL-5} & \multirow{2}{*}{IL-6} & \multirow{2}{*}{IL-7} & \multirow{2}{*}{IL-8} & \multirow{2}{*}{IL-9} & \multirow{2}{*}{IP-10} & \multirow{2}{*}{ \shortstack[c]{MCP-1 \\(MCAF)} } & \multirow{2}{*}{ \shortstack[c]{MIP \\ (1a)} } & \multirow{2}{*}{ \shortstack[c]{MIP \\ (1b)} } & \multirow{2}{*}{ \shortstack[c]{PDGF \\ (bb)} } & \multirow{2}{*}{ \shortstack[c]{RAN- \\ TES} } & \multirow{2}{*}{ \shortstack[c]{TNF \\ (a)} }  & \multirow{2}{*}{VEGF} & \multirow{2}{*}{ \shortstack[c]{FGF \\ basic}  } & \multirow{2}{*}{IL-17}\\ 
      &   &  &   &   &    &    &  &   &   &    &    &    &    &  & \\
        \hline 
    $d =  0.1$ &  98.6\% & 98.3\% & 98.8\% & 98.3\% & 98.4\% & 98.2\% & 98.3\% & 98.3\% & 98.0\% & 97.9\% & 98.5\% & 97.5\% & 98.5\% & 99.1\% & 99.3\%  \\ 
    $d =  0.01$ &  99.1\% & 99.1\% & 99.3\% & 99.1\% & 99.1\% & 98.6\% & 99.1\% & 99.0\% & 99.2\% & 98.9\% & 98.8\% & 99.2\% & 98.9\% & 99.3\% & 99.5\%\\ 
    \hline
    \hline
    \end{tabular}
    \end{adjustbox}
\end{table} 

\section*{S4. Additional Results for the Application to Microbiome Data (Section 4 of the main paper)}
\begin{figure}
    \centering
    \includegraphics[width=0.8\textwidth]{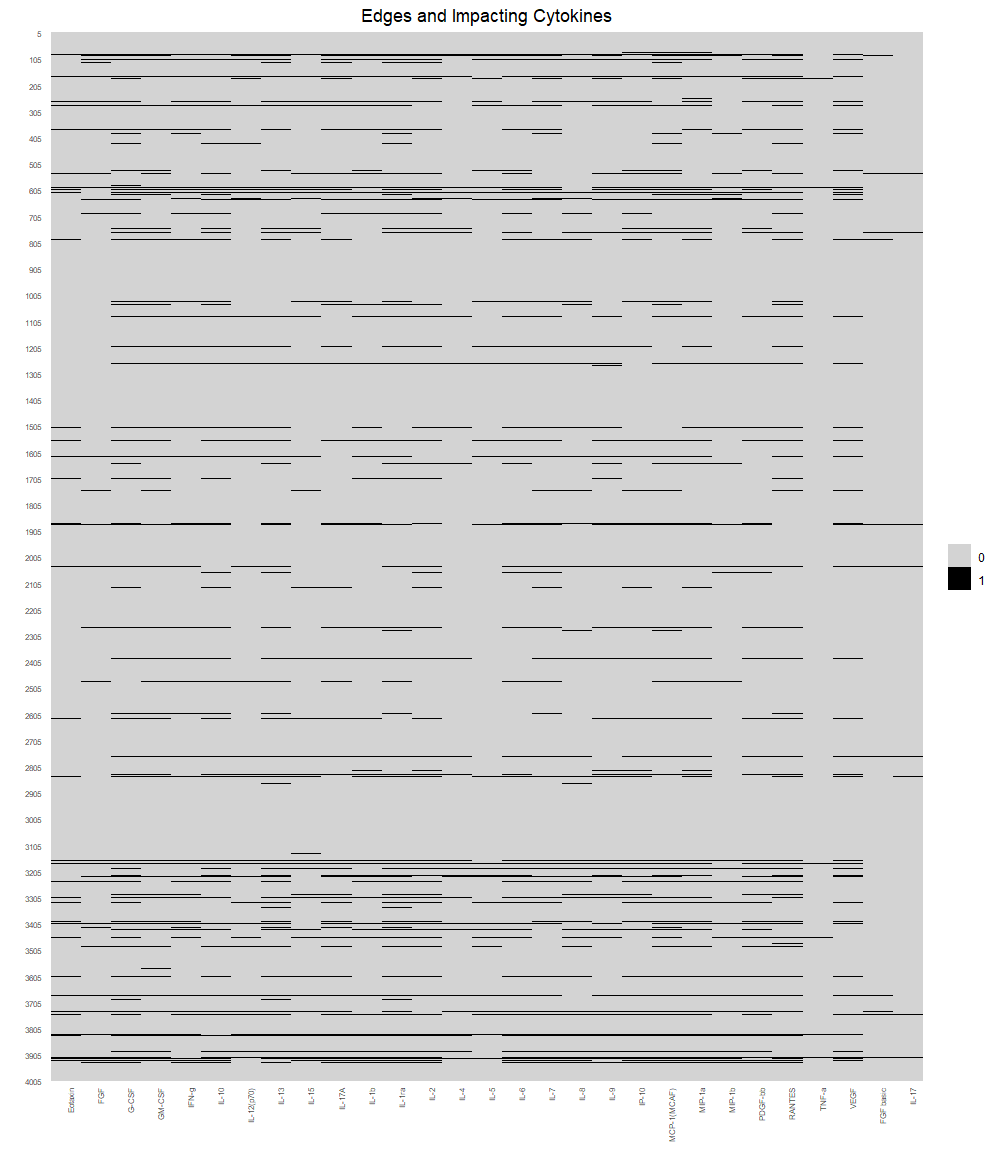}
    \caption{Cytokines Impacting Edges Selection}
    \label{fig:CIE}
\end{figure}

\begin{figure}
    \centering
    \includegraphics[width=0.8\textwidth]{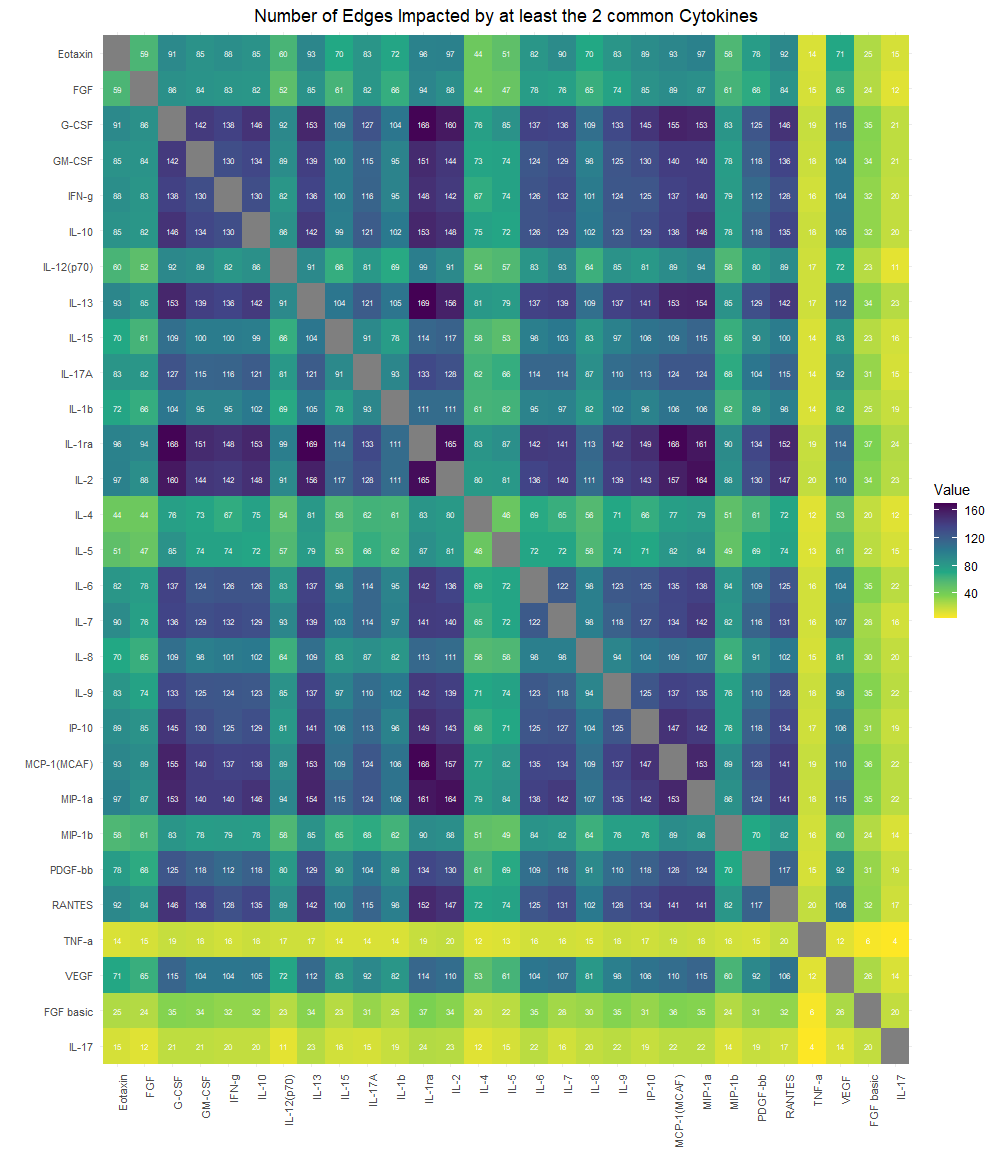}
    \caption{Numbers of Edges Impacted by at least the common two Cytokines.}
    \label{fig:EITwo}
\end{figure}

We plot the edges, along with the impacting cytokines, in Figure~\ref{fig:CIE} to present the detailed selection results in the main paper with $d_k = 0.05$. Given $p=90$ OTUs and $q = 29$ cytokines, the results correspond to a $4005$-by-$29$ matrix representing potential edges (node pairs) and impacting cytokines. In the figure, a black block indicates detected impact, while a gray block indicates no impact. Due to the high dimensionality, the plot is too large to discern specific details. However, it is still evident that some edges are simultaneously impacted by multiple cytokines, as the impacting cytokines for a given edge form a horizontal line in the figure. Additionally, we provide a table, referred to as the `edges cytokines table', along with reproducible code on the GitHub page. The table includes the edge ID, the pair of nodes defining each edge, and whether it is impacted by a cytokine, corresponding to the figure. For further illustration, Figure~\ref{fig:EITwo} shows the number of edges impacted by at least $k = 2$ common cytokines. For instance, the number $59$ in the block (Eotaxin, FGF) indicates that there are $59$ edges impacted by a set of cytokines that includes both Eotaxin and FGF in common. Among these $59$ edges, some may also be impacted by cytokines other than Eotaxin and FGF. The provided table allows for similar analyses for $k > 2$.

\label{lastpage}

\end{document}